\documentclass[notitlepage, fleqn]{revtex4-1}
\usepackage{epsfig}
\usepackage{amsmath}
\usepackage{graphicx}
\usepackage{color}
\usepackage[utf8]{inputenc}
\usepackage{amsfonts,amssymb}
\usepackage[english]{babel}
\usepackage{hyperref}
\usepackage{lipsum}
\usepackage{graphics}
\usepackage{array}
\usepackage{tabu}
\usepackage{bm}
\usepackage{float}
\usepackage{setspace}
\usepackage{mathtools}

\newcommand{\be}{\begin{equation}}
\newcommand{\ee}{\end{equation}}
\newcommand{\ba}{\begin{eqnarray}}
\newcommand{\ea}{\end{eqnarray}}

\newcommand{\non}{\nonumber}
\newcommand{\bra}[1]{\langle #1|}
\newcommand{\ket}[1]{|#1\rangle}

\begin{document}

\title{Many-body quantum thermal machines}

\author{Victor Mukherjee}
\email{mukherjeev@iiserbpr.ac.in}
\affiliation{Department of Physical Sciences, IISER Berhampur, Berhampur 760010, India}

\author{Uma Divakaran}
\email{uma@iitpkd.ac.in}
\affiliation{Department of Physics, Indian Institute of Technology Palakkad, Palakkad, 678557, India}

\begin{abstract}
Thermodynamics of quantum systems and {quantum thermal machines} are rapidly developing fields, which have already delivered several promising results, as well as raised many intriguing questions.  {Many-body quantum machines} present new opportunities stemming from many-body effects. At the same time, they pose new challenges related to many-body physics. In this short review we discuss some of the recent developments on technologies based on many-body quantum systems. We mainly focus on many-body effects in quantum thermal machines. We also briefly address the role played by many-body systems in the development of quantum batteries and quantum probes.

\end{abstract}

\maketitle

\tableofcontents

\section{Introduction}
\label{secintro}

Recent years have witnessed a plethora of theoretical and 
experimental studies of technologies at the microscopic 
scales \cite{kurizki15quantum, millen16the} . 
The laws of quantum mechanics in general play vital roles in 
dictating the behaviors of systems at the atomic 
scales \cite{sakurai17modern}. At the same time, 
the operation of thermal machines is guided by the laws of thermodynamics, as has been well known since  the last two centuries \cite{callen85thermodynamics, kondepudi15modern}. Naturally, studying the operation of {quantum thermal machines} poses the challenge of understanding the laws of thermodynamics in quantum regime, which has in turn led to the burgeoning field of quantum thermodynamics \cite{scovil59three, scully03extracting, gemmer2009quantum, kosloff13quantum, kosloff14quantum, brandao15the, klimovsky15chapter, vinjanampathy16quantum, bera17generalized, masanes17a, binder18book, tuncer20quantum, bhattacharjee20quantum_revw}. The recent remarkable progress in our understanding of the thermodynamics in quantum regime, and in realizing {thermal machines and related technologies} based on quantum systems, have been largely driven by the current expertise in experimentally probing and controlling systems at the microscopic scales \cite{dolde11electric, bason12high, kucksko13nanometre, laskar20observation, pal20experimental}. This advancements in experimental know-how has given us unprecedented ability to devise microscopic machines \cite{koski14experimental, rossnagel16a, klaers17squeezed, Peterson18, klatzow19experimental, maslennikov19quantum, peterson20implementation}, which may be guided by the laws of quantum thermodynamics. 

{One of the major aims of the field of quantum thermodynamics is the development of quantum machines which can prove to be significantly beneficial, when compared to the existing machines based on classical physics. Most of the works done till now in this field have addressed machines based on single or few-body quantum systems. On the other hand, achieving significant advantage through quantum machines in general demands scaling-up of such machines to many-body systems.} This is a highly non-trivial problem, owing to 
the exponentially increasing dimension of the Hilbert space with system size. However, this also gives us the opportunity to understand the fundamental physics of thermodynamics of many-body quantum systems. In order to address this challenge, we need to develop our understanding of the dynamics of many-body systems driven out of equilibrium, and in presence of dissipative thermal and non-thermal baths. Recently several works have addressed this issue. For example, master equations aimed at  studying the dynamics of many-body quantum systems in presence of dissipative environments have been developed \cite{keck17dissipation, xu19many, nathan20universal}. Such studies can help to answer  questions regarding several intriguing many-body effects in presence of dissipative baths, such as topological properties \cite{carmele15stretched, bandyopadhyay20dynamical}, phase transitions \cite{hoyos07effects, grandi10quench, wang20dissipative} and many-body localization \cite{fischer16dynamics, altman20critical, bloch17signatures} to name a few.

There are already several works which give the readers in-depth reviews of different aspects of quantum thermodynamics, principles of quantum thermal machines and related quantum technologies; for example, see Refs. \cite{gemmer2009quantum, kosloff13quantum, kosloff14quantum, klimovsky15chapter, vinjanampathy16quantum, binder18book, tuncer20quantum, bhattacharjee20quantum_revw}. In contrast, in this short non-exhaustive review, we specifically focus on the role played by many-body systems in different emerging quantum technologies. We shall mainly discuss quantum engines and refrigerators based on many-body working mediums (WM), and also briefly address the important role played by many-body systems in the studies of quantum probes and quantum batteries. The many-body effects may arise due to collective coupling between a many-body system and external dissipative baths \cite{niedenzu18cooperative, kloc19collective, latune20collective}, or due to inter-particle interactions in a many-body system \cite{jaramillo16quantum, Cakmak16, herrera17dft, skelt19many, yang19an, yunt19topological, zawadzki20work}. {Several works have focussed on utilizing these many-body effects to design novel quantum thermal machines \cite{jaramillo16quantum, campisi16the, Cakmak16,  niedenzu18cooperative, yang19an, yunt19topological, hartmann19many,  kloc19collective, latune20collective, hartmann20multispin, revathy20universal}, quantum batteries \cite{binder15quantacell, campaioli17enhancing, ferraro18high, campaioli18quantum, le18spin, rossini19quantum, andolina19extractable, andolina19quantum, rossini19many, ito20collectively, cakmak20ergotropy,  carrega20dissipative, crescente20charging} and quantum probes \cite{zanardi08quantum, rams18at, hovhannisyan18measuring, potts19fundamental, mok20optimal,  latune20collective, bayat20_arxiv, bayat21global}.}  In light of the recent rapid progress in experimental studies of quantum systems, one can envisage experimental realizations of such technologies in very near future, in several existing platforms, such as those based on Rydberg atoms \cite{kim18detailed, omran19generation}, ion traps \cite{rossnagel16a}, optical lattices \cite{schreiber15observation} and nitrogen vacancy centers in diamonds \cite{klatzow19experimental}.

This review is organized as follows: in Sec. \ref{sectech}, we present some of the technical details that would be helpful for the readers to follow this review, such as dissipative dynamics in presence of collective coupling in Sec. \ref{secdiscoll}, scaling theory in quantum critical systems in Sec. \ref{secscaling},  dissipative dynamics of free-Fermionic systems in Sec. \ref{secdissff}, shortcut to adiabaticity using counterdiabatic driving in Sec. \ref{sectechSTA} {and work and heat in quantum mechanics in Sec. \ref{secWH}}. Then we discuss  quantum engines based on collective coupling in Sec. \ref{seccoll}; we focus on Otto cycles with collective coupling in Sec. \ref{seccollotto}, while we consider continuous thermal machines in Sec. \ref{seccolcont}. We also discuss collective effects arising due to spin statistics in Sec. \ref{seccolss}. Section \ref{secint} deals with quantum thermal machines based on interacting many-body systems; we discuss the effect of criticality on the operation of quantum thermal machines in  Sec. \ref{seccritical}, shortcut to adiabaticity in many-body quantum engines in Sec. \ref{secsta}, quantum advantage in quantum engines in Sec. \ref{secQA}, quantum engines based on localized states in Sec. \ref{secmbl} and quantum Szilard engines in Sec. \ref{secszilard}. 
Next we briefly discuss  other quantum technologies in 
Sec. \ref{secOQT}, viz. quantum batteries 
in Sec. \ref{secQB} and quantum probes in Sec. \ref{secQP}.  
Finally, we conclude in Sec. \ref{seccon}.

\section{Technical details}
\label{sectech}

{In this review we shall address quantum thermal machines based on different many-body models. These thermal machines have been studied in different regimes, such as close to phase transitions, in presence of localization-delocalization transition, etc. Studying this broad subject requires several techniques related to many-body physics and physics of systems out of equilibrium. A detailed analysis of all the models and the associated techniques is beyond the scope of the current review. However, 
in this section we discuss some of the results and tools which can be useful for studying machines based on many-body quantum systems, which are addressed in this review. }

\subsection{Dissipative dynamics of a many-body quantum system collectively coupled to a bath}
\label{secdiscoll}

{Several recent works on many-body quantum thermal machines have focussed on a many-body system collectively coupled to thermal baths. The popularity of this model stems from the fact that it can be represented in terms of angular momentum operators, thereby simplifying the problem significantly. In addition, the non-trivial physics arising due to super-radiant phenomenon can lead to interesting properties of these quantum thermal machines, including the existence of quantum advantage.
Therefore in order to understand the physics of collective phenomenon in many-body quantum thermal machines discussed in Secs. \ref{seccollotto} and \ref{seccolcont}, let us first delve deeper into the dynamics and heat capacities of quantum systems collectively coupled to thermal baths.} We discuss below the dissipative dynamics of $N \geq 1$ spins; {we consider the system Hamiltonian $H_{\rm S}$ corresponding to the $N$ spins to be given by}
\ba
H_{\rm S} = \sum_{r=1}^N  \omega_r J_z^r,
\label{Hamilwm_coll1}
\ea
where $J_{\alpha}^r$ is the local angular momentum operator associated with the $r$-th spin, along $\alpha = x, y, z$ direction. In this review we take $\hbar$ and $k_{\rm B}$ as unity. Here we focus on the case of indistinguishable spins, brought about by $\omega_r = \omega$~$\forall~r$, in which case \eqref{Hamilwm_coll1} can be written as 
\ba
H_{\rm S} = \omega \mathcal{J}_z,
\label{Hamilwm_coll2}
\ea
where we have used the collective operators $\mathcal{J}_{\alpha} =  \sum_{r=1}^N J_{\alpha}^r$. We consider the system (Eq. \eqref{Hamilwm_coll2}) coupled collectively to a thermal bath through an interaction Hamiltonian given by
\ba
H_{int} = \lambda B\otimes \mathcal{J}_x,
\label{Hamilwm_coll3}
\ea
where $B$ denotes a Hermitian bath operator and $\lambda$ is the system-bath interaction strength. 

The collective operators $\mathcal{J}_{\alpha}$ gives rise to Dicke states $\ket{j, m}$, which are simultaneous eigenstates of $\mathcal{J}^2 = \mathcal{J}_x^2 + \mathcal{J}_y^2 + \mathcal{J}_z^2$ and $\mathcal{J}_z$:
\ba
\mathcal{J}^2\ket{j, m}_i = j(j+1)\ket{j,m}_i;~~~\mathcal{J}_z\ket{j,m}_i = m\ket{j, m}_i.
\label{statesjmi}
\ea
The
collective basis $\{\ket{j, m}_i\}$ satisfies the constraints: $j \in \left[j_0; Ns\right],~m \in \left[-j; j\right],~i \in \left[1;l_j \right]$, where $j_0 = 0$ for $s \geq 1$, while $j_0 = 1/2$ for $N$  odd and $s = 1/2$, $s$ being the dimension of each spin. In Eq. \eqref{statesjmi}, we use $\ket{j,m}_i$ to represent the degenerate eigenstates with eigenvalues $j(j+1)$ and $m$, with the degeneracy index $i$ running from $1$ to $l_j$; the integer $l_j$ denotes the degeneracy of the associated eigenspace (see below)  \cite{breuer02, mandel95optical, latune19thermodynamics}. We note that the largest possible spin $j = Ns$ is unique, 
formed by the totally-symmetric $N$-atom states.
Analogous to single particle operators, one can also define the raising and lowering operators $\mathcal{J}_{+}$ and $\mathcal{J}_{-}$, respectively, through the relations:
\ba
\mathcal{J}^{\pm} &=& \mathcal{J}_x \pm i\mathcal{J}_y \non\\
\mathcal{J}^{\pm}\ket{j, m}_i &=& \sqrt{\left(j \mp m \right)\left(j \pm m +1 \right)}\ket{j, m \pm 1}_i.
\label{eqcollblock}
\ea

In the limit of weak system-bath coupling ($|\lambda| \ll 1$), one can apply Born, Markov and secular approximations to arrive at a Markovian master equation in the interaction picture, given by \cite{latune19thermodynamics, latune20collective}
\ba
\dot{\rho} = \Gamma(\omega)\left(\mathcal{J}^{-} \rho \mathcal{J}^{+} - \mathcal{J}^{+}\mathcal{J}^{-}\rho \right) +  \Gamma(-\omega)\left(\mathcal{J}^{+} \rho \mathcal{J}^{-} - \mathcal{J}^{-}\mathcal{J}^{+}\rho \right) + \text{h.c.},
\label{meq_h}
\ea
where $\Gamma(\omega) = \lambda^2 \int_0^{\infty} \exp\left[i \omega u\right] {\rm Tr} \left(\rho_{\rm B} B(u) B \right) du$. Here $\rho_{\rm B}$ denotes the state of the bath at temperature $T = 1/\beta$, while $B(u) = \exp[i H_{\rm B} u ]B \exp[i H_{\rm B} u]$ is the bath operator in the interaction picture, with respect to the Bath Hamiltonian $H_{\rm B}$. 

As one can see from  Eq. \eqref{eqcollblock},  the raising and lowering operators $\mathcal{J}^{\pm}$ acting on a state $\ket{j, m}_i$ do not change the value of $j$. Consequently, for an initial state devoid of any correlation between different eigenspaces  of $\mathcal{J}^2$, the master equation \eqref{meq_h} keeps the dynamics confined within each $j$ eigenspace. Therefore for each $j$, the dynamics is same as that of the thermalization of a system consisting of  $2j+1$ non-degenerate energy levels.  One can use the master equation \eqref{meq_h} to arrive at dynamics of the populations $\rho_{j, m, i} = {_i}\bra{j, m}\rho\ket{j, m}_i$: 
\ba
\dot{\rho}_{j, m, i} &=& G(\omega)\left[(j-m)(j+m+1)\rho_{j, m+1, i} -  (j+m)(j-m+1)\rho_{j, m, i}\right] \non\\ &+& G(-\omega)\left[(j+m)(j-m+1)\rho_{j, m-1, i} -  (j-m)(j+m+1)\rho_{j, m, i}\right],
\label{eqmecol}
\ea
where $G(\omega) = \Gamma(\omega) + \Gamma^{*}(\omega)$. {We note that the coefficients in front of $\rho_{j, m, i}$ and $\rho_{j, m \pm 1, i}$ on the r.h.s. of Eq. (7) are of the order of $j$ for $m \approx \pm j$, while  these coefficients scale as $\sim j^2$ for $m \approx 0$ \cite{kloc19collective}.}

The system reaches a steady-state $\rho^{\rm ss}(\beta)$, defined by $\dot{\rho} = 0$ for $\rho = \rho^{\rm ss}(\beta)$, at long times. This steady-state  may not be a thermal state. For an initial  state satisfying the constraint
${_i}\langle j, m|\rho_0|j,m\rangle_{i^{\prime}} = 0$ for $i \neq i^{^{\prime}}$, we arrive at the steady-state \cite{latune20collective}
\ba
\rho_N^{\rm ss}(\beta) = \sum_{j = J_0}^{Ns}\sum_{i = 1}^{l_j} p_{j,i}\rho_{j, i}^{\rm th}(\beta).
\label{rhosscolN1}
\ea
{The steady-state \eqref{rhosscolN1} depends on the initial state $\rho_0$ through the probabilities 
\ba
p_{j,i} = \sum_{m = -j}^j {_i}\langle j, m|\rho_0|j,m\rangle_i,
\label{eqmultiplicity}
\ea
corresponding to an eigenspace of total spin $j$, where $\sum_{j=j_0}^{ns} \sum_{i=1}^{l_j} p_{j,i} = 1$ \cite{latune19thermodynamics}. We note that the master equation \eqref{eqmecol} does not mix states with $i \neq i^{\prime}$, such that $p_{j,i}$ are time-independent $\forall~j, i$. This leads to the steady-state \eqref{rhosscolN1} with
\ba
\rho_{j,i}^{\rm th}(\beta) = Z_j(\beta)^{-1} \sum_{m = -j}^{j} e^{-m   \omega \beta} \ket{j, m}_i {_i}\bra{j,m};~~~~~Z_j(\beta) = \sum_{m = -j}^{j} e^{-m   \omega \beta}.
\label{rhosscolN2}
\ea}
In contrast to the collective coupling scenario discussed above, $N$ such spins interacting independently {with} the thermal bath reaches the direct-product thermal steady state
{\ba
\rho^{\rm th}_{\rm ind} = \frac{e^{-   \omega \beta \mathcal{J}_z }}{{\rm Tr} e^{-  \omega \beta \mathcal{J}_z}} = \otimes_{r = 1}^N  \rho^{\rm th}_{{\rm ind},r}.
\label{rhossind1}
\ea
Here $ \rho^{\rm th}_{{\rm ind},r}$ is the Gibbs state reached by the $r$-th spin, which evolves in presence of the local Hamiltonian  $\omega J_z^r$ (see Eqs. \eqref{Hamilwm_coll1} and \eqref{Hamilwm_coll2}), and is given
 by \cite{greiner95thermodynamics, breuer02, latune19thermodynamics, latune20collective} 
\ba
\rho^{\rm th}_{{\rm ind}, r} = \frac{e^{-  \omega \beta J_z^r}}{\sum_{m=-s}^s e^{-m  \omega \beta}}.
\label{rhossind2}
\ea
The direct-product thermal steady state $\rho^{\rm th}_{\rm ind}$ is clearly different from the steady-state (Eqs. \eqref{rhosscolN1} - \eqref{rhosscolN2}) reached through collective coupling.}

{The difference in the steady-states reached due to collective coupling (Eqs. \eqref{rhosscolN1} - \eqref{rhosscolN2}) and independent coupling (Eqs. \eqref{rhossind1} - \eqref{rhossind2}) can lead to crucial differences in the behaviors of the systems, including in their heat capacities. As we shall show below in Sec. \ref{seccollotto}, heat capacities can play important roles in the operation of many-body heat engines \cite{latune20collective}. Further, previous studies have shown the importance of heat capacity in the context of critical heat engines \cite{campisi16the}. Therefore let us now briefly discuss the effect of collective and independent system-bath couplings on the corresponding heat capacities.

The heat capacity $C = \partial E/\partial T = -\beta^2 \partial E/\partial \beta$ of a system in a thermal state  $\rho$ at temperature $T$ quantifies the change in its mean energy $E = {\rm Tr}\left[H_{\rm S} \rho\right]$ as a function of the change in its temperature.}
The collective heat capacity $C^{\rm col}(\beta)$ of the spin ensemble in the steady state \eqref{rhosscolN1} is given by 
\ba
C^{\rm col}(\beta) = -\beta^2 \frac{\partial E^{ss}(\beta)}{\partial \beta} = \sum_{j = j_0}^{ns}\sum_{i=1}^{l_j} p_{j,i}C_j(\beta),
\label{spheatcol1}
\ea
where the steady-state energy $E^{ss}(\beta)$ of the spin ensemble is 
\ba
E^{ss}(\beta) = {\rm Tr}\left[H_{\rm S} \rho^{ss}_N(\beta)\right] =  \omega {\rm Tr}\left[\mathcal{J}_z \rho^{ss}_N(\beta)\right] = \sum_{j = j_0}^{ns}\sum_{i=1}^{l_j} p_{j,i} e_j(\beta);~~~
 e_j(\beta) =   \omega \sum_{m=-j}^j \frac{me^{-m  \omega \beta}}{Z_j(\beta)},
\label{Ess}
\ea
and 
\ba
C_j(\beta) = - \beta^2  \frac{\partial e_j (\beta)}{\partial \beta} =  (  \omega \beta)^2
 \left[\left(\frac{1}{2\sinh(  \omega \beta/2)} \right)^2 - \left(\frac{j+1/2}{\sinh(j+1/2)  \omega \beta} \right)^2 \right].
 \label{spheatcol2}
\ea
One can use Eqs. \eqref{spheatcol1} and \eqref{spheatcol2} to show that the largest heat capacity $C_+^{\rm col}(\beta) = C_{j=ns}(\beta)$ is obtained for $p_{j=ns} = 1$. {Here we note that the degeneracy $l_{j}$ equals unity for $j = ns$ \cite{mandel95optical, latune19thermodynamics}.} 

{Similarly, the
independent heat capacity $C^{\rm ind}(\beta)$, given by:
\ba
C^{\rm ind}(\beta) = - \beta^2 \frac{\partial E^{\rm th}(\beta)}{\partial \beta} = N C_{j = s}(\beta),
\label{spheatind}
\ea 
where 
\ba
E^{\rm th}(\beta) =   \omega {\rm Tr}(\mathcal{J}_z \rho^{\rm th}_{\rm ind} (\beta)) = n e_{J = s}(\beta).
\label{eth}
\ea

In the following we compare
the best case scenario, $C_+^{\rm col}(\beta)$ to the independent heat capacity $C^{\rm ind}(\beta)$.
Using Eqs. \eqref{spheatcol1} - \eqref{spheatind}, one can show that \cite{latune20collective}}
\ba
\lim_{  \omega |\beta| \gg 1}\frac{C_{+}^{\rm col}(\beta)}{C^{\rm ind}(\beta)} \sim  N^{-1},
\ea
while 
\ba
\lim_{  \omega |\beta| \ll 1}\frac{C_{+}^{\rm col}(\beta)}{C^{\rm ind}(\beta)} \sim  \frac{Ns + 1}{s + 1} + \mathcal{O}\left[N(N   \omega \beta)^2 \right].
\ea
The collective $C_+^{\rm col}(\beta_{cr} = 1/ T_{cr})$ becomes equal to the independent heat capacity $C^{\rm ind}(\beta_{cr} = 1/ T_{cr})$ at a critical bath temperature $T_{cr}$, given by  
\ba
\frac{ T_{cr}(n, s)}{  \omega} \simeq \left(\frac{4Ns(s+1) + 1}{12} \right)^{1/2}.
\label{TNcr}
\ea

\subsection{Scaling theory in critical systems}
\label{secscaling}

Till now we have considered many-body effects arising due to the collective coupling between a many-body system and a thermal bath. However, many-body effects can also arise due to the presence of interactions between the subsystems of a many-body system. An intriguing phenomenon arising in such interacting many-body systems, is that of phase transitions. Phase transitions are associated with phases characterized by different symmetries separated by critical points; classical phase transitions occur due to thermal fluctuations at non-zero temperatures \cite{chaikin95principles}, while quantum phase transitions result from quantum fluctuations at absolute zero temperature \cite{sachdev99quantum}.
{As discussed in Sec. \ref{seccritical}}, phase transitions have generated significant interest in the field of quantum thermodynamics, owing to the divergences of length and time scales 
 close to critical points \cite{campisi16the, ma17quantum, fadaie18topological, chand18critical, revathy20universal, fogarty20a}. These divergences in turn lead to the presence of universal features, through the general scaling relations described below.
Such a universality is also reflected in the non-equilibrium
dynamics arising due to the dynamics of systems driven through quantum phase transitions, which will be discussed in this section \cite{dziarmaga10dynamics, dutta15quantum}.

Quantum phase transitions are zero temperature phase transitions
where the nature of the ground state changes abruptly at a quantum critical point 
(QCP), due to change in some parameter $g$ characterizing the Hamiltonian 
of the system \cite{sachdev99quantum}.
The QCP can also be idenitified by
the vanishing of the gap between the ground state and the
first excited state at the QCP 
{$g = g_{\rm {cr}}$}. 
It can be shown that for a second order
quantum phase transition, the correlation length $\xi$
diverges following a power law
\begin{eqnarray}
\xi \sim |g-g_{\rm {cr}}|^{-\nu}
\label{eq_xi}
\end{eqnarray}
when the critical point $g_{\rm{cr}}$
is approached.
Similarly, the correlation time $\xi_{\tau}$ also diverges
with a power law as follows:
\begin{eqnarray}
\xi_{\tau} \sim |g-g_{\rm{cr}}|^{-\nu z}.
\end{eqnarray}
Here, $\nu$ and $z$ are the correlation length and dynamical
critical exponents associated with the
critical point.

In the last two decades, several works have addressed the non-equilibrium dynamics across a
QCP \cite{dziarmaga10dynamics, polkovnikov11colloquium, dutta15quantum}.
In particular, researchers started working on
understanding the effect of critical point when a system, initally
prepared in the ground state, is driven across a critical
point $g_{\rm{cr}}$ by varying $g$ linearly with speed $v$
\cite{zurek05dynamics,polkovnikov05universal}.
When the system
is far away from the critical point ($g\gg g_{\rm{cr}}$),
the correlation time
is small compared to the time scale in which the Hamiltonian
is varied. This enables the system to follow the instantaneous
ground state.
As soon as the relaxation time becomes comparable to this scale,
the system is no longer able to follow the instantaneous
ground state and gets excited. The universal relation connecting
density of excitations $n$  to the speed $v$
with which the Hamiltonian is varied and the critical exponents
associated with the QCP crossed is known
as Kibble Zurek scaling, and is given by
\begin{eqnarray}
n \sim v^{\frac{\nu d}{\nu z +1}}.
\end{eqnarray}
where $d$ is the dimensionality of the system. 
As expected, 
{the density of excitations $n$} decreases when the speed decreases.
{These excitations, also called defects, could be related to
the density of quasiparticles generated.
}
 The non-adiabatic excitations generated due to the dynamics 
close to QCP results in non-zero excitation energy 
$\mathcal{E}_{\rm ex}$ as well, i.e., the energy of the system 
in excess to its ground state energy. For quenches ending at 
the QCP, we have \cite{grandi10quench, fei2020work}
\ba
\mathcal{E}_{\rm ex} \sim v^{\frac{\nu (d+z)}{\nu z +1}}.
\ea
On the other hand, for quenches across the critical point, in general $\mathcal{E}_{\rm ex}$ does not follow universal scaling form. However, for systems in which the excitation energy is proportional to the density of defects, such as in the one-dimensional transverse Ising model discussed below, we have 
\ba
\mathcal{E}_{\rm ex} \sim v^{\frac{\nu d}{\nu z +1}}.
\ea
The universality seen in the non-equilibrium dynamics attracted
lot of attention of the scientists opening a plethora of papers
in the related subject
\cite{dziarmaga10dynamics,polkovnikov11colloquium,dutta15quantum,
zurek05dynamics,polkovnikov05universal,grandi10quench,fei2020work}
.

A prototypical model studied to exemplify critical
phenomena in quantum systems is transverse Ising model
given by
\begin{eqnarray}
H(t)=-J \sum_{i=1}^N \sigma_i^x \sigma_{i+1}^x - h(t) \sum_i \sigma_i^z.
\label{eq_tim}
\end{eqnarray}
Here $J$ denotes the interaction strength between any two nearest-neighbor spins, $h(t)$ is a uniform time-dependent magnetic field along the transverse ($z$) direction and $\sigma_i^{\alpha}$ denotes the Pauli matrix along $\alpha = x, y, z$ axis, corresponding to the spin at site $i$.
The excitation spectrum of this Hamiltonian can be obtained by
mapping the Pauli matrices to Jordan Wigner (JW) fermions $c_i$
\cite{lieb61two,pfeuty70the,bunder99effect}, where
the transformation equation is given by
\begin{eqnarray}
\sigma_i^-= \big(e^{i\pi \sum_{j<i}c_j^{\dagger}c_j}\big) c_i
\label{eq_JW1}
\end{eqnarray}
with $\sigma_i^-= (\sigma_i^x - \i \sigma_i^y$)/2.
Rewriting the above Hamiltonian in terms of JW fermions, we get
\begin{eqnarray}
H=-\sum_{i=1}^N J (c_i^{\dagger}c_{i+1}-c_i c_{i+1}^{\dagger})
+ J (c_i^{\dagger}c_{i+1}^{\dagger} - c_i c_{i+1})
+ h (c_i^{\dagger}c_i - c_i c_i^{\dagger}).
\label{eq_fermion}
\end{eqnarray}
In order to obtain the excitation spectrum, we re-write the
Hamiltonian in the Fourier space 
where $c_k$ is the Fourier
component of $c_i$, and is defined as
\begin{eqnarray}
c_k=\frac{1}{\sqrt{N}}\sum_j c_j e^{-ikj},
\end{eqnarray}
so that the Hamiltonian can be written as
(upto some constants)
\begin{eqnarray}
H=-\sum_k 2(h+ J\cos k)c_k^{\dagger}c_k
+ \sin k (c_k^{\dagger}c_{-k}^{\dagger} + c_k c_{-k})
\end{eqnarray}
The above Hamiltonian can now be diagonalized using
Bogoliubov rotation \cite{bunder99effect},
and the  excitation spectrum is given by
$\epsilon_k$=$2\sqrt{(h+ J\cos k)^2 + \sin^2k}$.
As mentioned before, the critical point is given by the
vanishing of the excitation spectrum  which happens
at $h=\pm J$ for $k=0,\pi$. It can further be shown that
$\nu=z=1$ for this model, which eventually leads to $n \sim v^{0.5}$ \cite{dziarmaga05dynamics}.

\subsection{Dissipative dynamics in free-Fermionic systems}
\label{secdissff}

\begin{figure}[h]
\includegraphics[width=0.51\columnwidth]{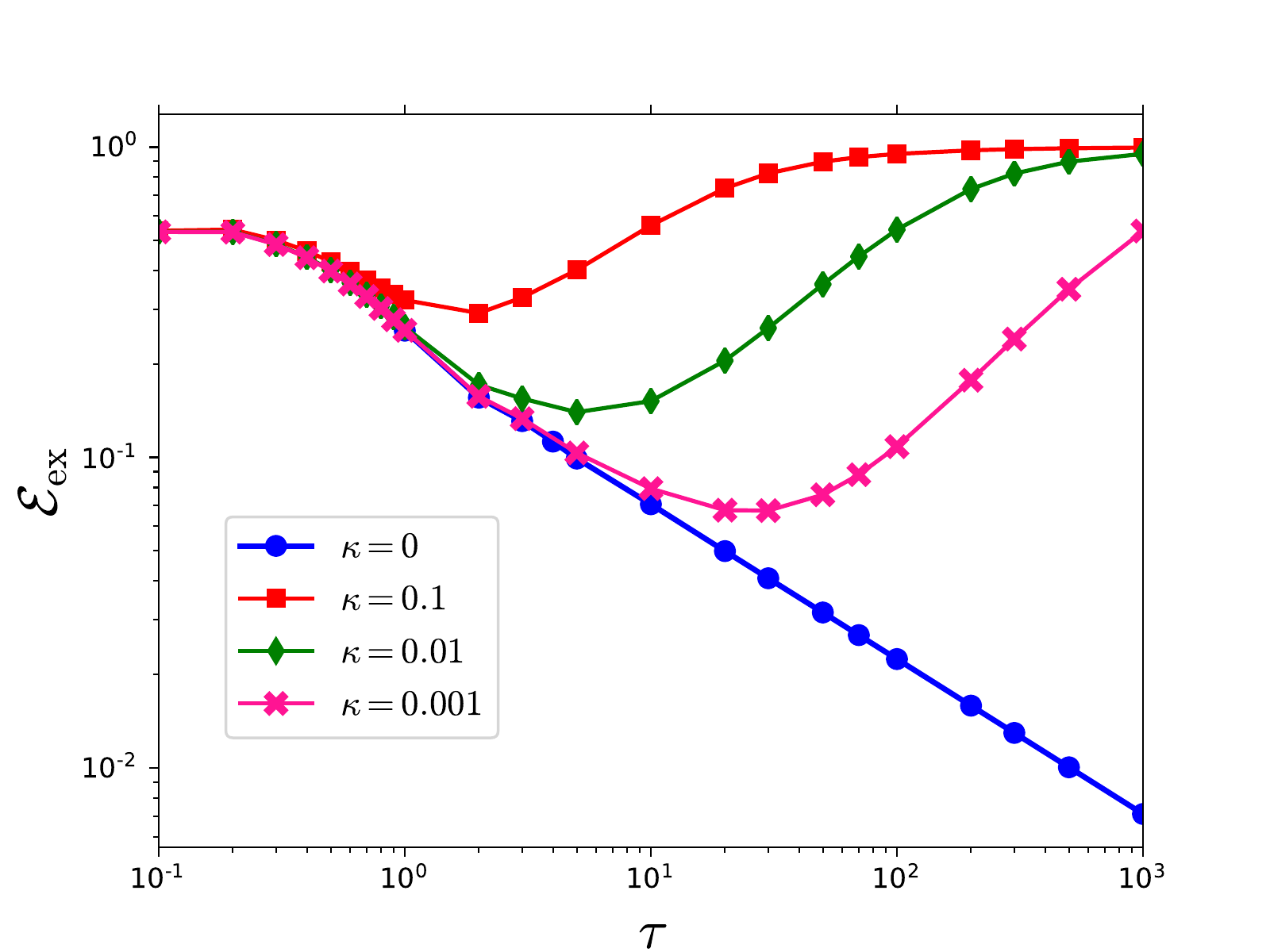}
\caption{The excitation energy $\mathcal{E}_{\rm ex}$ as a function of $\tau$.
The data with solid circles correspond to $\kappa=0$ case
(or no bath case) which shows
KZ scaling given by $\tau^{-1/2}$ with $\nu=1,~z=1$
for transverse Ising model. As $\kappa$
is increased (Cf, Eq. \eqref{eqFFdis}), more defects are generated
for large $\tau$ resulting in a minimum in the
$\mathcal{E}_{\rm ex}-\tau$ graph .
Reproduced from Ref. \cite{keck17dissipation}.
}
\label{fig_keck}
\end{figure}
Continuing the discussion in the previous section, we
now consider the dynamics of many-body 
free-Fermionic systems similar to that given in
Eq. \ref{eq_fermion},
but now in presence of a dissipative bath \cite{keck17dissipation}. Such an analysis
is important from the perspective of various many-body quantum technologies, such as adiabatic quantum
computation and quantum annealing where the interaction of
the time dependent Hamiltonian
with the environment is unavoidable due to longer time
scales involved, resulting in dissipation and decoherence.
{Also as we discuss in Sec. \ref{seccritical} below, dynamics of free Fermionic systems in presence of dissipative environment can be crucial for studying quantum engines operated close to quantum phase transitions. It is well known that many spin dependent Hamiltonians, such as
transverse Ising model, Kitaev model, XY model in presence
of transverse field, etc.
can be written in terms of fermions \cite{lieb61two, pfeuty70the, sengupta08exact, dutta15quantum}.
The most general time dependent form of such Hamiltonians
having the quadratic form in terms of fermions can be written as:}

\begin{eqnarray}
H(t)=\sum_{m,n}[
c_m^{\dagger} A_{m,n}(t)c_n + \frac{1}{2}
(c_m^{\dagger}B_{m,n}c_n^{\dagger} + \rm{hermitian~ conjugate})
]
\label{eq_freefermion}
\end{eqnarray}
where {$A_{m,n},~ B_{m,n}$} are symmetric 
and antisymmetric matrices,
respectively, with elements depending upon the parameters
of the original Hamiltonian. 

As expected, the general equation for evolution of the
density matrix 
{of such a system when coupled to an environment}
would include the unitary evolution as well as
a term involving system-bath coupling or the dissipative term
$\mathcal{D}(\rho)$:

\begin{eqnarray}
\frac {\partial \rho}{\partial t}
= -i[H(t),\rho] + \mathcal{D}(\rho)
\label{eqFFdis0}
\end{eqnarray}

We now discuss the effect of presence of a critical point
in the time evolution of
systems which are  coupled to
a bath. For simplicity the bath used is Markovian bath
where the dissipation term can be written in Lindblad form
\begin{eqnarray}
\mathcal{D}[L_n](\rho)=\sum_n \kappa_n (L_n \rho L_n^{\dagger}
-\frac{1}{2}\{\rho,L_n^{\dagger}L_n\})
\label{eqFFdis}
\end{eqnarray}
where $L_n$ are  local Lindblad operators describing the
environment, and $\kappa_n$ are the site dependent coefficients related to system-bath coupling strength \cite{breuer02}.

In Ref. \cite{keck17dissipation}, the authors
have considered three different types of Lindblad operators, namely,
(i)$L_n= c_n^{\dagger}$,
(ii) $L_n=c_n $,
(iii) $L_n=c_n^{\dagger}c_n$.
Here, we present the results corresponding to the first type of
bath with $L_n=c_n^{\dagger}$. 
Such a bath helps in studying the competition
between unitary dynamics and baths in an almost exact way.
As mentioned before, the competition arises because
defects due to Kibble Zurek will decrease when the quench time scale $\tau$
is increased. But this will cause the system to interact
for more time with the environment resulting in increased
defect generation due to environment.
One must also note that such a bath need not take the system
to a thermalized state, but to some steady state.
The example Hamiltonian used to demonstrate these competitions
is that of transverse Ising model given in Eq. \ref{eq_tim},
where the transverse field $h$ is varied as
$t/\tau$, so that the speed $v$ with which the quantum
critical point $h=J$ is crossed, is given by $1/\tau$.
{Eq. \ref{eqFFdis0} is integrated numerically
starting from the ground state of the initial Hamiltonian.
The quantity studied in Fig. \ref{fig_keck} is the excitation energy $\mathcal{E}_{\rm ex}$, which is the difference
between the energy at the final state reached after the
time evolution, given by $Tr[H\rho]$, and the ground state 
of the final Hamiltonian.
} 
For simplicity, $\kappa_n$ in
the dissipative term is taken as a constant $\kappa$
independent of site index $n$.
As seen in the figure and explained in the text above,
there is a non-monotonic behavior in the defects generated
with an optimal value of $\tau$ for which $\mathcal{E}_{\rm ex}$
takes its minimum value. Different scalings have been explained
assuming that the defects generated during the unitary evolution
and while interacting with the environment are unrelated,
i.e., the total defects created are sum of those due to
unitary evolution and interaction with the environment.

\subsection{Counterdiabatic driving}
\label{sectechSTA}

The performance of quantum thermal machines is quantified by efficiency and output power in case of engines, and rate of refrigeration in case of refrigerators. As has been known for classical as well as quantum thermal machines, the maximum efficiency is bounded by the Carnot limit, through the second law of thermodynamics \cite{callen85thermodynamics, kondepudi15modern, chenu18thermodynamics}. In general, such efficiencies are reached only in the absence of non-adiabatic excitations,  i.e., for long cycle time limit. However, in the absence of any control, such high efficiencies are achieved at the price of vanishing output power  or  refrigeration rate, which are inversely proportional to the total cycle period. Consequently,  application of shortcuts to adiabaticity (STA) to devise control protocols aimed at enhancing the power, refrigeration rate and efficiency in finite-time thermal machines through suppression of non-adiabatic excitations, has gained a lot of attention lately.  
STA has been developed and studied thoroughly in the context of closed  \cite{demirplak03adiabatic, berry09transitionless, campo12assisted, deffner14classical, patra17shortcuts, odelin19shortcuts, xu20effects, patra21semiclassical} and open \cite{vacanti14transitionless, alipour20shortcuts} quantum and classical systems in presence of time-dependent Hamiltonians, and has also been implemented experimentally \cite{ bason12high, an16shortcuts}. In the last few years, it has also been extended to the field of quantum thermodynamics  \cite{delcampo14more}, and proved to be immensely successful in enhancing the performance of a wide class of quantum thermal machines \cite{cakmak19spin, li18an, abah20shortcut}.
Application of STA to enhance the performance of quantum thermal machines have been accompanied by several interesting question regarding the cost of implementation of such control protocols  as well 
{(see Sec. \ref{secsta})} \cite{campbell17trade, abah18performance, abah19shortcut}.

Scaling-up of quantum technologies demands analysis of many-particle quantum machines. However, the exponentially increasing size of the Hilbert space significantly increases the complexity of the problem. This increased complexity is reflected in the application of STA in many-body quantum systems subjected to time-dependent Hamiltonians as well. For example, finding the exact STA protocol  may involve the knowledge of the complete energy spectrum of a system, which in general can be highly non-trivial for interacting many-body systems \cite{campo12assisted, mukherjee16local}. This issue can be tackled through the method of approximate counterdiabatic driving \cite{sels17minimizing, kolodrubetz17geometry, claeys19floquet},  which does not require explicit knowledge of the many-body eigenstates. Recently, STA through counterdiabatic driving has proved to be highly beneficial for enhancing the performance of many-body quantum thermal machines as well \cite{beau16scaling, hartmann19many, hartmann20multispin}.

Here following Refs. \cite{sels17minimizing} and \cite{kolodrubetz17geometry}, we discuss the derivation of the approximate gauge potential aimed at constructing the corresponding counderdiabatic Hamiltonian, which can completely eliminate, or significantly reduce, non-adiabatic excitations arising due to finite rate of change of the original Hamiltonian. To this end, we consider a state $\ket{\psi(t)}$ evolving under a time-dependent Hamiltonian $H_0(\vartheta(t))$ following the Schr\"{o}dinger equation:
\ba
i  \dfrac{\partial }{\partial t} \ket{\psi(t)} = H_0(\vartheta(t))\ket{\psi(t)}.
\label{eqSTA1}
\ea
Here the parameter $\vartheta(t)$ introduces time-dependence in the Hamiltonian. {We consider a frame rotating via a $\vartheta(t)$ dependent unitary transformation $U(\vartheta(t))$, such that the unitarily rotated Hamiltonian  $\tilde{H}_0(\vartheta(t)) = U^{\dagger}H_0(\vartheta(t))U$ is diagonal at all times. In  this rotating frame Eq. \eqref{eqSTA1} can be written as 
\ba
i  \dfrac{\partial }{\partial t}\left(U \ket{\tilde{\psi}(t)}\right) &=& H_0(\vartheta(t))U \ket{\tilde{\psi}(t)} \non\\
\implies i  \dfrac{\partial U }{\partial t}\ket{\tilde{\psi}(t)} + i  U\dfrac{\partial }{\partial t}\ket{\tilde{\psi}(t)} &=& H_0(\vartheta(t))U \ket{\tilde{\psi}(t)},
\label{eqSTA1a}
\ea
where
\ba
 \ket{\tilde{\psi}(t)} &=& U^{\dagger} \ket{\psi}.
\label{eqSTA3}
\ea
Therefore multiplying both sides of Eq. \eqref{eqSTA1a} by $U^{\dagger}$, we get
\ba
i  \dfrac{\partial}{\partial t}\ket{\tilde{\psi}(t)} &=& \tilde{H}_m \ket{\tilde{\psi}(t)}\non\\
\tilde{H}_m &=& \tilde{H}_0(\vartheta(t)) - \dot{\vartheta}\tilde{\mathcal{A}}_{\vartheta}.
\label{eqSTA2}
\ea}
The adiabatic gauge potential $\tilde{\mathcal{A}}_{\vartheta}$ is given by $\tilde{\mathcal{A}}_{\vartheta} = U^{\dagger} \mathcal{A}_{\vartheta} U$, where $\mathcal{A}_{\vartheta} = i  \partial_{\vartheta}$. According to the construction  above,
$\tilde{H}_0(\vartheta(t))$ is diagonal at all times. Consequently, any non-adiabatic excitation arises due to the term $\dot{\vartheta}\tilde{\mathcal{A}}_{\vartheta}$  in Eq. \eqref{eqSTA2}. Therefore in order to eliminate the non-adiabatic excitations, it suffices to add $\dot{\vartheta} \mathcal{A}_{\vartheta}$ to the original Hamiltonian $H_0(\vartheta(t))$, such that we finally arrive at the STA Hamiltonian
\ba
H_{\rm STA} = H_0 + H_{\rm CD},
\ea
where the counterdiabatic Hamiltonian is given by
\ba
H_{\rm CD} =  \dot{\vartheta} \mathcal{A}_{\vartheta}.
\label{eqHCD}
\ea
One can find the exact counterdiabatic
Hamiltonian, in terms of the instantaneous eigenbasis $\ket{m}$ and the respective eigenvalues $E_m$, through the relation
\ba
\bra{m} \mathcal{A}_{\vartheta} \ket{n} =  i   \frac{\bra{m}\partial_{\vartheta} H_0 \ket{n}}{E_n - E_m};~~~m\neq n.
\label{eqSTA4}
\ea
Equation \eqref{eqSTA4}  requires complete knowledge of the instantaneous eigenstates at all times, and therefore can be impractical to implement, specially in many-body systems. Therefore we aim to find an approximate solution $\mathcal{A}^*_{\vartheta}$ to $\mathcal{A}_{\vartheta}$, which would reduce the non-adiabatic excitations significantly, while being implementable in experimental setups. The choice of the specific form of $\mathcal{A}^*_{\vartheta}$ depends on the constraints involved, such as the range of interactions allowed in the control terms.  Accordingly, we define the operator $M_{\vartheta}$ through the relation
\ba
i  \partial_{\vartheta} H_0 = \left[\mathcal{A}_{\vartheta}, H_0 \right] - i  M_{\vartheta},
\label{eqSTA5}
\ea
such that 
\ba
M_{\vartheta} = -\sum_n \dfrac{\partial E_n (\vartheta)}{\partial \vartheta } \ket{n(\vartheta)}\bra{n(\vartheta)}.
\label{eqSTA6}
\ea
Now let us define the Hermitian
operator
\ba
G_{\vartheta}\left(\mathcal{A}_{\vartheta}^{*} \right) = \partial_{\vartheta} H_0 + i\left[\mathcal{A}_{\vartheta}^{*}, H_0 \right].
\label{eqSTA7}
\ea
As one can see from Eqs. \eqref{eqSTA5} and \eqref{eqSTA7}, $G_{\vartheta}\left(\mathcal{A}_{\vartheta} \right) = -M_{\vartheta}$. Next we employ the variational principle method; 
instead of solving for $\mathcal{A}_{\vartheta}$ directly, which would require detailed knowledge of the spectrum (see Eq. \eqref{eqSTA4}), we minimize the operator distance 
\ba
D^2(\mathcal{A}_{\vartheta}^{*}) = {\rm Tr}\left[\left(G_{\vartheta}(\mathcal{A}_{\vartheta}^{*}) + M_{\vartheta} \right)^2\right]
\label{eqSTA8}
\ea
between $G_{\vartheta}\left(\mathcal{A}_{\vartheta}^{*} \right)$ and $-M_{\vartheta}$, with respect to the parameter $\mathcal{A}_{\vartheta}^{*}$. Clearly, $D^2(\mathcal{A}_{\vartheta}^{*}) $ assumes the minimal value (zero) for $\mathcal{A}_{\vartheta}^{*} = \mathcal{A}_{\vartheta}$.
{As shown in Ref. \cite{kolodrubetz17geometry}}, minimizing this operator distance $D^2(\mathcal{A}_{\vartheta}^{*})$ is equivalent to minimizing
the term
\ba
\mathcal{S}(\mathcal{A}_{\vartheta}^*) = {\rm Tr}\left[G_{\vartheta}^2 (\mathcal{A}_{\vartheta}^*) \right],
\label{eqSTA9}
\ea
 i.e., finding the solution for the equation
\ba
\dfrac{\delta \mathcal{S}(\mathcal{A}_{\vartheta}^*)}{\delta \mathcal{A}_{\vartheta}^*}  = 0,
\label{eqSTA10}
\ea
where $\delta$ denotes the functional derivative. Therefore to summarize, finding an approximate counterdiabatic Hamiltonian (see Eq. \eqref{eqHCD}) boils down to finding the solution to the above equation \eqref{eqSTA10}.

\subsection{Work and heat}
\label{secWH}

A discussion about thermodynamics of quantum systems and quantum thermal machines (see Secs.\ref{seccoll} and \ref{secint}) necessitates the introduction of work and heat in quantum mechanics. Over the years, several related definitions of work and heat have been proposed \cite{alicki79the, vinjanampathy16quantum, alipour19unambiguous}. In this review, we shall use the definitions proposed in Ref. \cite{alicki79the}; we consider a system undergoing Markovian dynamics in presence of a thermal bath, and being driven by a Hamiltonian slowly changing in time. The change in energy $\Delta E$ of the system is given by 
\ba
\Delta E(t) = \int_0^t \frac{d}{dt^{\prime}} {\rm Tr}\left[ \rho(t^{\prime}) H(t^{\prime})\right]dt^{\prime} = \int_0^t {\rm Tr}[\rho(t^{\prime}) \dot{H}(t^{\prime})] dt^{\prime} + \int_0^t {\rm Tr}[\dot{\rho}(t^{\prime}) H(t^{\prime})] dt.
\label{eqWH}
\ea
Here $\rho(t)$ denotes the state of the system at time $t$. For a Hamiltonian slowly changing in time, one can relate the  first term on the r.h.s. of Eq. \eqref{eqWH} as the work $W(t)$ \cite{alicki79the}:
\ba
W(t) = \int_0^t {\rm Tr}[\rho(t^{\prime}) \dot{H}(t^{\prime})] dt^{\prime},
\label{eqW}
\ea 
while the heat flow $\mathcal{Q}(t)$ is given by
\ba
\mathcal{Q}(t) =  \int_0^t {\rm Tr}[\dot{\rho}(t^{\prime}) H(t^{\prime})] dt^{\prime}.
\label{eqQ}
\ea
In case of a system coupled to a thermal bath and in presence of a time-independent Hamiltonian $H_0$, such as during a non-unitary stroke of an Otto cycle (see Sec. \ref{seccollotto}), the work $W(t)$ is zero (see Eq. \eqref{eqW}). In this case the total energy change $\Delta E(t)$ of the system is due to the heat flow $\mathcal{Q}(t)$ between the system and the thermal bath, such that
\ba
\mathcal{Q}(t) = {\rm Tr}[\rho(t)H_0] - {\rm Tr}[\rho(0)H_0] = \Delta E(t).
\label{eqQotto}
\ea

We note that the presence of non-Markovian dynamics \cite{whitney18non}, or a non-thermal bath \cite{niedenzu18quantum}, may non-trivially affect the expressions of heat and work.

\section{Thermal machines with collective coupling}
\label{seccoll}

As with classical thermodynamics where studies on thermal machines 
such as heat engines and refrigerators \cite{scovil59three, scully03extracting, quan07quantum, cleuren12cooling, kolar12quantum, rossnagel14nanoscale, uzdin15equivalence, watanabe17quantum, friedenberger17when, paz17fundamental, niedenzu18quantum, ghosh18two, ghosh19are, erdman19maximum, mukherjee20anti} are inherenctly connected with fundamental principles of thermodynamics \cite{brandao15the, bera17generalized, masanes17a}, the corresponding quantum regime is no different. We shall now use the technical details presented above to discuss specific examples of some of the many-body quantum machines studied in the literature.
In this section we shall focus on many-body quantum machines where non-trivial co-operative effects arise due to collective coupling between the many-body WM and the thermal baths.

\subsection{Collective effects in Otto engines}
\label{seccollotto}

\begin{figure}[h]
\begin{center}
\includegraphics[width = 0.51\columnwidth]{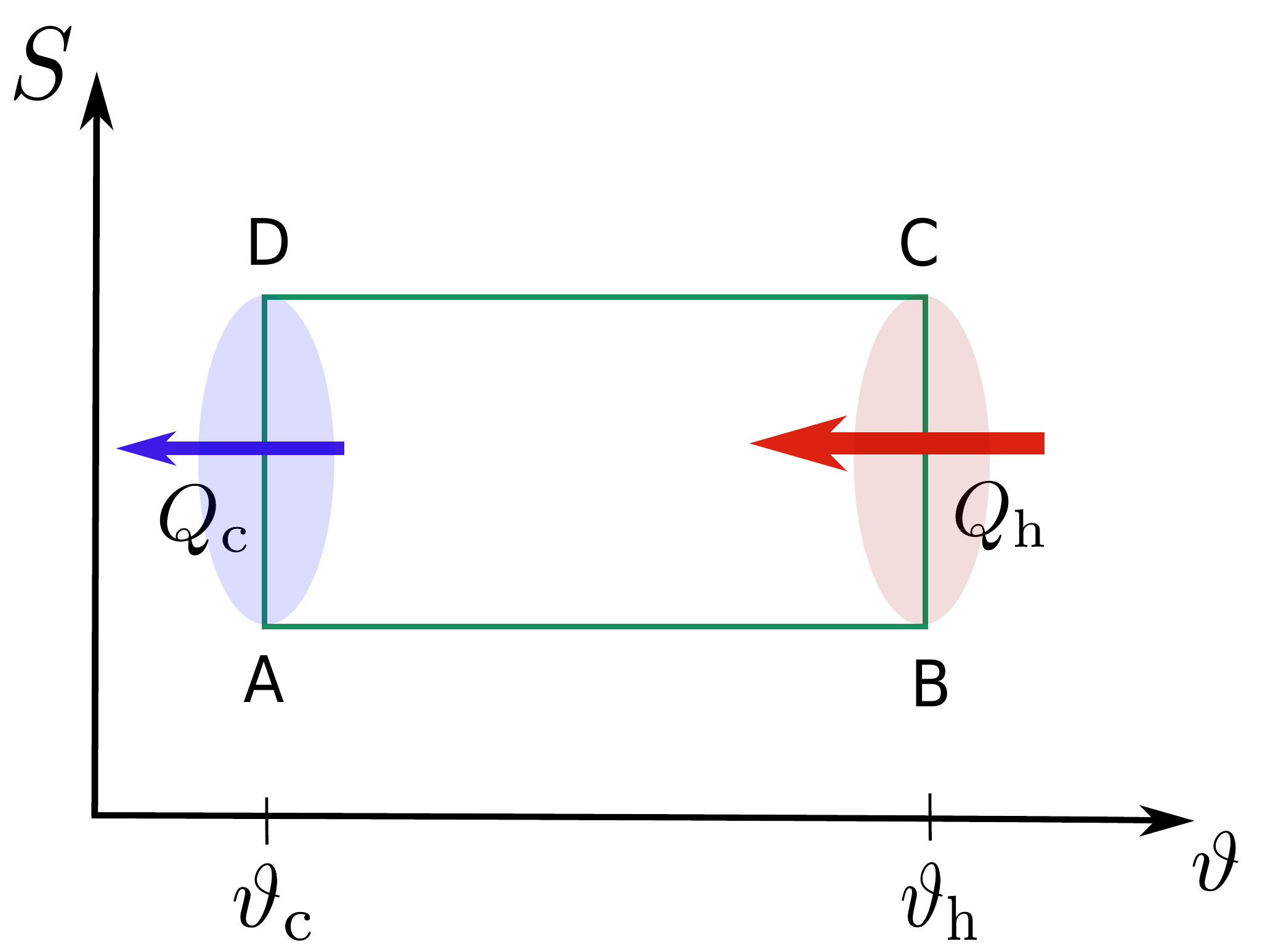}
\caption{Schematic diagram showing the Otto cycle in the entropy ($S$) - $\vartheta$ plane. A$\to$B and C$\to$D denote the unitary strokes, wherein the Hamiltonian parameter $\vartheta$ is tuned between $\vartheta_{\rm c}$ and $\vartheta_{\rm h}$. Heat $\mathcal{Q}_{\rm h}$ ($\mathcal{Q}_{\rm c}$) flows between the hot (cold) bath and the WM during the non-unitary stroke B$\to$C (D$\to$A).}
\label{fig_otto}
\end{center}
\end{figure}

Quantum engines in general consist of a central system, termed as the working medium (WM), which is subjected to time-dependent Hamiltonians, and coupled to a cold thermal bath at a temperature $T_{\rm c}$ and a hot thermal bath at a temperature $T_{\rm h} > T_{\rm c}$. The setup is engineered so as to convert a part of the thermal energy of the hot bath into usable output work, following the laws of thermodynamics \cite{callen85thermodynamics, gemmer2009quantum, kondepudi15modern}.   One of the most widely studied quantum thermal engines is the Otto engine, which is described by the following four strokes (see Fig. \ref{fig_otto}):\\
\begin{itemize}
 \setlength\itemsep{1em}
 \item {\it Stroke 1, A$\to$B:} We start with the WM in thermal equilibrium with a cold bath at temperature $T_{\rm c}$. The WM Hamiltonian $H_{\rm S}(\vartheta)$ is changed as a function of time from 
 $H_{\rm S}(\vartheta_{\rm c})$ to $H_{\rm S}(\vartheta_{\rm h})$ in a time $\tau_1$, where as before, $\vartheta$ is a parameter which characterizes the Hamiltonian. {The system is isolated from the bath, and evolves isentropically during this unitary stroke.}
 \item {\it Stroke 2, B$\to$C:} The WM is coupled to a hot thermal bath at temperature $T_{\rm h}$ for a time duration $\tau_2$. The Hamilotnian is kept constant during this non-unitary stroke. For a stroke duration $\tau_2$ of the order of the thermalization time $\tau_{\rm th}$ or longer, the WM reaches the steady state corresponding to the hot bath. The heat flow $\mathcal{Q}_{\rm h}$ between the WM and the hot bath is given by (see Eq. \eqref{eqQ})
\ba
\mathcal{Q}_{\rm h} = \mathcal{E}_{\rm C} - \mathcal{E}_{\rm B},
\label{heath}
\ea
where $\mathcal{E}_j$ is the energy of the WM at point $j = {\rm A, B, C, D}$ (see Fig. \ref{fig_otto}). 
\item {\it Stroke 3, C$\to$D:} The WM is decoupled from the thermal bath, following which, the Hamiltonian is changed from  $H_{\rm S}(\vartheta_{\rm h})$ to $H_{\rm S}(\vartheta_{\rm c})$ in a time duration $\tau_3$. As for the stroke 1, the entropy remains constant during this unitary stroke.
\item {\it Stroke 4, D$\to$A:} The WM is coupled to the cold thermal bath at temperature $T_{\rm c}$ for a time duration $\tau_4$. As for the second stroke, $\tau_4  \gtrsim \tau_{\rm th}$ allows the WM to reach the steady-state corresponding to the cold bath, thus completing the cycle. The heat exchanged $\mathcal{Q}_{\rm c}$ during this non-unitary stroke is given by 
\ba
\mathcal{Q}_{\rm c} = \mathcal{E}_{\rm A} - \mathcal{E}_{\rm D}.
\label{heatc}
\ea
\end{itemize}
The output of the thermal machine described above is quantified by the work {
\ba
W = -\left(\mathcal{Q}_{\rm h} + \mathcal{Q}_{\rm c}\right) = \left(\mathcal{E}_{\rm B} - \mathcal{E}_{\rm A}\right) + \left(\mathcal{E}_{\rm D} - \mathcal{E}_{\rm C}\right),
\label{eqwork}
\ea
}power 
\ba
\mathcal{P} = \frac{W}{\tau_{cyc}}
\label{eqpower}
\ea
and the efficiency
\ba
\eta = -\frac{W}{\mathcal{Q}_{\rm h}},
\label{eqeff}
\ea
where $\tau_{cyc} = \tau_{1} + \tau_{2} + \tau_{3} + \tau_{4}$ is the total cycle duration.

Following Ref. \cite{latune20collective}, we now discuss the operation of an Otto engine consisting of a N-spin WM, collectively coupled to hot and cold thermal baths, during the respective non-unitary strokes. The WM is  in presence of the Hamiltonian (see Eq. \eqref{Hamilwm_coll2})
\ba
H_{\rm S}(\vartheta(t)) = \vartheta(t)  \omega \mathcal{J}_z.
\ea 
Here $\omega$ is a time-independent constant, while the parameter $\vartheta(t)$  introduces time-dependence in the Hamiltonian during the unitary strokes, and  assumes the constant value $\vartheta_{\rm h}$ ($\vartheta_{\rm c}$) during stroke 2 (stroke 4). For the spins interacting collectively with the thermal baths during the non-unitary strokes $2$ and $4$ (see Eq. \eqref{Hamilwm_coll3}), the corresponding steady-states are described in Sec. \ref{secdiscoll} (See Eqs. \eqref{rhosscolN1} and \eqref{rhosscolN2}). For comparison, we also consider below the case where each spin interacts independently with a bath during a non-unitary stroke; in this case the spin ensemble reaches the usual thermal equilibrium state  $\rho_{\rm th} (T_x, \vartheta_x) = Z^{-1}(\beta_x, \vartheta_x) e^{-H_{\rm S}(\beta_x \vartheta_x)}$, with $\beta_x = 1/ T_x$, $x = {\rm h, c}$ and 
$Z(T_x, \vartheta_x) = {\rm Tr} e^{-\beta_x H_{\rm S}(\vartheta_x)}$, at the end of a non-unitary stroke.

In Ref. \cite{latune20collective}, the authors considered the Otto cycle described above to show that the work output per cycle can be related to the specific heat of the WM through the relation
\ba
W =  \Delta \eta \vartheta_{\rm h}^2 \left(\beta_c - \beta_h \right)\frac{C(\theta_h)}{ \theta_h^2} + \mathcal{O}\left(\Delta \eta^2 \right).
\label{eqWC}
\ea
Here $\theta_x = \vartheta_x \beta_x$, $\Delta \eta = \eta_c - \eta = \vartheta_{\rm c} / \vartheta_{\rm h} - \beta_h/\beta_c$ is the difference between the Carnot efficiency $\eta_c = 1 - \beta_h/\beta_c$ and the actual efficiency, which is considered to be small here, and $C(\theta_h)$ denotes the collective or the independent heat capacity, depending on whether
the spins interact collectively or independently with the baths.

As discussed in Sec, \ref{secdiscoll}, in case of collective WM-bath coupling, the steady-state $\rho_N^{\rm ss}(\beta)$, and hence the corresponding specific heat $C^{\rm col}(\beta)$, depend non-trivially on the initial state $\rho_0$. Therefore in order to focus on the maximum possible advantage obtained through collective coupling, we consider the best case scenario for the collective spin machine, which corresponds to $\rho_0$ belonging to the symmetrical subspace, i.e., $j = Ns$ and $p_{j=Ns} = 1$. Consequently we have $C_+^{\rm col}(\beta) = C_{j=Ns}(\beta)$, so that we need to compare $W_+^{\rm col} \approx  \Delta \eta \lambda_h^2 \left(\beta_c - \beta_h \right)\frac{C_{j = Ns}(\theta_h)}{ \theta_h^2} $ with $W^{\rm ind} \approx  \Delta \eta \lambda_h^2 \left(\beta_c - \beta_h \right)N\frac{C_{j = s}(\theta_h)}{ \theta_h^2} $ (see Eq. \eqref{eqWC}). Evaluating $C_j(\theta_h)/\theta_h^2$, one can show that keeping $\Delta \eta$, $\beta_c - \beta_h$ and $\vartheta_{\rm h}$ fixed, both $W^{\rm ind}$ and $W^{\rm col}$ assume maximum values for $\beta_h \to 0$, such that one arrives at the following relations:
\ba
W^{\rm ind} \leq W_{max}^{\rm ind} := \Delta \eta \vartheta_{\rm h}^2 \beta_c \frac{\omega^2}{12} N\left[\left(2s+1 \right)^2 - 1 \right]
\ea
and 
\ba
W^{\rm col}_+ \leq W_{max}^{\rm col} := \Delta \eta \vartheta_{\rm h}^2 \beta_c \frac{\omega^2}{12} \left[\left(2Ns + 1 \right)^2 - 1 \right] = \frac{Ns + 1}{s+1} W_{max}^{\rm ind},
\ea
where we have neglected terms of the order of $\mathcal{O}(\Delta \eta^2)$ or smaller.

Let us now focus on the more practical scenario of finite bath temperature $T_{\rm h}$; as one can infer from Eqs. \eqref{TNcr} and \eqref{eqWC}, for a fixed finite temperature $T_{\rm h}$, increasing the size $N$ of the WM increases the work output $W^{\rm col}$ for small $N$, {in which regime it can be more beneficial than an independent-spin engine}, until it reaches the critical number 
$N = N_{\rm cr}  \simeq \frac{3(T_{\rm h}/\omega \vartheta_{\rm h})^2 - 1/4}{s(s+1)}$.  Beyond this size of the WM, the independent-spin engine performs better than the collective-spin one, if we focus solely on the work output per cycle. 
{On a similar note, the collective engine performs better than the independent engine only for $\vartheta_{\rm h} < \vartheta_{\rm h,cr} \simeq T_{\rm h} \sqrt{12}/ \left[  \omega \sqrt{4Ns (s + 1) + 1} \right] $, in terms of output work per cycle.}

However, in order to understand the advantage offered by collective engines, one needs to focus on the output power, instead of output work. One can use Eq. \eqref{eqmecol} to show that collective coupling between the WM and a bath leads to faster dynamics.
For a system starting from a thermal state at inverse temperature $\beta_0$ with $  \omega |\beta_0| \gg 1$, collective system-bath interaction shortens the equilibration timescale by at least $N$ times, as compared to that obtained in presence of independent dissipation.
This collective-coupling induced speed-up in equilibration in turn translates to enhancement in power output of collective engines as compared to their independent counterpart, quantified by the ratio
 \ba
\lim_{T_{\rm h} \gg \vartheta_{\rm h} T_{\rm cr}(N,s)} \frac{\mathcal{P}_+^{\rm col}}{\mathcal{P}^{\rm ind}} \sim  \frac{N(Ns + 1)}{s+1}.
\label{powcolind4}
 \ea
On the other hand, for fixed $T_{\rm h}$ and $N \gg N_{\rm cr}(T_h, \lambda_h)$, the advantage offered by faster equilibration for a collective heat engine is cancelled out by the less work output per cycle, so that the output powers of the two machines become equivalent.

The above results are derived for Otto cycles with long durations of thermalizations strokes ($\tau_2, \tau_4 \gtrsim \tau_{\rm th}$), which allows the WM to reach steady states at the end of the non-unitary strokes. Interestingly, the advantage due to collective effects is present in finite-time collective heat engines as well, brought about by $\tau_{3}, \tau_4 < \tau_{\rm th}$ \cite{kloc19collective}. In this case, the short durations of the non-unitary strokes do not allow the WM to reach the corresponding steady-states. {Nevertheless, after a transient regime, the setup reaches a stable mode of operation, corresponding to a limit cycle, albeit with less output work per cycle.} However, the shorter duration of each cycle eventually leads to enhanced power output, for collective, as well as independent engines. For high enough temperature of the hot bath, the power scales as $N^2$ for the collective engine, as opposed to a linear scaling obtained in case of independent engines. In contrast, the shorter duration of the non-unitary strokes reduce the work output and heat input equally, thereby keeping the the efficiency unchanged. 
Consequently, in terms of power output, collective heat engines are more beneficial than independent heat engines for $\tau_{3}, \tau_4 \lesssim \tau_{\rm th}$. In general the regime $\tau_{3}, \tau_4 > \tau_{\rm th}$ is detrimental for operation of heat engines, since the work output per cycle increases with increasing $\tau_{3}$, $\tau_4$ only for $\tau_{3}, \tau_4 \leq \tau_{\rm th}$. 

{Finally, we note that in general high temperatures are detrimental to quantum features in systems. However, here the quantum effects arise due to collective coupling between the WM and the thermal baths (see \eqref{Hamilwm_coll3}), and we consider the symmetric subspace $j = Ns$. For a thermal bath at a very low temperature, the steady states correspond to the spins being in their respective ground states, such that the steady state arising due to independent coupling coincides with that obtained in case of collective coupling.  On the other hand, for baths at high temperatures, the steady states arising due to collective coupling and independent coupling can be appreciably different (see Eqs. \eqref{rhosscolN1} - \eqref{rhossind2})  \cite{niedenzu18cooperative, latune19thermodynamics}. Furthermore, for high enough temperatures, eigenstates with $m \approx 0$ become populated, which in turn gives rise to accelerated equilibration (see \eqref{eqmecol}) \cite{kloc19collective}. These eventually lead to enhancement in the performance of collective heat engines for high $T_{\rm h}$, as discussed above \cite{latune20collective}.}

\subsection{Collective effects in continuous thermal machines}
\label{seccolcont}

Till now we have considered dissipation in presence of time-independent Hamiltonians only. However, analysis of several thermal machines may involve time-dependent Hamiltonians as well. For example, continuous thermal machines operate in presence of periodically modulated WM Hamiltonian, while they are simultaneously coupled to a hot and a cold thermal bath \cite{kosloff13quantum, alicki14, klimovsky15chapter}. In contrast to the stroke thermal machines, one does not need to repeatedly couple and decouple the WM with the thermal baths, in order to operate the machine. Instead, here we consider spectral separation of baths, {such that the two baths interact with the WM at different ranges of energy (see Eq. \eqref{eqSpec} below)}, which eventually leads to non-zero output power \cite{klimovsky13minimal}.  Below we study the dissipative dynamics of a many-body quantum system subjected to a periodically modulated Hamiltonian, and collectively coupled to a hot and a cold thermal bath \cite{niedenzu18cooperative}.

As before (see Eq. \eqref{Hamilwm_coll2}), we consider indistinguishable spins in presence of the Hamiltonian
\ba
H_{\rm S} =    \omega(t) \mathcal{J}_z.
\ea
However, in contrast to the Otto cycle discussed in the previous
section, we now consider $\omega(t)$ modulated  periodically in time at a frequency $\Omega$, such that $\omega(t + \tau_{\rm cyc}) = \omega(t)$, $\tau_{\rm cyc} = 2\pi/\Omega$ being the time period of modulation.
Secular approximation demands the cycle period $\tau_{\rm cyc}$ to be much less than the thermalization time scale \cite{breuer02}. Further, we assume weak system-bath coupling, such that one can apply the Born, Markov and secular approximations to arrive at the master equation governing  the dynamics in the
interaction picture \cite{alicki12periodically, klimovsky13minimal, kosloff13quantum, alicki14, klimovsky15chapter}:
\ba
\dot{\rho} = \sum_{v\in\{c,h\}} \sum_{q\in \mathbb{Z}}  \mathcal{L}_{v,q} \rho,
\label{meqcont}
\ea
where {
\ba
 \mathcal{L}_{v,q} \rho &=& \frac{1}{2}P(q)G_v(\omega_0 + q\Omega) \mathcal{D}\left[\mathcal{J}_{-}\right]\rho + \frac{1}{2}P(q)G_v(\omega_0 + q\Omega) e^{-\beta_v \left(\omega_0 + q\Omega \right)} \mathcal{D}\left[\mathcal{J}_{+}\right] \rho \non\\
 &=& \frac{1}{2}P(q)G_v(\omega_0 + q\Omega)\left(\mathcal{J}^{-} \rho \mathcal{J}^{+} - \mathcal{J}^{+}\mathcal{J}^{-}\rho \right) +  \frac{1}{2}P(q)G_v(\omega_0 + q\Omega) e^{-\beta_v \left(\omega_0 + q\Omega \right)} \left(\mathcal{J}^{+} \rho \mathcal{J}^{-} - \mathcal{J}^{-}\mathcal{J}^{+}\rho \right) \non\\&+& \text{h.c.}.
\label{disseqflq}
\ea}
 Here $\omega_0 = \frac{1}{\tau_{\rm cyc}} \int_0^{\tau_{\rm cyc}} \omega(t) dt$ is the bare (unperturbed) frequency of each spin, $G_v(\nu)$ denotes the bath spectral function at frequency $\nu$, for the $v$-th ($v = h, c$) bath, and we have considered the {Kubo–Martin–Schwinger} condition $G_v(-\nu) = G_v(\nu) \exp\left[-\beta_v \nu \right]$ \cite{breuer02}. $P(q)$ denotes the weight corresponding to the $q$-th harmonic, and is given by
 \ba
 P(q) = \left|\frac{1}{\tau_{\rm cyc}} \int_0^{\tau_{\rm cyc}} \exp\left[i\int^t_0\left(\omega(t^{\prime}) - \omega_0 \right) dt^{\prime}\right] e^{-iq\Omega t} dt \right|^2.
 \label{pqflq}
 \ea
{Here we have used the Floquet method to arrive at the master equation \eqref{meqcont}. Comparing with the dissipative dynamics discussed in Secs. \ref{secdiscoll} and \ref{secdissff} (see Eqs. \eqref{meq_h}, \eqref{eqFFdis0} and \eqref{eqFFdis}), the additional index $q$-dependent pre-factors, corresponding to the q-th harmonic, arise in Eqs. \eqref{meqcont} - \eqref{pqflq} due to the periodic modulation considered here \cite{klimovsky15chapter}.}

In Ref, \cite{niedenzu18cooperative}, the authors considered a WM comprised of $N$ spin-$1/2$s, and spectral separation of baths through the condition
\ba
G_{\rm c}(\nu) &\approx& 0~~ \text{for} ~\nu \geq \omega_0, \non\\
G_{\rm h}(\nu) &\approx& 0~~ \text{for} ~\nu \leq \omega_0.
\label{eqSpec}
\ea
In such a continuous engine, the collective output power $\mathcal{P}_{\rm col}$ and its counterpart $\mathcal{P}_{\rm ind} := N\mathcal{P}\left(\frac{1}{2} \right)$, established by $N$ spin-$1/2$s independently coupled to the thermal baths, follow the relation
\ba
\lim_{\beta_{eff}\omega_0 \to \infty} \frac{\mathcal{P}_{\rm col}}{\mathcal{P}_{\rm ind}} = 1
\label{powratiohigh}
\ea
in the low-temperature regime, while
\ba
\lim_{\beta_{eff}\omega_0 \to 0} \frac{\mathcal{P}_{\rm col}}{\mathcal{P}_{\rm ind}} = \frac{N+2}{3}
\label{powratiolow}
\ea
in the high-temperature regime. Here the inverse effective temperature $\beta_{eff}$ is 
defined through the relation
\ba
\exp\left(-\beta_{eff} \omega_0 \right) := \frac{\sum_{v\in\{c,h\}}\sum_{q\in\mathbb{Z}} P(q) G_v(\omega_0 + q\Omega)e^{-\beta_i (\omega_0 + q\Omega)}}{\sum_{v \in\{c,h\} P(q) G_v(\omega_0 + q\Omega)}}.
\label{eqbetaeff}
\ea
As one can see from  Eq. \eqref{powratiolow}, a superradiant scaling behaviour $\mathcal{P}_{\rm col} \sim N\mathcal{P}_{\rm ind} = N^2 \mathcal{P}\left(\frac{1}{2}\right)$ is exhibited at sufficiently high effective temperatures, when the spin-$N/2$ particle is considerably excited.
Therefore as for stroke thermal machines, high temperatures are beneficial for the operation of continuous thermal machines in presence of collective system-bath coupling as well. For a given value of $\beta_{eff}$ (see Eq. \eqref{eqbetaeff}) one arrives at the saturation relation
\ba
\lim_{N\to \infty} \frac{\mathcal{P}_{\rm col}}{\mathcal{P}_{\rm ind}} = \coth\left(\frac{\beta_{eff} \omega_0}{2}\right).
\label{powsat}
\ea
As seen from Eq. \eqref{powsat}, $\mathcal{P}_{\rm col} \to \mathcal{P}_{\rm ind}$  in the low effective-temperature regime $\beta_{eff} \omega_0 \gg 1$, such that collective coupling does not provide any advantage as compared to the independent coupling in this case, even for large particle numbers. In contrast,  the rhs of Eq. \eqref{powsat}
diverges as $2(\beta_{eff} \omega_0)^{-1}$ in the high-temperature regime of $\beta_{eff} \omega_0 \to 0$, implying the existence of significant enhancement in power output in this regime.

Let us compare Eq. \eqref{powratiolow} with the equivalent result in case of stroke engines, viz. Eq. \eqref{powcolind4}. In the case of stroke engines, collective effects lead to an enhancement of the order of $N^2$ at high temperatures, which is $N$ times higher than the enhancement obtained in case of continuous engines (see Eq. \eqref{powratiolow}). The additional enhancement in case of stroke engines stem from the reduction in thermalization times by a factor of $N$, due to collective coupling between the WM and the thermal baths. This effect is not present in continuous engines, which lack any thermalization stroke \cite{latune20collective}.

Finally, one can show that the advantage offered by collective coupling is present in the refrigerator regime as well, where the collective coupling leads to enhanced rate of refrigeration of the cold bath.

\subsection{Collective effects due to spin statistics}
\label{seccolss}

Spin statistics of a many-body WM can also lead to non-trivial effects of quantum thermal machines \cite{zheng15quantum}. Recently it has been shown that collective effects arising due to the Bosonic statistics of indistinguishable particles can also lead to enhancement in the performance of quantum thermal machines \cite{watanabe20quantum}. 
When multiple indistinguishable Bosonic work resources are coupled to an external system, the output of such a setup, quantified by the internal
energy change of the external system, exhibits an enhancement, as compared to when the setup consists of distinguishable work resources. 

\section{Interacting many-body quantum thermal machines}
\label{secint}

In this section we focus on quantum thermal machines with WM comprised of interacting many-body systems. Different platforms have been used to study such machines, for example, Rydberg atoms \cite{carollo20nonequilibrium}, multiferroic chain \cite{azimi14quantum}, etc. Such WM allow us to study quantum thermal machines in presence of several many-body effects, including  topological phase transitions \cite{fadaie18topological, yunt20topological, kumar20a}, superfluid to insulating phase transition \cite{fogarty20a} and time-translation symmetry breaking \cite{carollo20nonequilibrium_arxiv}, to name a few. Here we discuss a few such quantum engines studied  recently, to highlight the non-trivial role played by inter-particle interactions in the operation of quantum machines.

\subsection{Critical engines}
\label{seccritical}

Criticality is a vibrant field of study both in classical as well as quantum condensed matter physics, owing to the divergence of different parameters close to critical points, and the resultant universality \cite{chaikin95principles, sachdev99quantum}. Therefore it is no surprise that quantum machines with critical WM  have received a significant amount of attention as well \cite{fusco16work, campisi16the, ma17quantum, chand18critical, fadaie18topological, abiuso20optimal, fogarty20a, revathy20universal}. 

As mentioned earlier, the maximum efficiency of an engine is bounded by the Carnot limit, which usually occurs in engines operating in infinite time, such that the Carnot limit is reached only at the cost of zero power output $\mathcal{P}$. Consequently, one of the major challenges in quantum thermodynamics, is to design quantum engines which can deliver finite power, even as the efficiency approaches the Carnot limit.  In Ref. \cite{campisi16the}, the authors addressed this problem by focussing on critical heat engines; they showed that the divergence of specific heat close to criticality can allow us to design quantum engines which can operate infinitesimally close to the Carnot efficiency $\eta_C$, but without sacrificing the output power. The authors quantified the operation of an engine as 
\ba
\dot{\Pi} = \frac{\mathcal{P}}{\Delta \eta},
\ea
where $\Delta \eta =  \eta_C - \eta$ is the deviation of the efficiency $\eta$ from the Carnot limit $\eta_C$. For $N$ identical engines working in parallel, the total output power $\mathcal{P} \sim N$, while $\eta \sim 1$ is independent of $N$. Consequently, we have $\dot{\Pi} \sim N$. However this linear scaling with system size $N$ is obtained at the cost of larger resources (larger number of independent engines), and therefore does not signify any advantage obtained due to many-body effects. 
On the other hand, let us consider a quantum engine with a WM comprised of $N$ interacting particles. A super-extensive scaling of $\dot{\Pi} \sim N^{1 + a}$ with the system size, signified by $a > 0$ would imply $\Delta \eta \sim N^{-a}$ for $\mathcal{P} \sim N$, thus implying one can reach Carnot efficiency by increasing $N$, without compromising on the average power per particle. This can indeed be the case for quantum machines operated close to a second order phase transitions. The total work output $W$ per cycle of an Otto engine is given by
\ba
W = \mathcal{Q}_2 + \mathcal{Q}_4,
\ea
where $\mathcal{Q}_j = C \Delta T_j$ is the heat exchange during the $j = 2,4$ (non-unitary) strokes, which depends on the specific heat $C$ and the effective change in temperature $ \Delta T_j$ during the $j$th stroke. In order to design an engine which can harness the advantage provided by second order phase transitions, the authors of Ref. \cite{campisi16the} considered an Otto cycle in which the temperature of the hot bath coincides with the critical temperature of the WM. A second order phase transition is accompanied by a diverging specific heat $C \sim N^{1 + \alpha/\nu d}$, and hence diverging work output. On the other hand, critical slowing down close to the phase transitions implies the time scale  $\tau_{\rm cyc}$ for the cycle varies as $\tau_{\rm cyc} \sim N^{z/d}$. Combining the above results, for $\Delta \eta \ll 1$, one arrives at the universal relation
\ba
\dot{\Pi}  &\sim& N^{1 + a} \non\\
a &=& \frac{\alpha - \nu z}{\nu d},
\label{equniv}
\ea
where $\alpha, \nu$ and $z$ are respectively, the specific heat, correlation length and dynamical critical exponents, while $d$ denotes the dimensionality of the WM. The universal relation Eq. \eqref{equniv} allows us to choose the WM wisely, such that one can operate it at the Carnot efficiency, without significantly compromising on the power. For example, $\nu$ and $d$ are positive. Therefore $\alpha > \nu z$ implies one can asymptotically approach the Carnot efficiency at non-zero $\mathcal{P}$, while the stronger condition $\alpha - \nu z \geq 1$, which can happen for example in Dy$_2$Ti$_2$O$_7$, ensures that one can reach the Carnot efficiency asymptotically, without compromising on the output power per particle \cite{campisi16the}. 

{A couple of comments are in order here regarding the critical heat engine described above. Firstly, here only power and efficiency have been used to quantify the operation of a heat engine. However, fluctuations in power can also be an important criteria \cite{holubec17work}. Ideally, a heat engine should operate with large power and high efficiency, while showing minimal fluctuations in its output. But as shown in Ref. \cite{pietzonka18universal} for systems exhibiting classical Markovian dynamics, a heat engine operating with finite power close to the Carnot efficiency can result in diverging fluctuations in power. Secondly, we note that no other quantum feature was considered in Ref. \cite{campisi16the}, besides the discreteness of the energy spectrum. Consequently, this raises interesting questions regarding the effect of phase transitions on the operation of classical heat engines as well, such as in steam engines. However, a detailed discussion regarding phase transitions in classical thermal machines is beyond the scope of the current review.}

\begin{figure}
\includegraphics[width = 0.5\textwidth]{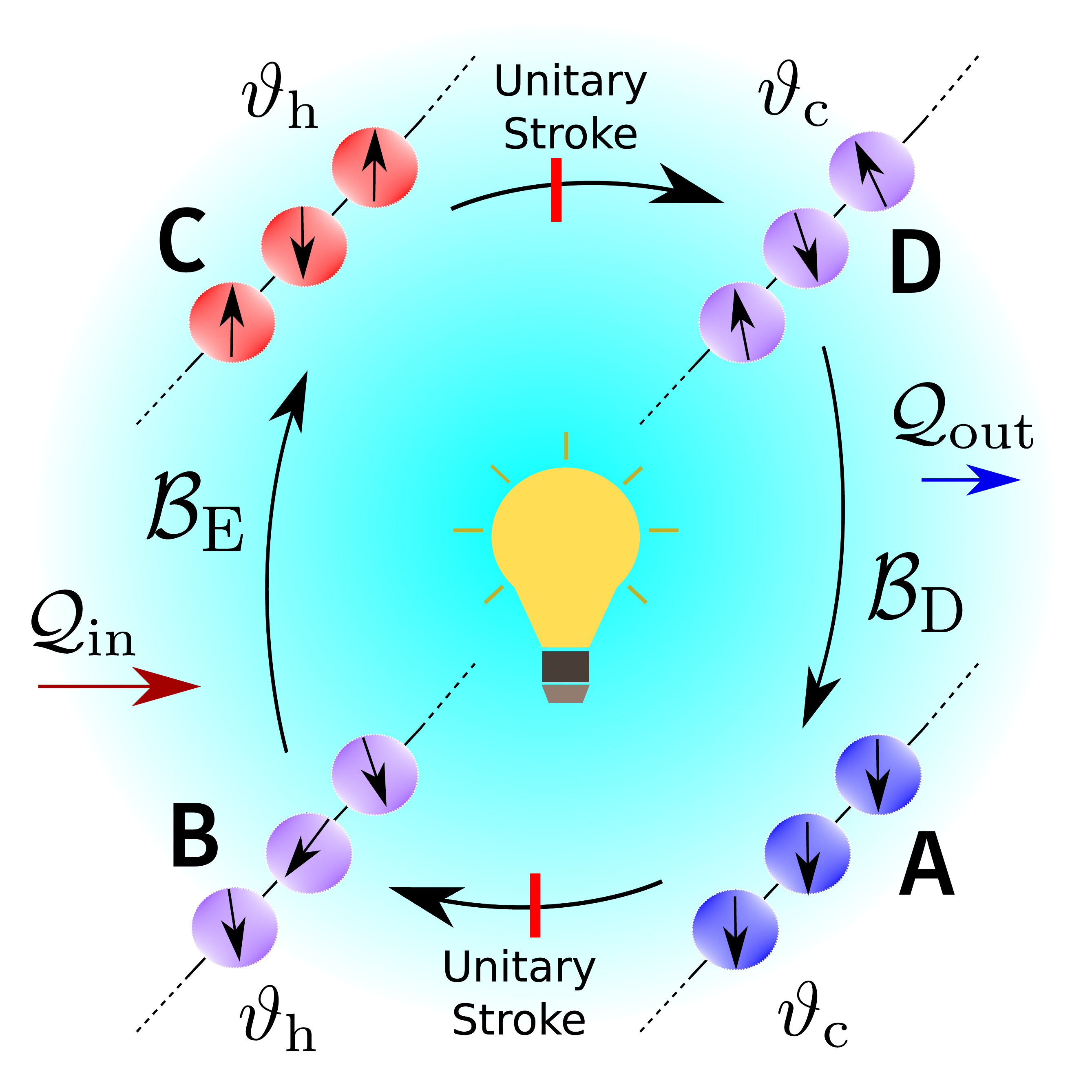}
\caption{Schematic diagram of a many-body Otto cycle close to quantum phase transition. A relaxing bath takes the WM close to its ground state at A (denoted by all spins anti-parallel to the $z$ axis, on blue spheres). Non-adiabatic excitations are generated as the WM is driven at a finite rate across  quantum critical point(s) during the unitary stroke A$\to$B (denoted by spins aligned along different directions on violet spheres, at B). Energy $\mathcal{Q}_{\rm in}$ flows from an energizing bath $\mathcal{B}_{\rm E}$ to the WM during the non-unitary stroke B$\to$C, which takes the system to a high-entropy mixed state at C (denoted by spins aligned parallel and anti-parallel to the $z$ axis, on red spheres). The unitary stroke from C$\to$D across  quantum critical point(s) again generates non-adiabatic excitations (denoted by spins aligned along different directions on violet spheres, at D). Energy $\mathcal{Q}_{\rm out}$ flows from the WM to a dissipative relaxing bath $\mathcal{B}_{\rm D}$ during the non-unitary stroke D$\to$A. (After Ref. \cite{revathy20universal})}
\label{figKZMQEschem}
\end{figure}

\begin{figure}
\includegraphics[width = 0.5\textwidth]{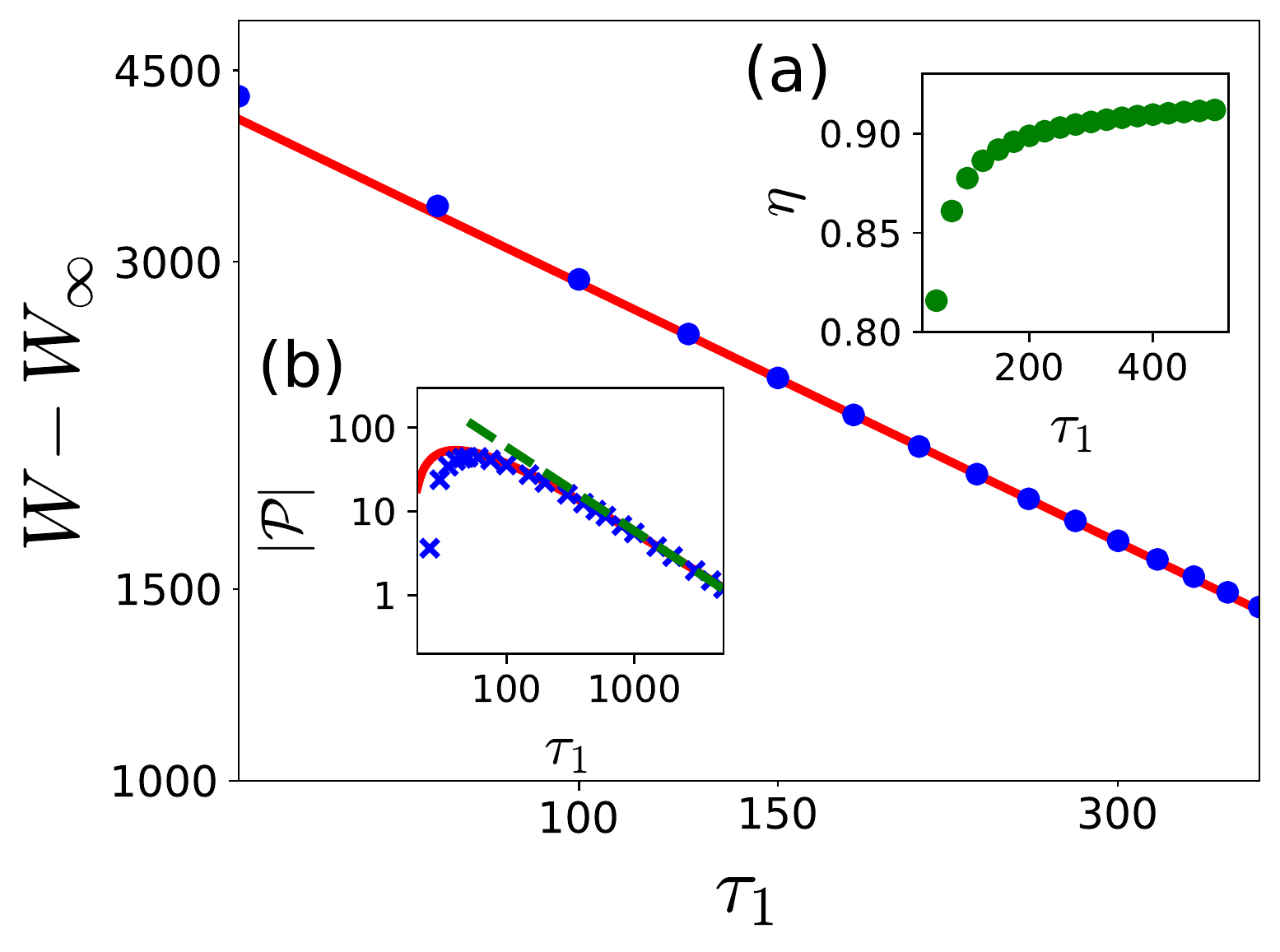}
\caption{Work output follows universal Kibble-Zurek scaling (Eq. \eqref{workkzm2}) in a quantum engine modelled with a transverse Ising chain WM, driven across QCPs.
The points are the numerical values and red solid line corresponds to
$\tau_1^{-1/2}$. For transverse Ising model, $d=\nu =z=1$.
Inset (a): Variation of $\eta$ with $\tau_1$.
(b) Variation of Power with
$\tau_1$. The green dashed line corresponds to $1/\tau_1$
scaling, points represent
numerical data and solid line is the analytical expression.
The parameters used are:
$L=100, h_1=70, h_2 = -5, \tau_2=0.01, \mu_{\rm E}'=1,
\mu_{\rm E}=0.995, \mu_{\rm R}'=0, \mu_{\rm R}=1$
with $W_{\infty}=$-6481.205. (After Ref. \cite{revathy20universal})}
\label{figKZMQE}
\end{figure}

Recently, universal behaviors in finite-time quantum engines have also been studied close to quantum phase transitions. Classical phase transitions occur due to thermal fluctuations. In contrast, quantum phase transitions are accompanied by quantum fluctuations, which arise due to vanishing energy gaps close to QCPs \cite{sachdev99quantum}. Consequently, signatures of quantum phase transitions, such as universality, are very fragile in presence of thermal fluctuations, at non-zero temperatures \cite{hoyos07effects, patane08adiabatic}. 
However, in Ref. \cite{revathy20universal} the authors used Kibble-Zurek mechanism \cite{kibble80some, zurek05dynamics, polkovnikov05universal,  damski06adiabatic} to show that universal behaviors may persist in appropriately designed finite-time many-body Otto engines operated close to criticality, in spite of the presence of thermal fluctuations during the non-unitary strokes (see Sec. \ref{secscaling}). The authors considered a many-body Otto engine, which is driven across QCPs during the unitary stroke 1 (A $\to$ B) at a rate $\tau_1^{-1}$  for a time interval $\tau_1$, followed by a non-unitary stroke 2 (B $\to$ C), during which input energy $\mathcal{Q}_{\rm in}$ is provided by a dissipative, but not necessarily thermal, energizing bath $\mathcal{B}_{\rm E}$; unitary stroke 3 (C $\to$ D) during which the WM is driven back across the QCPs at a rate $\tau_2^{-1}$ for a time interval $\tau_2$, where $\tau_2$ may not be equal to $\tau_1$; and finally, a non-unitary stroke 4 (D $\to$ A) to complete the cycle, during which energy $\mathcal{Q}_{\rm out}$ flows from the WM to a dissipative relaxing bath  $\mathcal{B}_{\rm D}$  (see Fig. \ref{figKZMQEschem}). The dynamics during the non-unitary strokes can be modelled following Sec. \ref{secdissff}. 
The output of such an engine shows universal scaling forms 
with respect to the duration of the unitary stroke 1 (A $\to$ B), provided the engine satisfies the following general conditions:
\begin{itemize}
\item The \emph{relaxing} bath $\mathcal{B}_{\rm D}$ takes the WM close to its ground state. For example, $\mathcal{B}_{\rm D}$  can be a cold thermal bath at temperature $T_{\rm c} \ll \mathcal{E}_{4}$, $\mathcal{E}_{4}$ being the typical energy scale of the WM during the stroke 4  (D $\to$ A).
\item The WM is driven at a finite rate $\tau_1^{-1}$ across a quantum critical point (or points) 
during the {unitary stroke {A} $\to$ {B}}, i.e., $\tau_1$ is finite.
\item The energizing bath $\mathcal{B}_{\rm E}$ takes the WM to a 
unique steady state, which does not depend of the state of the WM at the beginning of the stroke 2.
\item The state of the WM does not change appreciably during the unitary stroke 3  (C $\to$ D). This can result for example if the energizing bath $\mathcal{B}_{\rm E}$ takes the WM to a steady state with high entropy. Therefore analogous to $\mathcal{B}_{\rm D}$, $\mathcal{B}_{\rm E}$  can be modelled by a hot thermal bath at temperature $T_{\rm h} \gg \mathcal{E}_{2}$, where $\mathcal{E}_{2}$ is the typical energy scale of the WM during the stroke 2  (B $\to$ C). Alternatively we may consider fast quench during the stroke 3 (i.e., $\tau_2 \to 0$), such that the WM does not get time to evolve significantly during this stroke. 
\end{itemize}
For an Otto engine satisfying the general conditions discussed above, the output work shows universal scaling as a function of $\tau_1$, given by \cite{grandi10quench,fei2020work, revathy20universal}
\ba
W - W_{\infty} \sim  \tau_1^{-\frac{\nu (d+z)}{\nu z+1}},
\label{workkzm3}
\ea
where we have considered the stroke 1 (A $\to$ B) ends at a QCP at B. 
Here $W_{\infty}$ is the work output in infinite time Otto cycles operating in the limit $\tau_1 \to \infty$. Alternatively, for unitary strokes which cross critical points, the work output is not universal in general.  However, for systems and quench protocols in which the excitation energy $\mathcal{E}_{\rm ex}$ during the stroke 1 is
proportional to the {density of excitations (see Sec. \ref{secscaling})}, such as in transverse Ising model WM,
the scaling  \eqref{workkzm3} is modified as (see Fig. \ref{figKZMQE})
\ba
W - W_{\infty} \sim  \tau_1^{-\frac{\nu d}{\nu z+1}}.
\label{workkzm2}
\ea
Furthermore, in the limit where $\tau_1$ is the most dominant time scale of one cycle such that
$\tau_{\rm{cyc}} \approx \tau_1$, 
one can use \eqref{workkzm3} and \eqref{workkzm2} to derive a scaling
relation for the output power $\mathcal{P}$
\ba
\mathcal{P} = \frac{W}{\tau_{\rm cyc}} \approx \frac{W_{\infty}}{\tau_1}
+ R \tau_1^{-\frac{\nu d + x\nu z + 1}{\nu z+1}},
\label{eq_powerscaling}
\ea
where $R$ is a WM-dependent constant.
{Here we have introduced the parameter $x$ for brevity; $x$ can assume the following two values: $x=1$ corresponds to crossing the critical point, while $x=2$
is for B being a critical point.} The optimal quench rate $\tau_1^{-1} = \tau_{\rm opt}^{-1}$
delivering the maximum power can be found from the condition
\ba
\frac{\partial \mathcal{P}}{\partial \tau_1}\big|_{\tau_{\rm opt}} = 0
\label{tauopt_cond}
\ea
 which yields
\ba
 \tau_{\rm opt} &=& \left[\frac{R \left(\nu d + x\nu z + 1 \right)}{|W_{\infty}| \left(\nu z + 1 \right)} \right]^{\left(\nu z +1\right)/[\nu d+(x-1)\nu z]},
\label{tau_opt}
\ea
with the corresponding efficiency $\hat{\eta}$ at maximum power being
\ba
\hat{\eta} &=& -\frac{W_{\infty} +  \mathcal{E}_{\rm ex}( \tau_{\rm opt})}{\mathcal{Q}_{\rm in}^{\infty} - \mathcal{E}_{\rm ex}(\tau_{\rm opt})}.
\label{eq_opteff}
\ea
Here $\mathcal{Q}_{\rm in}^{\infty}$ is the input energy corresponding to the infinite time Otto engine limit of $\tau_1 \to \infty$. 
{We note that the scalings Eq. \eqref{workkzm3} and \eqref{workkzm2} remain valid even for $\tau_{\rm cyc} \gg \tau_1$, which can happen in case of non-unitary strokes of long durations. However, as can be seen from Eqs. \eqref{workkzm3} - \eqref{eq_powerscaling}, in this case the output power $\mathcal{P}$ ceases to show the clear scaling with $\tau_1$ (Eq. \eqref{eq_powerscaling}), even though it still depends on $\tau_1$ through $W$ and $\tau_{\rm cyc}$. Consequently, $\tau_{\rm opt}$ also does not follow the scaling form Eq. \eqref{tau_opt} in this regime. Nevertheless, one can still find $\tau_{\rm opt}$, and hence $\hat{\eta}$, through the condition Eq. \eqref{tauopt_cond}.}

One can use \eqref{workkzm2}-\eqref{eq_opteff} to choose WM with appropriate
critical exponents and dimensionality, so as to
design optimal 
many-body quantum critical engines. 
For example, other factors remaining constant, enhancement in  output power would demand a WM with large
dimension $d$ (see \eqref{workkzm2}).
Furthermore, in case of free-Fermionic WM operated in presence of locally thermal baths, one can also arrive at a maximum efficiency bound $\eta_{\rm max}$ which shows universal scaling with respect to the length $N$ of the WM, given by the relation 
\ba
1 - \eta_{\rm max} \propto N^{-z}.
\ea
Clearly, $\eta_{\rm max}$ increases with increasing $N$, thus indicating that many-body quantum engines can be more efficient than few-body ones.

Recently, many-body quantum machines with ultracold gas WM, driven across superfluid and insulating phase has also been studied. The authors have shown that the existence of many-body effects and the critical point can boost the performance of a $N$-particle many-body engine, as compared to $N$ single particle engines, through enhancement in efficiency and power. Shortcuts to adiabaticity can further improve the performance of such many-body quantum engines close to criticality \cite{fogarty20a}.  Quantum criticality has also been shown to improve the efficiency in {quantum heat engine (QHE)} \cite{ma17quantum} based on the Lipkin-Meshkov-Glick model \cite{lipkin65validity, meshkov65validity, glick65validity}, as well as maximize work in interaction driven QHEs, with interacting Bose gas as a WM \cite{yang19an}. Recently, effects of topological phase transitions on the performance of QHEs have been studied \cite{fadaie18topological}. In particular, studies on  Otto heat engine with a finite length Kitaev chain as the working medium has shown that topological phase transition can enhance the efficiency, as well as work output of such engines \cite{yunt20topological}. 

{Finally, we note that as mentioned above, fluctuations in the output of a quantum thermal machine can be a crucial criteria in determining the reliability of the machine. Studies on fluctuations can be even more important for quantum machines operated close to quantum phase transitions, owing to diverging length and time scales close to criticality. Recently, it was shown that many-body interactions can dramatically affect the statistics of energy fluctuations, and the resultant work distribution, in finite size
Hubbard chains driven across metal - Mott - insulator quantum phase transitions in finite time \cite{zawadzki20work}.  }

\subsection{Shortcuts to adiabaticity}
\label{secsta}

In this section, we focus on enhancing the performance of quantum thermal machines through STA. As we discussed in Sec. \ref{sectechSTA}, adiabatic QHEs operate with high efficiency, at the cost of vanishing power. On the other hand, non-adiabatic excitations in finite-time heat engines can enhance the power, but at the cost of reducing the efficiency. STA provides a solution to this conundrum, through suppression of non-adiabatic excitations in finite-time heat engines, thereby enhancing the power output, while keeping the efficiency close to the value obtained in its adiabatic counterpart. Following Ref. \cite{hartmann19many} here we discuss STA in many-body QHEs, through approximate counterdiabatic driving \cite{sels17minimizing, kolodrubetz17geometry, claeys19floquet}.

We consider an Otto cycle with a spin chain WM, described by the Hamiltonian
\ba
H_0(t)=- \sum_{i=1}^{N} h_i(t) \sigma_i^x - \sum_{i=1}^{N} b_i(t) \sigma_i^z - \sum_{i=1}^{N} J_i(t) \sigma_i^z \sigma_{i+1}^z.
\label{eq:IsingSpinModel}
\ea
Here  $N$ is the total number of spins, $J_i(t)$ denotes the interaction strength between the spins at sites $i$ and $i+1$, $b_i(t)$ is the longitudinal field strength along $z$ direction, while $h_i(t)$ is the transverse field strength along $x$ direction, at site $i$. We impose periodic boundary conditions, given by $\sigma_{N+1}=\sigma_{1}$. 
 The explicit time-dependence of the Hamiltonian parameters during the unitary strokes $1$ and $3$  are taken to be
\ba
h_i(t)&=h_{i,\mathrm{i}}+(h_{i,\mathrm{f}}-h_{i,\mathrm{i}})\sin^2\left[\frac{\pi}{2}\sin^2 \left(\frac{\pi t}{2\tau}\right)\right] \nonumber \\
b_i(t)&=b_{i,\mathrm{i}}+(b_{i,\mathrm{f}}-b_{i,\mathrm{i}})\sin^2\left[\frac{\pi}{2}\sin^2 \left(\frac{\pi t}{2\tau}\right)\right]  \nonumber \\
J_i(t)&=J_{i,\mathrm{i}}+(J_{i,\mathrm{f}}-J_{i, \mathrm{i}})\sin^2\left[\frac{\pi}{2}\sin^2 \left(\frac{\pi t}{2\tau}\right)\right],
\ea
where the index $\mathrm{i}$ ($\mathrm{f}$) refers to the initial (final) values of the parameters, and $\tau_1 = \tau_3 = \tau$ is the duration of a unitary stroke. As one can see, in general $\left[H_0(t), H_0(t^{\prime})\right] \neq 0$ for $t \neq t^{\prime}$. Consequently, in absence of control, non-adiabatic excitations are generated during the unitary strokes, which are in general detrimental to the operation of the heat engine \cite{feldmann06quantum, dann20quantum}. Therefore we aim to improve the performance of the above described finite-time heat engine, i.e., enhance the efficiency and power, through the application of counterdiabatic driving, following Sec. \ref{sectechSTA}. We achieve this through the  application of the STA Hamiltonian  
\ba
H_{\rm STA} = H_0 + H_{\rm CD},
\label{hamilcd}
\ea
where the counterdiabatic Hamiltonian $H_{\rm CD}$ is given by 
\ba
H_{\rm CD} = \dot{\vartheta}(t) \mathcal{A}_{\vartheta}(t).
\label{hamilcd}
\ea
The control
function
\ba
\vartheta(\vartheta_0,t) = \vartheta_0 \sin^2\left[\frac{\pi}{2} \sin^2 \left(\frac{\pi t}{2\tau}\right)\right]
\ea
is chosen so as to ensure smoothness at the beginning and end of the unitary
strokes, quantified by $\dot{\vartheta}(t = 0) = \dot{\vartheta}(t = \tau) = \ddot{\vartheta}(t = 0) = \ddot{\vartheta}(t = \tau)$. Here $\mathcal{A}_{\vartheta}(t)$ is the adiabatic gauge potential, obtained by following the protocol detailed in Sec.  \ref{sectechSTA} \cite{sels17minimizing, kolodrubetz17geometry, claeys19floquet} and $\vartheta_0$ is a global control parameter strength, introduced to tune the accuracy of the strokes  (see below).

Let us assume  $H_0$ and $H_{\rm CD}$ are supplied by two independent work reservoirs. {The counterdiabatic Hamiltonian  $H_{\rm CD}$  is designed so as to enhance the performance of the QHE, quantified by the work output and the efficiency. In this case the total work output is given by
\ba
W_{\rm STA} = \int_{0}^{\tau_{cyc}} {\rm Tr}\left[ \rho \dot{H}_{\rm STA}\right] dt = W_0 + W_{\rm CD},  
\label{stacd}
\ea
where the work performed on the work reservoir  implementing $H_0$ (termed as the “piston" or the “load") is
\ba
W_0 = \int_{0}^{\tau_{cyc}} {\rm Tr}\left[ \rho \dot{H}_{0} \right] dt,
\label{staw0}
\ea
while the work performed on the work reservoir  implementing $H_{\rm CD}$ (termed as the “controller") is
\ba
W_{\rm CD} =  \int_{0}^{\tau_{cyc}} {\rm Tr}\left[ \rho \dot{H}_{\rm CD} \right] dt.
\label{stawcd}
\ea
Here the work $W_{\rm CD}$ arising due to counterdiabatic driving is performed on the controller, and consequently is not available to the load. Therefore $W_{\rm CD}$ is considered as the {\it operational cost} of the counterdiabatic driving. Further, there can be additional {\it implementational costs} of the counterdiabatic driving as well which we do not consider here, since they in general depend on the details of the setup in question \cite{campbell17trade, abah18performance, abah19shortcut, hartmann19many, hartmann20multispin}. We note that in case of a single work reservoir implementing both $H_0$ and $H_{\rm CD}$,  the division Eq. \eqref{stacd} is in general not operationally relevant any more. 

The useful output power, obtained from the work performed on the load (see Eqs. \eqref{stacd} - \eqref{stawcd}),  is given by}
\ba
\mathcal{P} =   \frac{W_{0}}{\tau_{cyc}},
\label{powsta}
\ea
while the efficiency is
\ba
\eta  &=&   \frac{-W_{0}}{\mathcal{Q}_{\rm h}}~~~ \text{for}~ W_{\rm CD} \leq 0\non\\
&=&  \frac{-W_{0}}{\mathcal{Q}_{\rm h} + W_{\rm CD}}~~~ \text{for}~ W_{\rm CD} > 0.
\label{etasta}
\ea
Here
\ba
\mathcal{Q}_{\rm h} = \int_{0}^{\tau_{cyc}} {\rm Tr}\left[  \dot{\rho} H_{0} \right] dt
\ea
is the heat input during the non-unitary stroke 2.
{The dependence of  the expression of $\eta$ on the sign of $W_{\rm CD}$ (see Eq. \eqref{etasta}) stems from the fact that for $W_{\rm CD} > 0$, the setup acts as a hybrid thermomechanical engine, where part of the input energy comes from the hot bath, and the remaining comes from the external controller.}

We now use the local ansatz
\ba
H_{\rm CD}^*(t) = \dot{\vartheta}(\vartheta_0,t) \sum_{j=1}^{N} \zeta_j(t)  \sigma_j^y
\label{hcdstar}
\ea
to approximate the counterdiabatic Hamiltonian Eq. \eqref{hamilcd}.  Optimization of the control Hamiltonian following Sec. \ref{sectechSTA} leads to the following form of $\zeta_j(t)$ (see Eqs. \eqref{eqSTA7} - \eqref{eqSTA10})
\begin{equation}
\zeta_j(t)=\dfrac{1}{2}\dfrac{\dot{h}_j(t) b_j(t)-\dot{b}_j(t) h_j(t)}{h_j(t)^2+b_j(t)^2+J_{j-1}(t)^2 + J_{j}(t)^2}.
\label{eq:alphaspin}
\end{equation}
Consequently, the local STA Hamitonian assumes the form
\ba
H^*_{\mathrm{STA}}(t) = -\sum_{i=1}^{N} h_j(t) \sigma_j^x - \sum_{j=1}^{N} b_j(t) \sigma_j^z - \sum_{j=1}^{N} J_j(t) \sigma_j^z \sigma_{j+1}^z + H_{\rm CD}^*(t),
\label{eq:SpinCDHamiltonian}
\ea
with $H_{\rm CD}^*(t)$ given by Eqs. \eqref{hcdstar} and \eqref{eq:alphaspin}. Here the asterisk signifies that the Hamiltonian is inexact. 

\begin{figure}[h]
\begin{center}
\includegraphics[width = 0.71\columnwidth]{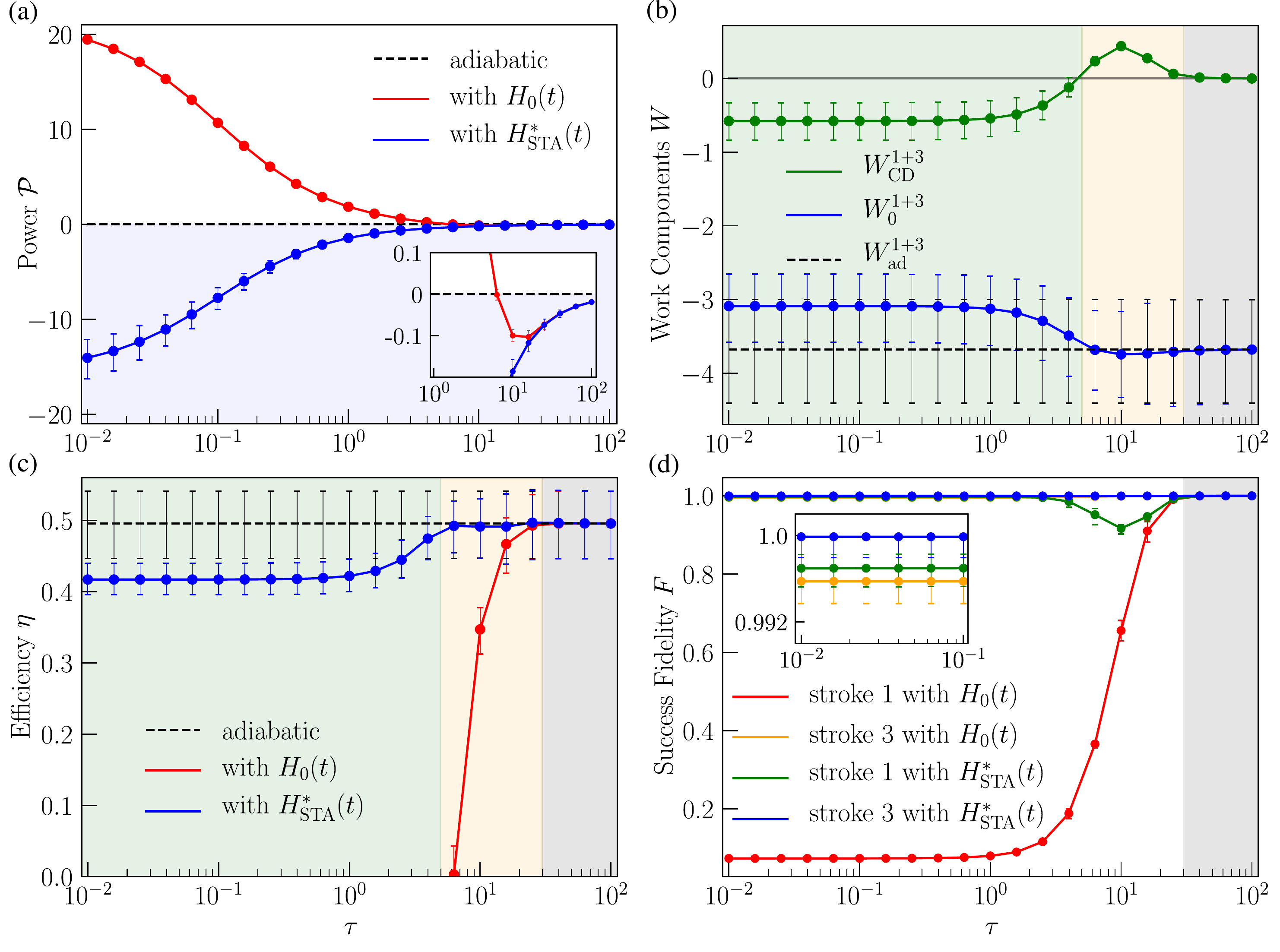}
\caption{.(a)~Power $\mathcal{P}$ of the sped-up Otto cycle governed by (i)~the original protocol $H_0(t)$ and (ii)~the shortcut-to-adiabaticity protocol $H^*_\mathrm{STA}(t)$  as a function of the isentropic-stroke duration $\tau=\tau_1=\tau_3$. The machine acts as an engine if $\mathcal{P}<0$ (blue-shaded area). Inset: Zoom on the region where the original protocol ceases to describe an engine for shorter cycle times. (b)~Work components $W_0$ and $W_{\rm CD}$ pertaining to the piston (load) and the external control device, respectively. The green (left) and yellow (middle) shaded areas depict the regions where the machine operates as a heat engine ($W_{\rm CD}<0$) and a thermo-mechanical engine ($W_{\rm CD}>0$), respectively. The grey (right) shaded area depicts the adiabatic limit region where $W_{\rm CD} < 10^{-3}$. (c)~Efficiency $\eta$ [for the heat-engine regime (green-shaded area) and for the hybrid thermo-mechanical regime (yellow-shaded area), respectively]. (d)~Success fidelities of the isentropic strokes with and without STA protocol. Inset: Zoom. Parameters: Duration of the isentropic strokes: $\tau_1=\tau_3=\tau$, duration of the thermalization strokes: $\tau_2=\tau_4=0.1$. The other parameters are $T_{\rm c} = 0.22$, $T_{\rm h} = 22$, $h_{j,\mathrm{i}} = 0.5$, $b_{j,\mathrm{i}}=0$, $h_{j,\mathrm{f}}=0$, $b_{z,\mathrm{f}}=1$, $J_{j,\mathrm{i}}=0$ for each spin. Disorderness is introduced in the interaction strengths where the $100$ final interaction strengths $J_{j,\mathrm{f}}$ are randomly chosen from a Gaussian distribution with standard deviation $\sigma=0.1$ and zero mean. The counterdiabatic drive is optimized by a control parameter $\vartheta_0$ bounded in $[0,1]$. The vertical bars denote the largest and lowest values of power, work, efficiency and success fidelity, respectively. (After Ref. \cite{hartmann19many})}
\label{fig_STA}
\end{center}
\end{figure}
The approximate counterdiabatic Hamiltonian  $H_{\rm CD}^*(t)$ drives the WM to a state $\rho_B^{\prime}$ ($\rho^{\prime}_D$) at point B (D) (see Fig.  \ref{fig_otto}), which is in general different from the state $\rho_B$ ($\rho_D$) obtained in the adiabatic limit. However, one can increase the reliability of $H_{\rm CD}^*(t)$ by optimizing  the global control parameter strength  $\vartheta_0$, subject to the constraint $0 \leq \vartheta_0 \leq 1$, so as to maximize the Fidelity 
\ba
\mathcal{F}(\rho, \rho^{\prime}) = {\rm Tr}\sqrt{\sqrt{\rho}\rho^{\prime}\sqrt{\rho}}.
\label{eqfid}
\ea 

We note that restricting the approximate counterdiabatic Hamiltonian $H^*_{\rm CD}$ to the single spin form (Cf. Eq. \eqref{eq:SpinCDHamiltonian}) simplifies the analysis and makes it experimentally implementable. At the same time, as shown in Fig. \ref{fig_STA}, even this  single body ansatz  leads to significant enhancement in the performance of the heat engine. The approximate counterdiabatic driving allows the setup to act as a heat engine with large output power (see Fig. \ref{fig_STA}a), work (see Fig. \ref{fig_STA}b) and efficiency (see Fig. \ref{fig_STA}c), while maintaining a high fidelity $\mathcal{F}$ (Cf. Eq. \eqref{eqfid}) with the actual cycle (see Fig. \ref{fig_STA}d)  even for small cycle period $\tau$; this is in sharp contrast to the results obtained in absence of control, where the setup fails to act as a heat engine for small $\tau$.
Similar advantages can be obtained in case of many-body quantum refrigerators as well, in which case STA protocol enhances the refrigeration rate and the efficiency of refrigeration, above their non-adiabatic counterparts.
{Furthermore, remarkably, as discussed in Ref.  \cite{hartmann20multispin}, one can use the expression for exact $H_{\rm CD}$ (Eqs. \eqref{eqHCD} and \eqref{eqSTA4}) to show that ${\rm Tr}\left[\rho(t) \dot{H}_{\rm CD}(t) \right] = 0~\forall~t$, if the
additional adiabatic gauge potential and consequently the counterdiabatic
Hamiltonian $H_{\rm CD}(t)$ is exact, i.e., $\mathcal{A}^*_{\vartheta}(t) = \mathcal{A}_{\vartheta}(t)$ and $H_{\rm CD}^*(t) = H_{\rm CD}(t)$, $\forall~t$. Consequently, the work component $W_{\rm CD}$ (Eq. \eqref{stawcd}) vanishes for exact counterdiabatic driving.}

\subsection{Quantum advantage in many-body thermal machines}
\label{secQA}

Designing quantum machines which show quantum advantage, i.e, outperform equivalent classical machines, is one of the main aims of the field of quantum technology. In Ref. \cite{jaramillo16quantum}, the authors showed that many-particle quantum effects, coupled with non-adiabatic effects, in a finite-time many-body Otto engine can be beneficial in this respect. The authors considered a WM comprising of $N$ interacting bosons, described by the Hamiltonian
\ba
H = \sum_{i = 1}^N \left[-\frac{1}{2m}\nabla^2 + \frac{1}{2}m \vartheta(t)^2 {\bf r}_i^2 \right] + \sum_{i < j} V\left({\bf r_i} -  {\bf r_j}\right).
\label{eq:njp1}
\ea
 Here the inter-particle interaction
assumes the following scaling form: $V({\bf r}/b) = b^2 V({\bf r})$.

One can model an Otto cycle using the many-body WM as discussed in Sec. \ref{seccollotto}.
In this case, the work output $W_{\rm na}$ per cycle is given by 
\ba
W_{\rm na} = \left(Q_{AB}^* \frac{\vartheta_{\rm h}}{\vartheta_{\rm c}} - 1 \right) \mathcal{E}_{\rm A} +  \left(Q_{\rm CD}^* \frac{\vartheta_{\rm c}}{\vartheta_{\rm h}} - 1 \right) \mathcal{E}_{\rm C}.
\label{worknap}
\ea
Here as before, the frequency $\vartheta(t) = \vartheta_{\rm h}$ ($\vartheta_{\rm c}$) during the second (fourth) stroke, during which the WM thermalizes with the hot (cold) thermal bath, while
 $\mathcal{E}_j$ denotes the average energy of the WM at point $j = $ A, C. $Q_{\rm AB}^* \geq 1$ ($Q_{\rm CD}^* \geq 1$) denotes the {\it non-adiabaticity parameter} for the unitary stroke A $\to$ B (C $\to$ D), which is related to the non-adiabatic excitations generated due to the finite rate of driving during the unitary strokes. These  non-adiabaticity parameters reduce to unity in the adiabatic limit of slow driving, quantified by $\dot{\vartheta} \to 0$, or equivalently, infinite durations of the unitary strokes 1 and 3 ($\tau_1, \tau_3 \to \infty$).  {Comparing Eq. \eqref{worknap} with Eq. \eqref{eqwork}, one can equate $Q_{AB}^* \frac{\vartheta_{\rm h}}{\vartheta_{\rm c}} \mathcal{E}_{\rm A}$ and $Q_{\rm CD}^* \frac{\vartheta_{\rm c}}{\vartheta_{\rm h}} \mathcal{E}_{\rm C}$ with $\mathcal{E}_{\rm B}$ and $\mathcal{E}_{\rm D}$, respectively. Further, the heat flow $\mathcal{Q}_{\rm h}$ is given by $\mathcal{Q}_{\rm h} = \mathcal{E}_{\rm C} - \mathcal{E}_{\rm B} = \mathcal{E}_{\rm C} - Q_{AB}^* \frac{\vartheta_{\rm h}}{\vartheta_{\rm c}} \mathcal{E}_{\rm A}$. Consequently, 
the efficiency of the engine is given by (see Eqs. \eqref{heath} - \eqref{eqeff})
\ba
\eta_{\rm na} = -\frac{W_{\rm na}}{\mathcal{Q}_{\rm h}} = 1 - \frac{\vartheta_{\rm c}}{\vartheta_{\rm h}}\frac{Q_{\rm CD}^* \mathcal{E}_{\rm C} -   \frac{\vartheta_{\rm h}}{\vartheta_{\rm c}}\mathcal{E}_{\rm A} }{\mathcal{E}_{\rm C} - Q_{AB}^* \frac{\vartheta_{\rm h}}{\vartheta_{\rm c}}\mathcal{E}_{\rm A}}.
\label{eff_nonad}
\ea}
As can be seen from Eq. \eqref{eff_nonad}, $\eta_{\rm na}$ is maximum, and equals the Otto efficiency $1 - \vartheta_{\rm c}/\vartheta_{\rm h}$,  in the adiabatic limit, when $Q_{AB}^*, Q_{\rm CD}^* = 1$. However, this in turn leads to vanishing output power $\mathcal{P}_{\rm na} = W_{\rm na} /\tau_{\rm cyc}$. 

Interestingly, the authors showed that non-adiabatic effects arising due to finite $\tau_{\rm cyc}$ can lead to quantum advantage. In order to study quantum non-adiabatic effects induced enhancement in the performance of many-body engines, the authors considered the specific example of a one-dimensional WM with the inter-particle interaction given by
\ba
V(z_i, z_j) =  \frac{1}{m} \sum_{i<j}\sum_{i<j}\frac{\theta\left(\theta - 1 \right)}{\left(z_i - z_j \right)^2},
\ea 
where $\theta \geq 0$ denotes the  inter-particle interaction strength. {The many-particle effects can be quantified using the following ratios
\ba
r_{\rm na}^{(N, \theta)} \equiv \frac{\mathcal{P}_{\rm na}^{(N, \theta)}}{N\mathcal{P}_{\rm na}^{(1, \theta)}},~~~~~~ \rho_{\rm na}^{(N,\theta)} \equiv \frac{\eta_{\rm na}^{N, \theta}}{\eta_{\rm na}^{(1, \theta)}}.
\label{eq:QA}
\ea
Here $r_{\rm na}^{(N, \theta)}$  compares the optimal output power $\mathcal{P}_{\rm na}^{(N, \lambda)}$ of a
$N$-particle QHE with that of $N$ single-particle QHEs, while $\rho_{\rm na}^{(N,\theta)}$ does the same, but for the efficiency at optimal output power.
The output power is optimized through tuning the ratio $\vartheta_{\rm c}/\vartheta_{\rm h}$, for fixed values of $\vartheta_{\rm c}$, $T_{\rm c}$, $T_{\rm h}$, $\theta$ and $N$.  As shown in \cite{jaramillo16quantum}, quantum fluctuations, arising due to interplay of many-particle and non-adiabatic effects, can enhance the power output and efficiency of a $N$-particle QHE, as compared to a single-particle one. Consequently, we consider the QHE  exhibits quantum advantage when both ratios given in Eq. \eqref{eq:QA} exceed unity.} Remarkably, it was shown that judicious choice of the parameters $\lambda, T_{\rm c}, T_{\rm h}, N$ and $\theta$, for example, high temperatures of the hot bath and not very low temperatures of the cold bath, allows the finite-time QHE described above to achieve quantum advantage \cite{jaramillo16quantum}.

\subsection{Quantum engines based on localized states}
\label{secmbl}

The unique features of many-body localization (MBL) can be beneficial for designing quantum machines. Many-body localization prevents systems from thermalizing under their intrinsic dynamics \cite{aleiner10a, nandkishore15many, alet18many, decker20floquet}. On the other hand, thermalizing (or weakly localized) systems obey the {eigenstate thermalization hypothesis (ETH)}  \cite{rigol07relaxation, rigol08thermalization}.  In Ref. \cite{halpern2019quantum}, the authors used the difference in energy level statistics of these two kinds of systems to design a  QHE  with a many-body WM exhibiting a MBL phase. 

In order to understand the operation of such a heat engine, let us look at the energy-level statsistics followed by the systems in the above two regimes. For systems in the MBL
 regime, the probability $P(\delta)$ of an energy gap assuming a size $\delta$, approximately obey the Poisson statistics \cite{oganesyan07localization, pal10many} :
\ba
P_{\rm MBL}^{(E)}(\delta) \approx \frac{1}{\langle \delta \rangle_E} e^{-\delta/\langle \delta \rangle_E}.
\label{pmbl}
\ea
 Here $\langle \delta \rangle_E$ is the average gap at the energy $E$. As can be seen from Eq. \eqref{pmbl}, an energy gap
$\delta$ has a finite probability of vanishing, given by $P_{\rm MBL}^{(E)}(\delta \to 0) \approx 1/\langle \delta \rangle_E > 0$. On the other hand, energy gaps in systems in the thermalizing
regime obey the Gaussian
orthogonal ensemble (GOE)  statistics \cite{oganesyan07localization}:
\ba
P_{\rm GOE}^{(E)}(\delta) \approx \frac{\pi}{2}\frac{\delta}{\langle \delta \rangle_E^2} e^{-\frac{\pi}{4}\delta^2/\langle \delta \rangle_E^2}.
\label{pgoe}
\ea
In contrast to the MBL spectra Eq. \eqref{pmbl}, small gaps appear with vanishing probability in the thermalizing regime, i.e., $P_{\rm GOE}^{(E)}(\delta \to 0) \to 0$.

In Ref.  \cite{halpern2019quantum}, the authors considered an interacting many-body WM, described by the Hamiltonian
\ba
H_{\rm meso}(t) = \frac{\varepsilon}{\kappa(\vartheta(t))}\left[\left(1 - \vartheta(t) \right)H_{\rm GOE} + \vartheta(t) H_{\rm MBL} \right].
\label{hmbl}
\ea
Here the average energy density per site  $\varepsilon$ sets the unit of energy. One can change the qualitative nature of the WM by tuning the parameter $\vartheta(t) \in \left[0,1\right]$. 
For $\vartheta(t) = 0$, the Hamiltonian Eq. \eqref{hmbl} reduces to $H_{\rm GOE}$ with a spectrum such that the energy gaps $\delta$ are distributed according to GOE statistics Eq. \eqref{pgoe}. On the other hand, 
 $\vartheta(t) = 1$ results in $H_{\rm meso}$ (Eq. \eqref{hmbl}) taking the form of $H_{\rm MBL}$, whose gaps
follow the  Poisson statistics Eq.  \eqref{pmbl}.  The renormalization factor $\kappa(\vartheta(t))$ is chosen such that the average energy gap $\langle \delta \rangle_E$ is kept constant, thereby emphasizing the effect of level-statistics on the operation of the heat engine. 

One can exploit the difference in level statistics described above  to design an Otto cycle which delivers a net output work. To this end, 
we consider a cold bath with bandwidth $\mathcal{W}_{\rm b} \ll \langle \delta \rangle$, such that that energy flow between the WM and the cold bath occurs only for gaps $\delta \leq \mathcal{W}_{\rm b} \ll \langle \delta \rangle$. We note that such
anomalously small gaps can appear with finite probabilities in the MBL regime (Cf. Eq. \eqref{pmbl}), while this is not the case for the ETH regime (Cf. Eq. \eqref{pgoe}).

We consider the following Otto cycle using the many-body WM (Eq. \eqref{hmbl}): the cycle starts with the WM in the ETH phase ($\vartheta(t) = \vartheta_{\rm h} = 0$), when the WM 
is in thermal equilibrium with the hot bath at temperature $T_{\rm h}$. During the (unitary) 
stroke 1,  $\vartheta(t)$ is tuned from zero to $\vartheta = \vartheta_{\rm c} = 1$, such that $H(t)$ changes from $H_{\rm GOE}$ to $H_{\rm MBL}$. We assume
the tuning is slow enough so as to result in adiabatic dynamics during the unitary strokes, i.e.,  non-adiabatic excitations are negligible. The WM is coupled to  a cold bath at temperature $T_c$ during the second
  (non-unitary) stroke. As per the MBL level statistics Eq. \eqref{pmbl}, the gap is small enough with
probability $\sim \frac{\mathcal{W}_{\rm b}}{\langle\delta\rangle}$, so as to allow thermalization with the cold bath. In the third (unitary stroke), $\vartheta(t)$ is tuned to
zero such that we  arrive at $H_{\rm GOE}$.
Finally, the WM thermalizes with the hot bath during the fourth stroke, thereby completing the cycle. 

In the adiabatic limit of long cycle times, $T_{\rm c} \ll \mathcal{W}_{\rm b} \ll \langle \delta \rangle$, $T_{\rm h} = \infty$ and the gap distributions Eqs. \eqref{pmbl}, \eqref{pgoe} and the average gap $\langle \delta \rangle_E$ being independent of energy $E$, one can show that the average work output for the cycle described above is
\ba
W_{\rm tot} \approx -\mathcal{W}_{\rm b} + \frac{2\ln 2}{\beta_C},
\ea
whereas the efficiency is
\ba
\eta \approx 1 - \frac{\mathcal{W}_{\rm b}}{2\langle \delta\rangle}.
\ea
Clearly, a small $\mathcal{W}_{\rm b}$ allows one to operate the engine with high efficiency, while producing a non-zero output work 
$W_{\rm tot}$, brought about by 
the different energy gap statistics of $H_{\rm MBL}$ and $H_{\rm GOE}$.   Furthermore, one can robustly scale up the engine described above to the thermodynamic limit, wherein effectively independent subengines run in parallel without affecting each other, owing to the finite localization length of MBL systems.  We note that non-adiabatic excitations arising due to finite rate of driving during the unitary strokes can reduce the work output of the above engine. {Furthermore, instead of the Gaussian orthogonal ensemble considered above, one can model the engine by considering Gaussian unitary or Gaussian symplectic ensembles as well. However, since different ensembles model different symmetries, their performance as an engine
needs to be analyzed separately \cite{halpern2019quantum}.}

On a related note, mobility edges separating localized and delocalized states have also been used to design QHEs. In Ref. \cite{chiaracane20quasiperiodic}, the authors considered a central WM coupled to a hot left bath with temperature $T_{\rm h}$ and chemical potential $\mu_{\rm h}$, and a cold right bath with temperature $T_{\rm c} < T_{\rm h}$ and chemical potentials $\mu_{\rm h} \neq \mu_{c}$.
The WM was taken to be a generalized Aubry-Andre-Harper (GAAH) model, given by the Hamiltonian
\ba
H = \sum_{i=1}^{N-1} t a_i^{\dagger} a_{i+1} + \text{h.c.} + \sum_{i=1}^N V_i a_i^{\dagger} a_{i},
\ea
where $t$ is the tunneling
constant, $a_i$ is the Fermionic annihilation operator at
site $i$ and the onsite potential at site $i$ is given by the quasiperiodic function
\ba
V_i = \frac{2\theta \left(2\pi \upsilon i + \phi \right)}{1 - \alpha\cos\left(2\pi \upsilon i + \phi \right)}.
\ea
Here $\theta$ is the strength of the potential, the phase $\phi$ shifts the origin of the potential, $\upsilon$ is an irrational number
and  $\alpha \in \left(-1, 1\right)$.  {The energy $E_{\rm c}$ of the mobility edge, separating the delocalized states with eigenenergies $< E_{\rm c}$ from the localized states with eigenenergies $> E_{\rm c}$, can be tuned through the parameter $\alpha$, and is given by  \cite{ganeshan15nearest,archak17quasi}
\ba
E_c = \frac{1}{\alpha}\text{sign}\left(\theta\right)\left(|t| - |\theta| \right).
\ea}
The WM is coupled at the boundaries to the two non-interacting baths. The authors showed that  the  presence of a mobility edge introduces an energy filter, which results in asymmetry in the dynamics of particles and holes. This in turn leads to  non-zero steady-state heat current and power output, such that the setup can act as an autonomous thermoelectric heat engine.

\subsection{Quantum Szilard engine with interacting Bosons}
\label{secszilard}

In contrast to the heat engines described above, which convert heat energy into useful output work, Szilard engines operate isothermally, and use information to do the same \cite{kim11quantum, lutz15information, mohammady17a, bengtsson18quantum, bengtsson18bosonic, berut12experimental}. Szilard engines in the microscopic regime have been realized experimentally  \cite{toyabe10experimental, roldan14universal, koski14experimental, koski15on}. The setup of a Szilard engine comprises of $N \geq 1$ particles in a box of length $L$. The operation of such an engine can be described by the following quasistatic steps \cite{kim11quantum}: (i) insertion
of a partition in the box, at position $0 \leq l \leq L$, (ii) measurement of the particle
number $n \leq N$ on one side (left, say) of the partition (iii) reversibly
translating the partition to its final position $0 \leq l^{\prime} \leq L$ and
(iv) removal of the partition at $l^{\prime} $, thereby completing the cycle.  Here we use the information about the particle numbers $n$ and $N - n$ to determine $l^{\prime} $, which eventually allows us to generate output work. The engine is operated so that the setup is in thermal equilibrium with a bath at temperature $T$ at all times.  The quasi-static insertion and removal of the partition in steps (i) and (iii) respectively, can be assumed not to incur any cost, as long as we restrict ourselves to the classical regime.  However, the situation changes dramatically for a quantum Szilard engine - in this case, the insertion of a partition non-trivially changes the wavefunction of the particles inside the box, thereby making it necessary to ensure that the insertion and removal of the partition are done isothermally as well. Interestingly, in Ref. \cite{bengtsson18quantum}, the authors showed that the performance of a quantum Szilard engine can benefit from many-body effects. A quantum Szilard engine with attractively interacting Bosons was shown to  enhance the conversion from information to work, as compared to non-interacting or repulsive Bosons.

\section{Quantum batteries and quantum probes}
\label{secOQT}

In addition to quantum engines and refrigerators, the broad 
field of quantum technologies \cite{kurizki15quantum, millen16the} 
also encompasses studies on several related setups, including  quantum simulators \cite{bernien17probing, ebadi20quantum}, quantum cloaks \cite{constantinos20perfect}, quantum batteries \cite{campaioli18quantum}, quantum probes \cite{giovannetti11}, quantum transistors \cite{joulain16quantum}, heat rectifiers \cite{silva20heat} and quantum clocks \cite{buzek99optimal, erker17autonomous}. While they are vibrant fields of research, and subjects deserving detailed discussions on their own, in this section we briefly discuss the role played by many-body systems in the development of a couple of such technologies closely related to the  thermodynamics of quantum systems {and quantum thermal machines}, viz. quantum batteries \cite{campaioli18quantum} and quantum probes \cite{giovannetti11}.

\subsection{Quantum batteries}
\label{secQB}

An engine converts energy from a dissipative reservoir into usable output power. On the other hand, batteries can be used to store energy, which can be extracted and used later, at an opportune time. Consequently the field of quantum batteries, i.e., batteries based on quantum systems and following the laws of quantum mechanics, is closely related to the field of quantum engines and refrigerators. 

A discussion regarding quantum batteries necessitates the 
introduction of the concepts of passive states and ergotropy; 
a passive state is a state whose energy cannot be reduced 
any further through cyclic unitary transformation - such 
states are diagonal in the energy eigenbasis, with populations 
of energy levels decreasing with increasing energy eigenvalues. 
On the other hand, ergotropy is the maximum amount of work that 
can be  extracted from a non-passive state, through unitary 
transformations \cite{pusz78, lenard78}. Naturally, 
charging a quantum system (battery) involves imparting 
ergotropy to the system, so as to take it to a non-passive state. 
Thereafter, when needed, one can extract energy, i.e., discharge 
the battery through cyclic unitary transformations. 
The performance 
of a quantum battery can be quantified through the rate of charging / discharging and the maximum amount of work that can 
be extracted \cite{crescente20ultrafast}. 
Previous studies have suggested that entanglement can be beneficial for the performance of such quantum batteries \cite{alicki13entanglement}. 
{In case of a quantum battery comprised of multiple qubits, 
global entangling operations can enhance the performance of such a 
battery, through significant enhancement in charging power per qubit  \cite{binder15quantacell, campaioli18quantum}.} 
Collective effects have also been used to charge many-body 
quantum batteries \cite{andolina19quantum} through quantum 
heat engines \cite{ito20collectively} and through dissipative 
thermal baths \cite{cakmak20ergotropy}, and has been shown to yield quantum advantage 
\cite{campaioli17enhancing, ferraro18high}. 
Many-body systems have also been shown to be beneficial for 
designing quantum batteries, for example, in presence of 
many-body localization \cite{rossini19many}, through  
enhanced charging power \cite{le18spin, rossini19quantum}, 
and {ultra-stable charging characterized by low fluctuations in the stored energy} \cite{rosa20ultra}, and can raise 
interesting questions regarding the role of correlations 
in the performance of the batteries \cite{andolina19extractable}.
Charging and discharging of a quantum battery subject to static
driving and time dependent classical source is also
a recent addition to the growing literature 
\cite{carrega20dissipative,crescente20charging}.

\subsection{Quantum probes}
\label{secQP}

Studies of quantum systems in presence of dissipative baths, or designing quantum machines, usually involve precise estimation of system and bath properties, which then leads us to the field of quantum metrology. Quantum metrology \cite{giovannetti11, paris09quantum} deals with probes based on quantum systems, such as quantum thermometers \cite{brunelli11qubit, correa15individual, correa17enhancement, brunner17quantum,  hovhannisyan18measuring, depasquale18quantum,  potts19fundamental,  mukherjee19enhanced} and magnetometers \cite{bhattacharjee20quantum, levy20single}. The precision of estimation of a parameter $x$ (such as temperature) of a system is quantified by the relative error of measurement
\ba
e_{\rm rel} = \frac{\delta x}{x},
\ea
which is bounded by the quantum Fisher information (QFI) $\mathcal{G}$ through the Cramer-Rao bound \cite{paris09quantum}
\ba
e_{\rm rel} \geq \frac{1}{x\sqrt{\mathcal{M}\mathcal{G}}},
\label{eqCR}
\ea
where $\delta x$ denotes the absolute error in measurement and $\mathcal{M}$ is the number of measurements. As can be seen from Eq. \eqref{eqCR}, $e_{\rm rel}$ decreases with increasing $\mathcal{G}$. Consequently, one of the  major aims of research in this field has been to find ways to enhance the QFI, so as to result in high precision measurements. For example, several works have addressed the issue of high-accuracy estimation of low temperatures, through achievement of large QFI  \cite{correa15individual, campbell18precision, mukherjee19enhanced}. To this end, one generally considers a probe interacting with a thermal bath at temperature $T$, such that the state of the probe at some optimal time (which is usually taken to be the steady state) is given by $\rho(T)$. In this case, the QFI is  given by 
\ba
\mathcal{G}(T) = \lim_{\delta \to 0} \frac{\partial^2 \mathcal{F}(\rho(T)\rho(T+\delta))}{\partial \delta^2}.
\ea
Probes based on many-body quantum systems have been shown to advantageous in this context \cite{hovhannisyan18measuring, mok20optimal}; critical points are associated with divergences in QFI, thereby raising the possibility of high-precision quantum metrology using many-body systems \cite{zanardi08quantum, rams18at, potts19fundamental}. In addition, as for QHEs, collective effects have also been shown to aid in high-precision quantum thermometry \cite{latune20collective}.

\section{Discussion and outlook}
\label{seccon}

Harnessing many-body effects to design high-performing quantum 
technologies is a rapidly progressing area of research.
{Many-body effects may allow us to develop novel 
quantum machines, which can offer us significant insights into the thermodynamics of such many-body systems.} 
At the same time, studies of many-body systems can be accompanied 
by significant challenges, owing to the diverging size of the 
associated Hilbert space. These challenges have necessitated 
the development of several techniques focussed on dealing with 
the dynamics of closed and open many-body quantum systems \cite{fischer16dynamics, kolodrubetz17geometry, sels17minimizing, keck17dissipation, claeys19floquet, nathan20universal}. 
Such techniques have in turn enabled researchers to design and 
study different quantum technologies  based on many-body systems, 
in the last few years.  In this short review, we have discussed 
some of the recent literature on this fascinating subject. 

{An intriguing question is, can these quantum machines present us with significant advantages 
as compared to equivalent classical machines? As discussed in Sec. \ref{secQA}, careful modelling of quantum machines can present us with non-trivial quantum advantages \cite{binder15quantacell, jaramillo16quantum, mukherjee20anti}. Such results suggest significant advancement in technologies may be achieved by replacing the currently existing classical machines by their quantum counterparts, in near future.  However, fabricating these machines can incur additional implementational costs, which can significantly restrict the practical benefits of such quantum machines. Therefore, the extent to which quantum machines can offer technological improvement, inside laboratories as well as for everyday use, is an open question, which demands further rigorous research.}

Many-body effects in different forms can aid in the performance of quantum technologies. For example, cooperative effects arising due to collective coupling between many-body systems and dissipative baths have been shown to be beneficial for designing quantum engines \cite{niedenzu18cooperative, kloc19collective, latune20collective}, quantum thermometers \cite{latune20collective} and quantum batteries \cite{ito20collectively}. In parallel, machines based on interacting many-body systems give rise to rich physics as well, for example in the form of criticality \cite{revathy20universal, fogarty20a}, and can necessitate the introduction of shortcuts to adiabaticity for enhancing the performance of such machines \cite{hartmann19many, hartmann20multispin}. Many-body systems have also been shown to aid in fast charging of batteries \cite{campaioli18quantum, rossini19quantum}. Here we have mainly focussed on quantum engines and briefly addressed quantum batteries and quantum probes. However, studies on many-body systems to develop other quantum technologies, such as quantum clocks \cite{peres80measurement, buzek99optimal, erker17autonomous} and quantum transistors \cite{joulain16quantum}, may also lead to interesting results. 

The recent advances in experimental know-how have made the realization of several quantum technologies a possibility in various platforms. For example, interacting spin-chains models exhibiting phase transitions have been experimentally realized using quantum simulators \cite{bernien17probing, zhang17observation, ebadi20quantum}, trapped ions \cite{cui2016experimental} and quantum annealer \cite{bando20probing}.
 In the recent years single or few-particle engines have already been realized experimentally, for example, using single ions \cite{rossnagel16a}, mechanical oscillators \cite{klaers17squeezed}, nitrogen vacancy centers in diamonds \cite{klatzow19experimental}, Rydberg atoms \cite{kim18detailed, omran19generation} and optical lattices \cite{bason12high, schreiber15observation, kaufman16}.  WM based on strontium \cite{norcia18cavity, barberena19driven, tucker20facilitating} or rubidium \cite{karg20light} atoms may be used for realizing collective effects in QHEs. With the rapid advancement in development and control of systems in the quantum regime, such many-body quantum technologies can be expected to be realized experimentally in the near future.

\begin{acknowledgments}
It is a pleasure to acknowledge Adolfo del Campo, Andreas Hartmann, Wolfgang Lechner, Glen Bigan Mbeng, Wolfgang Niedenzu and Revathy B. S.  for related collaborative works.
VM acknowledges SERB, India for Start-up Research Grant SRG/2019/000411 and IISER Berhampur for Seed grant.
UD acknowledges DST, India for INSPIRE Research grant. 
\end{acknowledgments}


\begin{thebibliography}{214}%
\makeatletter
\providecommand \@ifxundefined [1]{%
 \@ifx{#1\undefined}
}%
\providecommand \@ifnum [1]{%
 \ifnum #1\expandafter \@firstoftwo
 \else \expandafter \@secondoftwo
 \fi
}%
\providecommand \@ifx [1]{%
 \ifx #1\expandafter \@firstoftwo
 \else \expandafter \@secondoftwo
 \fi
}%
\providecommand \natexlab [1]{#1}%
\providecommand \enquote  [1]{``#1''}%
\providecommand \bibnamefont  [1]{#1}%
\providecommand \bibfnamefont [1]{#1}%
\providecommand \citenamefont [1]{#1}%
\providecommand \href@noop [0]{\@secondoftwo}%
\providecommand \href [0]{\begingroup \@sanitize@url \@href}%
\providecommand \@href[1]{\@@startlink{#1}\@@href}%
\providecommand \@@href[1]{\endgroup#1\@@endlink}%
\providecommand \@sanitize@url [0]{\catcode `\\12\catcode `\$12\catcode
  `\&12\catcode `\#12\catcode `\^12\catcode `\_12\catcode `\%12\relax}%
\providecommand \@@startlink[1]{}%
\providecommand \@@endlink[0]{}%
\providecommand \url  [0]{\begingroup\@sanitize@url \@url }%
\providecommand \@url [1]{\endgroup\@href {#1}{\urlprefix }}%
\providecommand \urlprefix  [0]{URL }%
\providecommand \Eprint [0]{\href }%
\providecommand \doibase [0]{http://dx.doi.org/}%
\providecommand \selectlanguage [0]{\@gobble}%
\providecommand \bibinfo  [0]{\@secondoftwo}%
\providecommand \bibfield  [0]{\@secondoftwo}%
\providecommand \translation [1]{[#1]}%
\providecommand \BibitemOpen [0]{}%
\providecommand \bibitemStop [0]{}%
\providecommand \bibitemNoStop [0]{.\EOS\space}%
\providecommand \EOS [0]{\spacefactor3000\relax}%
\providecommand \BibitemShut  [1]{\csname bibitem#1\endcsname}%
\let\auto@bib@innerbib\@empty
\bibitem [{\citenamefont {Kurizki}\ \emph {et~al.}(2015)\citenamefont
  {Kurizki}, \citenamefont {Bertet}, \citenamefont {Kubo}, \citenamefont
  {M{\o}lmer}, \citenamefont {Petrosyan}, \citenamefont {Rabl},\ and\
  \citenamefont {Schmiedmayer}}]{kurizki15quantum}%
  \BibitemOpen
  \bibfield  {author} {\bibinfo {author} {\bibfnamefont {G.}~\bibnamefont
  {Kurizki}}, \bibinfo {author} {\bibfnamefont {P.}~\bibnamefont {Bertet}},
  \bibinfo {author} {\bibfnamefont {Y.}~\bibnamefont {Kubo}}, \bibinfo {author}
  {\bibfnamefont {K.}~\bibnamefont {M{\o}lmer}}, \bibinfo {author}
  {\bibfnamefont {D.}~\bibnamefont {Petrosyan}}, \bibinfo {author}
  {\bibfnamefont {P.}~\bibnamefont {Rabl}}, \ and\ \bibinfo {author}
  {\bibfnamefont {J.}~\bibnamefont {Schmiedmayer}},\ }\href {\doibase
  10.1073/pnas.1419326112} {\bibfield  {journal} {\bibinfo  {journal}
  {Proceedings of the National Academy of Sciences}\ }\textbf {\bibinfo
  {volume} {112}},\ \bibinfo {pages} {3866} (\bibinfo {year} {2015})},\ \Eprint
  {http://arxiv.org/abs/https://www.pnas.org/content/112/13/3866.full.pdf}
  {https://www.pnas.org/content/112/13/3866.full.pdf} \BibitemShut {NoStop}%
\bibitem [{\citenamefont {Millen}\ and\ \citenamefont
  {Xuereb}(2016)}]{millen16the}%
  \BibitemOpen
  \bibfield  {author} {\bibinfo {author} {\bibfnamefont {J.}~\bibnamefont
  {Millen}}\ and\ \bibinfo {author} {\bibfnamefont {A.}~\bibnamefont
  {Xuereb}},\ }\href {\doibase 10.1088/2058-7058/29/1/30} {\bibfield  {journal}
  {\bibinfo  {journal} {Physics World}\ }\textbf {\bibinfo {volume} {29}},\
  \bibinfo {pages} {23} (\bibinfo {year} {2016})}\BibitemShut {NoStop}%
\bibitem [{\citenamefont {Sakurai}\ and\ \citenamefont
  {Napolitano}(2017)}]{sakurai17modern}%
  \BibitemOpen
  \bibfield  {author} {\bibinfo {author} {\bibfnamefont {J.~J.}\ \bibnamefont
  {Sakurai}}\ and\ \bibinfo {author} {\bibfnamefont {J.}~\bibnamefont
  {Napolitano}},\ }\href {\doibase 10.1017/9781108499996} {\emph {\bibinfo
  {title} {Modern Quantum Mechanics}}},\ \bibinfo {edition} {2nd}\ ed.\
  (\bibinfo  {publisher} {Cambridge University Press},\ \bibinfo {year}
  {2017})\BibitemShut {NoStop}%
\bibitem [{\citenamefont {Callen}(1985)}]{callen85thermodynamics}%
  \BibitemOpen
  \bibfield  {author} {\bibinfo {author} {\bibfnamefont {H.~B.}\ \bibnamefont
  {Callen}},\ }\href@noop {} {\emph {\bibinfo {title} {Thermodynamics and An
  Introduction to Thermostatistics}}}\ (\bibinfo  {publisher} {John Wiley \&
  Sons Inc, New York, 1985},\ \bibinfo {year} {1985})\BibitemShut {NoStop}%
\bibitem [{\citenamefont {Kondepudi}\ and\ \citenamefont
  {Prigogine}(2015)}]{kondepudi15modern}%
  \BibitemOpen
  \bibfield  {author} {\bibinfo {author} {\bibfnamefont {D.}~\bibnamefont
  {Kondepudi}}\ and\ \bibinfo {author} {\bibfnamefont {I.}~\bibnamefont
  {Prigogine}},\ }\href@noop {} {\emph {\bibinfo {title} {Modern
  Thermodynamics}}}\ (\bibinfo  {publisher} {John Wiley \& Sons Ltd,
  Chichester, 2nd edn},\ \bibinfo {year} {2015})\BibitemShut {NoStop}%
\bibitem [{\citenamefont {Scovil}\ and\ \citenamefont
  {Schulz-DuBois}(1959)}]{scovil59three}%
  \BibitemOpen
  \bibfield  {author} {\bibinfo {author} {\bibfnamefont {H.~E.~D.}\
  \bibnamefont {Scovil}}\ and\ \bibinfo {author} {\bibfnamefont {E.~O.}\
  \bibnamefont {Schulz-DuBois}},\ }\href {\doibase 10.1103/PhysRevLett.2.262}
  {\bibfield  {journal} {\bibinfo  {journal} {Phys. Rev. Lett.}\ }\textbf
  {\bibinfo {volume} {2}},\ \bibinfo {pages} {262} (\bibinfo {year}
  {1959})}\BibitemShut {NoStop}%
\bibitem [{\citenamefont {Scully}\ \emph {et~al.}(2003)\citenamefont {Scully},
  \citenamefont {Zubairy}, \citenamefont {Agarwal},\ and\ \citenamefont
  {Walther}}]{scully03extracting}%
  \BibitemOpen
  \bibfield  {author} {\bibinfo {author} {\bibfnamefont {M.~O.}\ \bibnamefont
  {Scully}}, \bibinfo {author} {\bibfnamefont {M.~S.}\ \bibnamefont {Zubairy}},
  \bibinfo {author} {\bibfnamefont {G.~S.}\ \bibnamefont {Agarwal}}, \ and\
  \bibinfo {author} {\bibfnamefont {H.}~\bibnamefont {Walther}},\ }\href
  {\doibase 10.1126/science.1078955} {\bibfield  {journal} {\bibinfo  {journal}
  {Science}\ }\textbf {\bibinfo {volume} {299}},\ \bibinfo {pages} {862}
  (\bibinfo {year} {2003})},\ \Eprint
  {http://arxiv.org/abs/https://science.sciencemag.org/content/299/5608/862.full.pdf}
  {https://science.sciencemag.org/content/299/5608/862.full.pdf} \BibitemShut
  {NoStop}%
\bibitem [{\citenamefont {Gemmer}\ \emph {et~al.}(2009)\citenamefont {Gemmer},
  \citenamefont {Michel},\ and\ \citenamefont {Mahler}}]{gemmer2009quantum}%
  \BibitemOpen
  \bibfield  {author} {\bibinfo {author} {\bibfnamefont {J.}~\bibnamefont
  {Gemmer}}, \bibinfo {author} {\bibfnamefont {M.}~\bibnamefont {Michel}}, \
  and\ \bibinfo {author} {\bibfnamefont {G.}~\bibnamefont {Mahler}},\
  }\href@noop {} {\emph {\bibinfo {title} {Quantum thermodynamics: Emergence of
  thermodynamic behavior within composite quantum systems}}},\ Vol.\ \bibinfo
  {volume} {784}\ (\bibinfo  {publisher} {Springer},\ \bibinfo {year}
  {2009})\BibitemShut {NoStop}%
\bibitem [{\citenamefont {Kosloff}(2013)}]{kosloff13quantum}%
  \BibitemOpen
  \bibfield  {author} {\bibinfo {author} {\bibfnamefont {R.}~\bibnamefont
  {Kosloff}},\ }\href {\doibase 10.3390/e15062100} {\bibfield  {journal}
  {\bibinfo  {journal} {Entropy}\ }\textbf {\bibinfo {volume} {15}},\ \bibinfo
  {pages} {2100} (\bibinfo {year} {2013})}\BibitemShut {NoStop}%
\bibitem [{\citenamefont {Kosloff}\ and\ \citenamefont
  {Levy}(2014)}]{kosloff14quantum}%
  \BibitemOpen
  \bibfield  {author} {\bibinfo {author} {\bibfnamefont {R.}~\bibnamefont
  {Kosloff}}\ and\ \bibinfo {author} {\bibfnamefont {A.}~\bibnamefont {Levy}},\
  }\href {\doibase 10.1146/annurev-physchem-040513-103724} {\bibfield
  {journal} {\bibinfo  {journal} {Annual Review of Physical Chemistry}\
  }\textbf {\bibinfo {volume} {65}},\ \bibinfo {pages} {365} (\bibinfo {year}
  {2014})}\BibitemShut {NoStop}%
\bibitem [{\citenamefont {Brand{\~a}o}\ \emph {et~al.}(2015)\citenamefont
  {Brand{\~a}o}, \citenamefont {Horodecki}, \citenamefont {Ng}, \citenamefont
  {Oppenheim},\ and\ \citenamefont {Wehner}}]{brandao15the}%
  \BibitemOpen
  \bibfield  {author} {\bibinfo {author} {\bibfnamefont {F.}~\bibnamefont
  {Brand{\~a}o}}, \bibinfo {author} {\bibfnamefont {M.}~\bibnamefont
  {Horodecki}}, \bibinfo {author} {\bibfnamefont {N.}~\bibnamefont {Ng}},
  \bibinfo {author} {\bibfnamefont {J.}~\bibnamefont {Oppenheim}}, \ and\
  \bibinfo {author} {\bibfnamefont {S.}~\bibnamefont {Wehner}},\ }\href
  {\doibase 10.1073/pnas.1411728112} {\bibfield  {journal} {\bibinfo  {journal}
  {Proceedings of the National Academy of Sciences}\ }\textbf {\bibinfo
  {volume} {112}},\ \bibinfo {pages} {3275} (\bibinfo {year} {2015})},\ \Eprint
  {http://arxiv.org/abs/https://www.pnas.org/content/112/11/3275.full.pdf}
  {https://www.pnas.org/content/112/11/3275.full.pdf} \BibitemShut {NoStop}%
\bibitem [{\citenamefont {Gelbwaser-Klimovsky}\ \emph
  {et~al.}(2015)\citenamefont {Gelbwaser-Klimovsky}, \citenamefont {Niedenzu},\
  and\ \citenamefont {Kurizki}}]{klimovsky15chapter}%
  \BibitemOpen
  \bibfield  {author} {\bibinfo {author} {\bibfnamefont {D.}~\bibnamefont
  {Gelbwaser-Klimovsky}}, \bibinfo {author} {\bibfnamefont {W.}~\bibnamefont
  {Niedenzu}}, \ and\ \bibinfo {author} {\bibfnamefont {G.}~\bibnamefont
  {Kurizki}},\ }\href {\doibase https://doi.org/10.1016/bs.aamop.2015.07.002}
  {\bibfield  {journal} {\bibinfo  {journal} {Advances In Atomic, Molecular,
  and Optical Physics}\ }\textbf {\bibinfo {volume} {64}},\ \bibinfo {pages}
  {329 } (\bibinfo {year} {2015})}\BibitemShut {NoStop}%
\bibitem [{\citenamefont {Vinjanampathy}\ and\ \citenamefont
  {Anders}(2016)}]{vinjanampathy16quantum}%
  \BibitemOpen
  \bibfield  {author} {\bibinfo {author} {\bibfnamefont {S.}~\bibnamefont
  {Vinjanampathy}}\ and\ \bibinfo {author} {\bibfnamefont {J.}~\bibnamefont
  {Anders}},\ }\href {\doibase 10.1080/00107514.2016.1201896} {\bibfield
  {journal} {\bibinfo  {journal} {Contemporary Physics}\ }\textbf {\bibinfo
  {volume} {57}},\ \bibinfo {pages} {545} (\bibinfo {year} {2016})},\ \Eprint
  {http://arxiv.org/abs/https://doi.org/10.1080/00107514.2016.1201896}
  {https://doi.org/10.1080/00107514.2016.1201896} \BibitemShut {NoStop}%
\bibitem [{\citenamefont {Bera}\ \emph {et~al.}(2017)\citenamefont {Bera},
  \citenamefont {Riera}, \citenamefont {Lewenstein},\ and\ \citenamefont
  {Winter}}]{bera17generalized}%
  \BibitemOpen
  \bibfield  {author} {\bibinfo {author} {\bibfnamefont {M.~N.}\ \bibnamefont
  {Bera}}, \bibinfo {author} {\bibfnamefont {A.}~\bibnamefont {Riera}},
  \bibinfo {author} {\bibfnamefont {M.}~\bibnamefont {Lewenstein}}, \ and\
  \bibinfo {author} {\bibfnamefont {A.}~\bibnamefont {Winter}},\ }\href
  {\doibase 10.1038/s41467-017-02370-x} {\bibfield  {journal} {\bibinfo
  {journal} {Nat. Commun.}\ }\textbf {\bibinfo {volume} {8}},\ \bibinfo {pages}
  {2180} (\bibinfo {year} {2017})}\BibitemShut {NoStop}%
\bibitem [{\citenamefont {Masanes}\ and\ \citenamefont
  {Oppenheim}(2017)}]{masanes17a}%
  \BibitemOpen
  \bibfield  {author} {\bibinfo {author} {\bibfnamefont {L.}~\bibnamefont
  {Masanes}}\ and\ \bibinfo {author} {\bibfnamefont {J.}~\bibnamefont
  {Oppenheim}},\ }\href {\doibase 10.1038/ncomms14538} {\bibfield  {journal}
  {\bibinfo  {journal} {Nat. Commun.}\ }\textbf {\bibinfo {volume} {8}},\
  \bibinfo {pages} {14538} (\bibinfo {year} {2017})}\BibitemShut {NoStop}%
\bibitem [{\citenamefont {Binder}\ \emph {et~al.}(2018)\citenamefont {Binder},
  \citenamefont {Correa}, \citenamefont {Gogolin}, \citenamefont {Anders},\
  and\ \citenamefont {Adesso}}]{binder18book}%
  \BibitemOpen
  \bibinfo {editor} {\bibfnamefont {F.}~\bibnamefont {Binder}}, \bibinfo
  {editor} {\bibfnamefont {L.~A.}\ \bibnamefont {Correa}}, \bibinfo {editor}
  {\bibfnamefont {C.}~\bibnamefont {Gogolin}}, \bibinfo {editor} {\bibfnamefont
  {J.}~\bibnamefont {Anders}}, \ and\ \bibinfo {editor} {\bibfnamefont
  {G.}~\bibnamefont {Adesso}},\ eds.,\ \href@noop {} {\emph {\bibinfo {title}
  {Thermodynamics in the quantum regime}}}\ (\bibinfo  {publisher} {Springer
  International Publishing},\ \bibinfo {year} {2018})\BibitemShut {NoStop}%
\bibitem [{\citenamefont {Tuncer}\ and\ \citenamefont
  {M\"{u}stecaplıo\={g}lu}(2020)}]{tuncer20quantum}%
  \BibitemOpen
  \bibfield  {author} {\bibinfo {author} {\bibfnamefont {A.}~\bibnamefont
  {Tuncer}}\ and\ \bibinfo {author} {\bibfnamefont {O.~E.}\ \bibnamefont
  {M\"{u}stecaplıo\={g}lu}},\ }\href {\doibase 10.3906/fiz-2009-12} {\bibfield
   {journal} {\bibinfo  {journal} {Turk J Phys}\ }\textbf {\bibinfo {volume}
  {44}},\ \bibinfo {pages} {404} (\bibinfo {year} {2020})}\BibitemShut
  {NoStop}%
\bibitem [{\citenamefont {Bhattacharjee}\ and\ \citenamefont
  {Dutta}(2020)}]{bhattacharjee20quantum_revw}%
  \BibitemOpen
  \bibfield  {author} {\bibinfo {author} {\bibfnamefont {S.}~\bibnamefont
  {Bhattacharjee}}\ and\ \bibinfo {author} {\bibfnamefont {A.}~\bibnamefont
  {Dutta}},\ }\href@noop {} {\enquote {\bibinfo {title} {Quantum thermal
  machines and batteries},}\ } (\bibinfo {year} {2020}),\ \Eprint
  {http://arxiv.org/abs/2008.07889} {arXiv:2008.07889 [quant-ph]} \BibitemShut
  {NoStop}%
\bibitem [{\citenamefont {Dolde}\ \emph {et~al.}(2011)\citenamefont {Dolde},
  \citenamefont {Fedder}, \citenamefont {Doherty}, \citenamefont {N{\"o}bauer},
  \citenamefont {Rempp}, \citenamefont {Balasubramanian}, \citenamefont {Wolf},
  \citenamefont {Reinhard}, \citenamefont {Hollenberg}, \citenamefont
  {Jelezko},\ and\ \citenamefont {Wrachtrup}}]{dolde11electric}%
  \BibitemOpen
  \bibfield  {author} {\bibinfo {author} {\bibfnamefont {F.}~\bibnamefont
  {Dolde}}, \bibinfo {author} {\bibfnamefont {H.}~\bibnamefont {Fedder}},
  \bibinfo {author} {\bibfnamefont {M.~W.}\ \bibnamefont {Doherty}}, \bibinfo
  {author} {\bibfnamefont {T.}~\bibnamefont {N{\"o}bauer}}, \bibinfo {author}
  {\bibfnamefont {F.}~\bibnamefont {Rempp}}, \bibinfo {author} {\bibfnamefont
  {G.}~\bibnamefont {Balasubramanian}}, \bibinfo {author} {\bibfnamefont
  {T.}~\bibnamefont {Wolf}}, \bibinfo {author} {\bibfnamefont {F.}~\bibnamefont
  {Reinhard}}, \bibinfo {author} {\bibfnamefont {L.~C.~L.}\ \bibnamefont
  {Hollenberg}}, \bibinfo {author} {\bibfnamefont {F.}~\bibnamefont {Jelezko}},
  \ and\ \bibinfo {author} {\bibfnamefont {J.}~\bibnamefont {Wrachtrup}},\
  }\href {\doibase 10.1038/nphys1969} {\bibfield  {journal} {\bibinfo
  {journal} {Nature Physics}\ }\textbf {\bibinfo {volume} {7}},\ \bibinfo
  {pages} {459} (\bibinfo {year} {2011})}\BibitemShut {NoStop}%
\bibitem [{\citenamefont {Bason}\ \emph {et~al.}(2012)\citenamefont {Bason},
  \citenamefont {Viteau}, \citenamefont {Malossi}, \citenamefont {Huillery},
  \citenamefont {Arimondo}, \citenamefont {Ciampini}, \citenamefont {Fazio},
  \citenamefont {Giovannetti}, \citenamefont {Mannella},\ and\ \citenamefont
  {Morsch}}]{bason12high}%
  \BibitemOpen
  \bibfield  {author} {\bibinfo {author} {\bibfnamefont {M.~G.}\ \bibnamefont
  {Bason}}, \bibinfo {author} {\bibfnamefont {M.}~\bibnamefont {Viteau}},
  \bibinfo {author} {\bibfnamefont {N.}~\bibnamefont {Malossi}}, \bibinfo
  {author} {\bibfnamefont {P.}~\bibnamefont {Huillery}}, \bibinfo {author}
  {\bibfnamefont {E.}~\bibnamefont {Arimondo}}, \bibinfo {author}
  {\bibfnamefont {D.}~\bibnamefont {Ciampini}}, \bibinfo {author}
  {\bibfnamefont {R.}~\bibnamefont {Fazio}}, \bibinfo {author} {\bibfnamefont
  {V.}~\bibnamefont {Giovannetti}}, \bibinfo {author} {\bibfnamefont
  {R.}~\bibnamefont {Mannella}}, \ and\ \bibinfo {author} {\bibfnamefont
  {O.}~\bibnamefont {Morsch}},\ }\href {\doibase 10.1038/nphys2170} {\bibfield
  {journal} {\bibinfo  {journal} {Nature Physics}\ }\textbf {\bibinfo {volume}
  {8}},\ \bibinfo {pages} {147} (\bibinfo {year} {2012})}\BibitemShut {NoStop}%
\bibitem [{\citenamefont {Kucsko}\ \emph {et~al.}(2013)\citenamefont {Kucsko},
  \citenamefont {Maurer}, \citenamefont {Yao}, \citenamefont {Kubo},
  \citenamefont {Noh}, \citenamefont {Lo}, \citenamefont {Park},\ and\
  \citenamefont {Lukin}}]{kucksko13nanometre}%
  \BibitemOpen
  \bibfield  {author} {\bibinfo {author} {\bibfnamefont {G.}~\bibnamefont
  {Kucsko}}, \bibinfo {author} {\bibfnamefont {P.~C.}\ \bibnamefont {Maurer}},
  \bibinfo {author} {\bibfnamefont {N.~Y.}\ \bibnamefont {Yao}}, \bibinfo
  {author} {\bibfnamefont {M.}~\bibnamefont {Kubo}}, \bibinfo {author}
  {\bibfnamefont {H.~J.}\ \bibnamefont {Noh}}, \bibinfo {author} {\bibfnamefont
  {P.~K.}\ \bibnamefont {Lo}}, \bibinfo {author} {\bibfnamefont
  {H.}~\bibnamefont {Park}}, \ and\ \bibinfo {author} {\bibfnamefont {M.~D.}\
  \bibnamefont {Lukin}},\ }\href {\doibase 10.1038/nature12373} {\bibfield
  {journal} {\bibinfo  {journal} {Nature}\ }\textbf {\bibinfo {volume} {500}},\
  \bibinfo {pages} {54} (\bibinfo {year} {2013})}\BibitemShut {NoStop}%
\bibitem [{\citenamefont {Laskar}\ \emph {et~al.}(2020)\citenamefont {Laskar},
  \citenamefont {Adhikary}, \citenamefont {Mondal}, \citenamefont {Katiyar},
  \citenamefont {Vinjanampathy},\ and\ \citenamefont
  {Ghosh}}]{laskar20observation}%
  \BibitemOpen
  \bibfield  {author} {\bibinfo {author} {\bibfnamefont {A.~W.}\ \bibnamefont
  {Laskar}}, \bibinfo {author} {\bibfnamefont {P.}~\bibnamefont {Adhikary}},
  \bibinfo {author} {\bibfnamefont {S.}~\bibnamefont {Mondal}}, \bibinfo
  {author} {\bibfnamefont {P.}~\bibnamefont {Katiyar}}, \bibinfo {author}
  {\bibfnamefont {S.}~\bibnamefont {Vinjanampathy}}, \ and\ \bibinfo {author}
  {\bibfnamefont {S.}~\bibnamefont {Ghosh}},\ }\href {\doibase
  10.1103/PhysRevLett.125.013601} {\bibfield  {journal} {\bibinfo  {journal}
  {Phys. Rev. Lett.}\ }\textbf {\bibinfo {volume} {125}},\ \bibinfo {pages}
  {013601} (\bibinfo {year} {2020})}\BibitemShut {NoStop}%
\bibitem [{\citenamefont {Pal}\ \emph {et~al.}(2020)\citenamefont {Pal},
  \citenamefont {Saryal}, \citenamefont {Segal}, \citenamefont {Mahesh},\ and\
  \citenamefont {Agarwalla}}]{pal20experimental}%
  \BibitemOpen
  \bibfield  {author} {\bibinfo {author} {\bibfnamefont {S.}~\bibnamefont
  {Pal}}, \bibinfo {author} {\bibfnamefont {S.}~\bibnamefont {Saryal}},
  \bibinfo {author} {\bibfnamefont {D.}~\bibnamefont {Segal}}, \bibinfo
  {author} {\bibfnamefont {T.~S.}\ \bibnamefont {Mahesh}}, \ and\ \bibinfo
  {author} {\bibfnamefont {B.~K.}\ \bibnamefont {Agarwalla}},\ }\href {\doibase
  10.1103/PhysRevResearch.2.022044} {\bibfield  {journal} {\bibinfo  {journal}
  {Phys. Rev. Research}\ }\textbf {\bibinfo {volume} {2}},\ \bibinfo {pages}
  {022044} (\bibinfo {year} {2020})}\BibitemShut {NoStop}%
\bibitem [{\citenamefont {Koski}\ \emph {et~al.}(2014)\citenamefont {Koski},
  \citenamefont {Maisi}, \citenamefont {Pekola},\ and\ \citenamefont
  {Averin}}]{koski14experimental}%
  \BibitemOpen
  \bibfield  {author} {\bibinfo {author} {\bibfnamefont {J.~V.}\ \bibnamefont
  {Koski}}, \bibinfo {author} {\bibfnamefont {V.~F.}\ \bibnamefont {Maisi}},
  \bibinfo {author} {\bibfnamefont {J.~P.}\ \bibnamefont {Pekola}}, \ and\
  \bibinfo {author} {\bibfnamefont {D.~V.}\ \bibnamefont {Averin}},\ }\href
  {\doibase 10.1073/pnas.1406966111} {\bibfield  {journal} {\bibinfo  {journal}
  {Proceedings of the National Academy of Sciences}\ }\textbf {\bibinfo
  {volume} {111}},\ \bibinfo {pages} {13786} (\bibinfo {year} {2014})},\
  \Eprint
  {http://arxiv.org/abs/https://www.pnas.org/content/111/38/13786.full.pdf}
  {https://www.pnas.org/content/111/38/13786.full.pdf} \BibitemShut {NoStop}%
\bibitem [{\citenamefont {Ro{\ss}nagel}\ \emph {et~al.}(2016)\citenamefont
  {Ro{\ss}nagel}, \citenamefont {Dawkins}, \citenamefont {Tolazzi},
  \citenamefont {Abah}, \citenamefont {Lutz}, \citenamefont {Schmidt-Kaler},\
  and\ \citenamefont {Singer}}]{rossnagel16a}%
  \BibitemOpen
  \bibfield  {author} {\bibinfo {author} {\bibfnamefont {J.}~\bibnamefont
  {Ro{\ss}nagel}}, \bibinfo {author} {\bibfnamefont {S.~T.}\ \bibnamefont
  {Dawkins}}, \bibinfo {author} {\bibfnamefont {K.~N.}\ \bibnamefont
  {Tolazzi}}, \bibinfo {author} {\bibfnamefont {O.}~\bibnamefont {Abah}},
  \bibinfo {author} {\bibfnamefont {E.}~\bibnamefont {Lutz}}, \bibinfo {author}
  {\bibfnamefont {F.}~\bibnamefont {Schmidt-Kaler}}, \ and\ \bibinfo {author}
  {\bibfnamefont {K.}~\bibnamefont {Singer}},\ }\href {\doibase
  10.1126/science.aad6320} {\bibfield  {journal} {\bibinfo  {journal}
  {Science}\ }\textbf {\bibinfo {volume} {352}},\ \bibinfo {pages} {325}
  (\bibinfo {year} {2016})}\BibitemShut {NoStop}%
\bibitem [{\citenamefont {Klaers}\ \emph {et~al.}(2017)\citenamefont {Klaers},
  \citenamefont {Faelt}, \citenamefont {Imamoglu},\ and\ \citenamefont
  {Togan}}]{klaers17squeezed}%
  \BibitemOpen
  \bibfield  {author} {\bibinfo {author} {\bibfnamefont {J.}~\bibnamefont
  {Klaers}}, \bibinfo {author} {\bibfnamefont {S.}~\bibnamefont {Faelt}},
  \bibinfo {author} {\bibfnamefont {A.}~\bibnamefont {Imamoglu}}, \ and\
  \bibinfo {author} {\bibfnamefont {E.}~\bibnamefont {Togan}},\ }\href
  {\doibase 10.1103/PhysRevX.7.031044} {\bibfield  {journal} {\bibinfo
  {journal} {Phys. Rev. X}\ }\textbf {\bibinfo {volume} {7}},\ \bibinfo {pages}
  {031044} (\bibinfo {year} {2017})}\BibitemShut {NoStop}%
\bibitem [{\citenamefont {Peterson}\ \emph {et~al.}(2019)\citenamefont
  {Peterson}, \citenamefont {Batalh\~ao}, \citenamefont {Herrera},
  \citenamefont {Souza}, \citenamefont {Sarthour}, \citenamefont {Oliveira},\
  and\ \citenamefont {Serra}}]{Peterson18}%
  \BibitemOpen
  \bibfield  {author} {\bibinfo {author} {\bibfnamefont {J.~P.~S.}\
  \bibnamefont {Peterson}}, \bibinfo {author} {\bibfnamefont {T.~B.}\
  \bibnamefont {Batalh\~ao}}, \bibinfo {author} {\bibfnamefont
  {M.}~\bibnamefont {Herrera}}, \bibinfo {author} {\bibfnamefont {A.~M.}\
  \bibnamefont {Souza}}, \bibinfo {author} {\bibfnamefont {R.~S.}\ \bibnamefont
  {Sarthour}}, \bibinfo {author} {\bibfnamefont {I.~S.}\ \bibnamefont
  {Oliveira}}, \ and\ \bibinfo {author} {\bibfnamefont {R.~M.}\ \bibnamefont
  {Serra}},\ }\href {\doibase 10.1103/PhysRevLett.123.240601} {\bibfield
  {journal} {\bibinfo  {journal} {Phys. Rev. Lett.}\ }\textbf {\bibinfo
  {volume} {123}},\ \bibinfo {pages} {240601} (\bibinfo {year}
  {2019})}\BibitemShut {NoStop}%
\bibitem [{\citenamefont {Klatzow}\ \emph {et~al.}(2019)\citenamefont
  {Klatzow}, \citenamefont {Becker}, \citenamefont {Ledingham}, \citenamefont
  {Weinzetl}, \citenamefont {Kaczmarek}, \citenamefont {Saunders},
  \citenamefont {Nunn}, \citenamefont {Walmsley}, \citenamefont {Uzdin},\ and\
  \citenamefont {Poem}}]{klatzow19experimental}%
  \BibitemOpen
  \bibfield  {author} {\bibinfo {author} {\bibfnamefont {J.}~\bibnamefont
  {Klatzow}}, \bibinfo {author} {\bibfnamefont {J.~N.}\ \bibnamefont {Becker}},
  \bibinfo {author} {\bibfnamefont {P.~M.}\ \bibnamefont {Ledingham}}, \bibinfo
  {author} {\bibfnamefont {C.}~\bibnamefont {Weinzetl}}, \bibinfo {author}
  {\bibfnamefont {K.~T.}\ \bibnamefont {Kaczmarek}}, \bibinfo {author}
  {\bibfnamefont {D.~J.}\ \bibnamefont {Saunders}}, \bibinfo {author}
  {\bibfnamefont {J.}~\bibnamefont {Nunn}}, \bibinfo {author} {\bibfnamefont
  {I.~A.}\ \bibnamefont {Walmsley}}, \bibinfo {author} {\bibfnamefont
  {R.}~\bibnamefont {Uzdin}}, \ and\ \bibinfo {author} {\bibfnamefont
  {E.}~\bibnamefont {Poem}},\ }\href {\doibase 10.1103/PhysRevLett.122.110601}
  {\bibfield  {journal} {\bibinfo  {journal} {Phys. Rev. Lett.}\ }\textbf
  {\bibinfo {volume} {122}},\ \bibinfo {pages} {110601} (\bibinfo {year}
  {2019})}\BibitemShut {NoStop}%
\bibitem [{\citenamefont {Maslennikov}\ \emph {et~al.}(2019)\citenamefont
  {Maslennikov}, \citenamefont {Ding}, \citenamefont {Habl{\"u}tzel},
  \citenamefont {Gan}, \citenamefont {Roulet}, \citenamefont {Nimmrichter},
  \citenamefont {Dai}, \citenamefont {Scarani},\ and\ \citenamefont
  {Matsukevich}}]{maslennikov19quantum}%
  \BibitemOpen
  \bibfield  {author} {\bibinfo {author} {\bibfnamefont {G.}~\bibnamefont
  {Maslennikov}}, \bibinfo {author} {\bibfnamefont {S.}~\bibnamefont {Ding}},
  \bibinfo {author} {\bibfnamefont {R.}~\bibnamefont {Habl{\"u}tzel}}, \bibinfo
  {author} {\bibfnamefont {J.}~\bibnamefont {Gan}}, \bibinfo {author}
  {\bibfnamefont {A.}~\bibnamefont {Roulet}}, \bibinfo {author} {\bibfnamefont
  {S.}~\bibnamefont {Nimmrichter}}, \bibinfo {author} {\bibfnamefont
  {J.}~\bibnamefont {Dai}}, \bibinfo {author} {\bibfnamefont {V.}~\bibnamefont
  {Scarani}}, \ and\ \bibinfo {author} {\bibfnamefont {D.}~\bibnamefont
  {Matsukevich}},\ }\href {\doibase 10.1038/s41467-018-08090-0} {\bibfield
  {journal} {\bibinfo  {journal} {Nature Communications}\ }\textbf {\bibinfo
  {volume} {10}},\ \bibinfo {pages} {202} (\bibinfo {year} {2019})}\BibitemShut
  {NoStop}%
\bibitem [{\citenamefont {Peterson}\ \emph {et~al.}(2020)\citenamefont
  {Peterson}, \citenamefont {Sarthour},\ and\ \citenamefont
  {Laflamme}}]{peterson20implementation}%
  \BibitemOpen
  \bibfield  {author} {\bibinfo {author} {\bibfnamefont {J.~P.~S.}\
  \bibnamefont {Peterson}}, \bibinfo {author} {\bibfnamefont {R.~S.}\
  \bibnamefont {Sarthour}}, \ and\ \bibinfo {author} {\bibfnamefont
  {R.}~\bibnamefont {Laflamme}},\ }\href@noop {} {\enquote {\bibinfo {title}
  {Implementation of a quantum engine fuelled by information},}\ } (\bibinfo
  {year} {2020}),\ \Eprint {http://arxiv.org/abs/2006.10136} {arXiv:2006.10136
  [quant-ph]} \BibitemShut {NoStop}%
\bibitem [{\citenamefont {Keck}\ \emph {et~al.}(2017)\citenamefont {Keck},
  \citenamefont {Montangero}, \citenamefont {Santoro}, \citenamefont {Fazio},\
  and\ \citenamefont {Rossini}}]{keck17dissipation}%
  \BibitemOpen
  \bibfield  {author} {\bibinfo {author} {\bibfnamefont {M.}~\bibnamefont
  {Keck}}, \bibinfo {author} {\bibfnamefont {S.}~\bibnamefont {Montangero}},
  \bibinfo {author} {\bibfnamefont {G.~E.}\ \bibnamefont {Santoro}}, \bibinfo
  {author} {\bibfnamefont {R.}~\bibnamefont {Fazio}}, \ and\ \bibinfo {author}
  {\bibfnamefont {D.}~\bibnamefont {Rossini}},\ }\href {\doibase
  10.1088/1367-2630/aa8cef} {\bibfield  {journal} {\bibinfo  {journal} {New
  Journal of Physics}\ }\textbf {\bibinfo {volume} {19}},\ \bibinfo {pages}
  {113029} (\bibinfo {year} {2017})}\BibitemShut {NoStop}%
\bibitem [{\citenamefont {Xu}\ \emph {et~al.}(2019)\citenamefont {Xu},
  \citenamefont {Thingna}, \citenamefont {Guo},\ and\ \citenamefont
  {Poletti}}]{xu19many}%
  \BibitemOpen
  \bibfield  {author} {\bibinfo {author} {\bibfnamefont {X.}~\bibnamefont
  {Xu}}, \bibinfo {author} {\bibfnamefont {J.}~\bibnamefont {Thingna}},
  \bibinfo {author} {\bibfnamefont {C.}~\bibnamefont {Guo}}, \ and\ \bibinfo
  {author} {\bibfnamefont {D.}~\bibnamefont {Poletti}},\ }\href {\doibase
  10.1103/PhysRevA.99.012106} {\bibfield  {journal} {\bibinfo  {journal} {Phys.
  Rev. A}\ }\textbf {\bibinfo {volume} {99}},\ \bibinfo {pages} {012106}
  (\bibinfo {year} {2019})}\BibitemShut {NoStop}%
\bibitem [{\citenamefont {Nathan}\ and\ \citenamefont
  {Rudner}(2020)}]{nathan20universal}%
  \BibitemOpen
  \bibfield  {author} {\bibinfo {author} {\bibfnamefont {F.}~\bibnamefont
  {Nathan}}\ and\ \bibinfo {author} {\bibfnamefont {M.~S.}\ \bibnamefont
  {Rudner}},\ }\href {\doibase 10.1103/PhysRevB.102.115109} {\bibfield
  {journal} {\bibinfo  {journal} {Phys. Rev. B}\ }\textbf {\bibinfo {volume}
  {102}},\ \bibinfo {pages} {115109} (\bibinfo {year} {2020})}\BibitemShut
  {NoStop}%
\bibitem [{\citenamefont {Carmele}\ \emph {et~al.}(2015)\citenamefont
  {Carmele}, \citenamefont {Heyl}, \citenamefont {Kraus},\ and\ \citenamefont
  {Dalmonte}}]{carmele15stretched}%
  \BibitemOpen
  \bibfield  {author} {\bibinfo {author} {\bibfnamefont {A.}~\bibnamefont
  {Carmele}}, \bibinfo {author} {\bibfnamefont {M.}~\bibnamefont {Heyl}},
  \bibinfo {author} {\bibfnamefont {C.}~\bibnamefont {Kraus}}, \ and\ \bibinfo
  {author} {\bibfnamefont {M.}~\bibnamefont {Dalmonte}},\ }\href {\doibase
  10.1103/PhysRevB.92.195107} {\bibfield  {journal} {\bibinfo  {journal} {Phys.
  Rev. B}\ }\textbf {\bibinfo {volume} {92}},\ \bibinfo {pages} {195107}
  (\bibinfo {year} {2015})}\BibitemShut {NoStop}%
\bibitem [{\citenamefont {Bandyopadhyay}\ \emph {et~al.}(2020)\citenamefont
  {Bandyopadhyay}, \citenamefont {Bhattacharjee},\ and\ \citenamefont
  {Dutta}}]{bandyopadhyay20dynamical}%
  \BibitemOpen
  \bibfield  {author} {\bibinfo {author} {\bibfnamefont {S.}~\bibnamefont
  {Bandyopadhyay}}, \bibinfo {author} {\bibfnamefont {S.}~\bibnamefont
  {Bhattacharjee}}, \ and\ \bibinfo {author} {\bibfnamefont {A.}~\bibnamefont
  {Dutta}},\ }\href {\doibase 10.1103/PhysRevB.101.104307} {\bibfield
  {journal} {\bibinfo  {journal} {Phys. Rev. B}\ }\textbf {\bibinfo {volume}
  {101}},\ \bibinfo {pages} {104307} (\bibinfo {year} {2020})}\BibitemShut
  {NoStop}%
\bibitem [{\citenamefont {Hoyos}\ \emph {et~al.}(2007)\citenamefont {Hoyos},
  \citenamefont {Kotabage},\ and\ \citenamefont {Vojta}}]{hoyos07effects}%
  \BibitemOpen
  \bibfield  {author} {\bibinfo {author} {\bibfnamefont {J.~A.}\ \bibnamefont
  {Hoyos}}, \bibinfo {author} {\bibfnamefont {C.}~\bibnamefont {Kotabage}}, \
  and\ \bibinfo {author} {\bibfnamefont {T.}~\bibnamefont {Vojta}},\ }\href
  {\doibase 10.1103/PhysRevLett.99.230601} {\bibfield  {journal} {\bibinfo
  {journal} {Phys. Rev. Lett.}\ }\textbf {\bibinfo {volume} {99}},\ \bibinfo
  {pages} {230601} (\bibinfo {year} {2007})}\BibitemShut {NoStop}%
\bibitem [{\citenamefont {De~Grandi}\ \emph {et~al.}(2010)\citenamefont
  {De~Grandi}, \citenamefont {Gritsev},\ and\ \citenamefont
  {Polkovnikov}}]{grandi10quench}%
  \BibitemOpen
  \bibfield  {author} {\bibinfo {author} {\bibfnamefont {C.}~\bibnamefont
  {De~Grandi}}, \bibinfo {author} {\bibfnamefont {V.}~\bibnamefont {Gritsev}},
  \ and\ \bibinfo {author} {\bibfnamefont {A.}~\bibnamefont {Polkovnikov}},\
  }\href {\doibase 10.1103/PhysRevB.81.012303} {\bibfield  {journal} {\bibinfo
  {journal} {Phys. Rev. B}\ }\textbf {\bibinfo {volume} {81}},\ \bibinfo
  {pages} {012303} (\bibinfo {year} {2010})}\BibitemShut {NoStop}%
\bibitem [{\citenamefont {Wang}\ and\ \citenamefont
  {Fazio}(2020)}]{wang20dissipative}%
  \BibitemOpen
  \bibfield  {author} {\bibinfo {author} {\bibfnamefont {P.}~\bibnamefont
  {Wang}}\ and\ \bibinfo {author} {\bibfnamefont {R.}~\bibnamefont {Fazio}},\
  }\href@noop {} {\enquote {\bibinfo {title} {Dissipative phase transitions in
  the fully-connected ising model with $p$-spin interaction},}\ } (\bibinfo
  {year} {2020}),\ \Eprint {http://arxiv.org/abs/2008.10045} {arXiv:2008.10045
  [cond-mat.quant-gas]} \BibitemShut {NoStop}%
\bibitem [{\citenamefont {Fischer}\ \emph {et~al.}(2016)\citenamefont
  {Fischer}, \citenamefont {Maksymenko},\ and\ \citenamefont
  {Altman}}]{fischer16dynamics}%
  \BibitemOpen
  \bibfield  {author} {\bibinfo {author} {\bibfnamefont {M.~H.}\ \bibnamefont
  {Fischer}}, \bibinfo {author} {\bibfnamefont {M.}~\bibnamefont {Maksymenko}},
  \ and\ \bibinfo {author} {\bibfnamefont {E.}~\bibnamefont {Altman}},\ }\href
  {\doibase 10.1103/PhysRevLett.116.160401} {\bibfield  {journal} {\bibinfo
  {journal} {Phys. Rev. Lett.}\ }\textbf {\bibinfo {volume} {116}},\ \bibinfo
  {pages} {160401} (\bibinfo {year} {2016})}\BibitemShut {NoStop}%
\bibitem [{\citenamefont {Lenar\ifmmode \check{c}\else
  \v{c}\fi{}i\ifmmode~\check{c}\else \v{c}\fi{}}\ \emph
  {et~al.}(2020)\citenamefont {Lenar\ifmmode \check{c}\else
  \v{c}\fi{}i\ifmmode~\check{c}\else \v{c}\fi{}}, \citenamefont {Alberton},
  \citenamefont {Rosch},\ and\ \citenamefont {Altman}}]{altman20critical}%
  \BibitemOpen
  \bibfield  {author} {\bibinfo {author} {\bibfnamefont {Z.}~\bibnamefont
  {Lenar\ifmmode \check{c}\else \v{c}\fi{}i\ifmmode~\check{c}\else
  \v{c}\fi{}}}, \bibinfo {author} {\bibfnamefont {O.}~\bibnamefont {Alberton}},
  \bibinfo {author} {\bibfnamefont {A.}~\bibnamefont {Rosch}}, \ and\ \bibinfo
  {author} {\bibfnamefont {E.}~\bibnamefont {Altman}},\ }\href {\doibase
  10.1103/PhysRevLett.125.116601} {\bibfield  {journal} {\bibinfo  {journal}
  {Phys. Rev. Lett.}\ }\textbf {\bibinfo {volume} {125}},\ \bibinfo {pages}
  {116601} (\bibinfo {year} {2020})}\BibitemShut {NoStop}%
\bibitem [{\citenamefont {L\"uschen}\ \emph {et~al.}(2017)\citenamefont
  {L\"uschen}, \citenamefont {Bordia}, \citenamefont {Hodgman}, \citenamefont
  {Schreiber}, \citenamefont {Sarkar}, \citenamefont {Daley}, \citenamefont
  {Fischer}, \citenamefont {Altman}, \citenamefont {Bloch},\ and\ \citenamefont
  {Schneider}}]{bloch17signatures}%
  \BibitemOpen
  \bibfield  {author} {\bibinfo {author} {\bibfnamefont {H.~P.}\ \bibnamefont
  {L\"uschen}}, \bibinfo {author} {\bibfnamefont {P.}~\bibnamefont {Bordia}},
  \bibinfo {author} {\bibfnamefont {S.~S.}\ \bibnamefont {Hodgman}}, \bibinfo
  {author} {\bibfnamefont {M.}~\bibnamefont {Schreiber}}, \bibinfo {author}
  {\bibfnamefont {S.}~\bibnamefont {Sarkar}}, \bibinfo {author} {\bibfnamefont
  {A.~J.}\ \bibnamefont {Daley}}, \bibinfo {author} {\bibfnamefont {M.~H.}\
  \bibnamefont {Fischer}}, \bibinfo {author} {\bibfnamefont {E.}~\bibnamefont
  {Altman}}, \bibinfo {author} {\bibfnamefont {I.}~\bibnamefont {Bloch}}, \
  and\ \bibinfo {author} {\bibfnamefont {U.}~\bibnamefont {Schneider}},\ }\href
  {\doibase 10.1103/PhysRevX.7.011034} {\bibfield  {journal} {\bibinfo
  {journal} {Phys. Rev. X}\ }\textbf {\bibinfo {volume} {7}},\ \bibinfo {pages}
  {011034} (\bibinfo {year} {2017})}\BibitemShut {NoStop}%
\bibitem [{\citenamefont {Niedenzu}\ and\ \citenamefont
  {Kurizki}(2018)}]{niedenzu18cooperative}%
  \BibitemOpen
  \bibfield  {author} {\bibinfo {author} {\bibfnamefont {W.}~\bibnamefont
  {Niedenzu}}\ and\ \bibinfo {author} {\bibfnamefont {G.}~\bibnamefont
  {Kurizki}},\ }\href {\doibase 10.1088/1367-2630/aaed55} {\bibfield  {journal}
  {\bibinfo  {journal} {New Journal of Physics}\ }\textbf {\bibinfo {volume}
  {20}},\ \bibinfo {pages} {113038} (\bibinfo {year} {2018})}\BibitemShut
  {NoStop}%
\bibitem [{\citenamefont {Kloc}\ \emph {et~al.}(2019)\citenamefont {Kloc},
  \citenamefont {Cejnar},\ and\ \citenamefont {Schaller}}]{kloc19collective}%
  \BibitemOpen
  \bibfield  {author} {\bibinfo {author} {\bibfnamefont {M.}~\bibnamefont
  {Kloc}}, \bibinfo {author} {\bibfnamefont {P.}~\bibnamefont {Cejnar}}, \ and\
  \bibinfo {author} {\bibfnamefont {G.}~\bibnamefont {Schaller}},\ }\href
  {\doibase 10.1103/PhysRevE.100.042126} {\bibfield  {journal} {\bibinfo
  {journal} {Phys. Rev. E}\ }\textbf {\bibinfo {volume} {100}},\ \bibinfo
  {pages} {042126} (\bibinfo {year} {2019})}\BibitemShut {NoStop}%
\bibitem [{\citenamefont {Latune}\ \emph {et~al.}(2020)\citenamefont {Latune},
  \citenamefont {Sinayskiy},\ and\ \citenamefont
  {Petruccione}}]{latune20collective}%
  \BibitemOpen
  \bibfield  {author} {\bibinfo {author} {\bibfnamefont {C.~L.}\ \bibnamefont
  {Latune}}, \bibinfo {author} {\bibfnamefont {I.}~\bibnamefont {Sinayskiy}}, \
  and\ \bibinfo {author} {\bibfnamefont {F.}~\bibnamefont {Petruccione}},\
  }\href {\doibase 10.1088/1367-2630/aba463} {\bibfield  {journal} {\bibinfo
  {journal} {New Journal of Physics}\ }\textbf {\bibinfo {volume} {22}},\
  \bibinfo {pages} {083049} (\bibinfo {year} {2020})}\BibitemShut {NoStop}%
\bibitem [{\citenamefont {Jaramillo}\ \emph {et~al.}(2016)\citenamefont
  {Jaramillo}, \citenamefont {Beau},\ and\ \citenamefont {del
  Campo}}]{jaramillo16quantum}%
  \BibitemOpen
  \bibfield  {author} {\bibinfo {author} {\bibfnamefont {J.}~\bibnamefont
  {Jaramillo}}, \bibinfo {author} {\bibfnamefont {M.}~\bibnamefont {Beau}}, \
  and\ \bibinfo {author} {\bibfnamefont {A.}~\bibnamefont {del Campo}},\ }\href
  {http://stacks.iop.org/1367-2630/18/i=7/a=075019} {\bibfield  {journal}
  {\bibinfo  {journal} {New J. Phys.}\ }\textbf {\bibinfo {volume} {18}},\
  \bibinfo {pages} {075019} (\bibinfo {year} {2016})}\BibitemShut {NoStop}%
\bibitem [{\citenamefont {{\c{C}}akmak}\ \emph {et~al.}(2016)\citenamefont
  {{\c{C}}akmak}, \citenamefont {Altintas},\ and\ \citenamefont
  {E.~M{\"u}stecaplioglu}}]{Cakmak16}%
  \BibitemOpen
  \bibfield  {author} {\bibinfo {author} {\bibfnamefont {S.}~\bibnamefont
  {{\c{C}}akmak}}, \bibinfo {author} {\bibfnamefont {F.}~\bibnamefont
  {Altintas}}, \ and\ \bibinfo {author} {\bibfnamefont {{\"O}.}~\bibnamefont
  {E.~M{\"u}stecaplioglu}},\ }\href {\doibase 10.1140/epjp/i2016-16197-0}
  {\bibfield  {journal} {\bibinfo  {journal} {The European Physical Journal
  Plus}\ }\textbf {\bibinfo {volume} {131}},\ \bibinfo {pages} {197} (\bibinfo
  {year} {2016})}\BibitemShut {NoStop}%
\bibitem [{\citenamefont {Herrera}\ \emph {et~al.}(2017)\citenamefont
  {Herrera}, \citenamefont {Serra},\ and\ \citenamefont
  {D'Amico}}]{herrera17dft}%
  \BibitemOpen
  \bibfield  {author} {\bibinfo {author} {\bibfnamefont {M.}~\bibnamefont
  {Herrera}}, \bibinfo {author} {\bibfnamefont {R.~M.}\ \bibnamefont {Serra}},
  \ and\ \bibinfo {author} {\bibfnamefont {I.}~\bibnamefont {D'Amico}},\ }\href
  {\doibase 10.1038/s41598-017-04478-y} {\bibfield  {journal} {\bibinfo
  {journal} {Scientific Reports}\ }\textbf {\bibinfo {volume} {7}},\ \bibinfo
  {pages} {4655} (\bibinfo {year} {2017})}\BibitemShut {NoStop}%
\bibitem [{\citenamefont {Skelt}\ \emph {et~al.}(2019)\citenamefont {Skelt},
  \citenamefont {Zawadzki},\ and\ \citenamefont {D'Amico}}]{skelt19many}%
  \BibitemOpen
  \bibfield  {author} {\bibinfo {author} {\bibfnamefont {A.~H.}\ \bibnamefont
  {Skelt}}, \bibinfo {author} {\bibfnamefont {K.}~\bibnamefont {Zawadzki}}, \
  and\ \bibinfo {author} {\bibfnamefont {I.}~\bibnamefont {D'Amico}},\ }\href
  {\doibase 10.1088/1751-8121/ab4fb6} {\bibfield  {journal} {\bibinfo
  {journal} {Journal of Physics A: Mathematical and Theoretical}\ }\textbf
  {\bibinfo {volume} {52}},\ \bibinfo {pages} {485304} (\bibinfo {year}
  {2019})}\BibitemShut {NoStop}%
\bibitem [{\citenamefont {Chen}\ \emph {et~al.}(2019)\citenamefont {Chen},
  \citenamefont {Watanabe}, \citenamefont {Yu}, \citenamefont {Guan},\ and\
  \citenamefont {del Campo}}]{yang19an}%
  \BibitemOpen
  \bibfield  {author} {\bibinfo {author} {\bibfnamefont {Y.-Y.}\ \bibnamefont
  {Chen}}, \bibinfo {author} {\bibfnamefont {G.}~\bibnamefont {Watanabe}},
  \bibinfo {author} {\bibfnamefont {Y.-C.}\ \bibnamefont {Yu}}, \bibinfo
  {author} {\bibfnamefont {X.-W.}\ \bibnamefont {Guan}}, \ and\ \bibinfo
  {author} {\bibfnamefont {A.}~\bibnamefont {del Campo}},\ }\href {\doibase
  10.1038/s41534-019-0204-5} {\bibfield  {journal} {\bibinfo  {journal} {npj
  Quantum Information}\ }\textbf {\bibinfo {volume} {5}},\ \bibinfo {pages}
  {88} (\bibinfo {year} {2019})}\BibitemShut {NoStop}%
\bibitem [{\citenamefont {Yunt}\ \emph {et~al.}(2019)\citenamefont {Yunt},
  \citenamefont {Fadaie},\ and\ \citenamefont {Özgür
  E.~Müstecaplıoğlu}}]{yunt19topological}%
  \BibitemOpen
  \bibfield  {author} {\bibinfo {author} {\bibfnamefont {E.}~\bibnamefont
  {Yunt}}, \bibinfo {author} {\bibfnamefont {M.}~\bibnamefont {Fadaie}}, \ and\
  \bibinfo {author} {\bibnamefont {Özgür E.~Müstecaplıoğlu}},\ }\href@noop
  {} {\enquote {\bibinfo {title} {Topological and finite size effects in a
  kitaev chain heat engine},}\ } (\bibinfo {year} {2019}),\ \Eprint
  {http://arxiv.org/abs/1908.02643} {arXiv:1908.02643 [cond-mat.stat-mech]}
  \BibitemShut {NoStop}%
\bibitem [{\citenamefont {Zawadzki}\ \emph {et~al.}(2020)\citenamefont
  {Zawadzki}, \citenamefont {Serra},\ and\ \citenamefont
  {D'Amico}}]{zawadzki20work}%
  \BibitemOpen
  \bibfield  {author} {\bibinfo {author} {\bibfnamefont {K.}~\bibnamefont
  {Zawadzki}}, \bibinfo {author} {\bibfnamefont {R.~M.}\ \bibnamefont {Serra}},
  \ and\ \bibinfo {author} {\bibfnamefont {I.}~\bibnamefont {D'Amico}},\ }\href
  {\doibase 10.1103/PhysRevResearch.2.033167} {\bibfield  {journal} {\bibinfo
  {journal} {Phys. Rev. Research}\ }\textbf {\bibinfo {volume} {2}},\ \bibinfo
  {pages} {033167} (\bibinfo {year} {2020})}\BibitemShut {NoStop}%
\bibitem [{\citenamefont {Campisi}\ and\ \citenamefont
  {Fazio}(2016)}]{campisi16the}%
  \BibitemOpen
  \bibfield  {author} {\bibinfo {author} {\bibfnamefont {M.}~\bibnamefont
  {Campisi}}\ and\ \bibinfo {author} {\bibfnamefont {R.}~\bibnamefont
  {Fazio}},\ }\href {http://dx.doi.org/10.1038/ncomms11895} {\bibfield
  {journal} {\bibinfo  {journal} {Nat. Commun.}\ }\textbf {\bibinfo {volume}
  {7}},\ \bibinfo {pages} {11895} (\bibinfo {year} {2016})}\BibitemShut
  {NoStop}%
\bibitem [{\citenamefont {Hartmann}\ \emph
  {et~al.}(2020{\natexlab{a}})\citenamefont {Hartmann}, \citenamefont
  {Mukherjee}, \citenamefont {Niedenzu},\ and\ \citenamefont
  {Lechner}}]{hartmann19many}%
  \BibitemOpen
  \bibfield  {author} {\bibinfo {author} {\bibfnamefont {A.}~\bibnamefont
  {Hartmann}}, \bibinfo {author} {\bibfnamefont {V.}~\bibnamefont {Mukherjee}},
  \bibinfo {author} {\bibfnamefont {W.}~\bibnamefont {Niedenzu}}, \ and\
  \bibinfo {author} {\bibfnamefont {W.}~\bibnamefont {Lechner}},\ }\href
  {\doibase 10.1103/PhysRevResearch.2.023145} {\bibfield  {journal} {\bibinfo
  {journal} {Phys. Rev. Research}\ }\textbf {\bibinfo {volume} {2}},\ \bibinfo
  {pages} {023145} (\bibinfo {year} {2020}{\natexlab{a}})}\BibitemShut
  {NoStop}%
\bibitem [{\citenamefont {Hartmann}\ \emph
  {et~al.}(2020{\natexlab{b}})\citenamefont {Hartmann}, \citenamefont
  {Mukherjee}, \citenamefont {Mbeng}, \citenamefont {Niedenzu},\ and\
  \citenamefont {Lechner}}]{hartmann20multispin}%
  \BibitemOpen
  \bibfield  {author} {\bibinfo {author} {\bibfnamefont {A.}~\bibnamefont
  {Hartmann}}, \bibinfo {author} {\bibfnamefont {V.}~\bibnamefont {Mukherjee}},
  \bibinfo {author} {\bibfnamefont {G.~B.}\ \bibnamefont {Mbeng}}, \bibinfo
  {author} {\bibfnamefont {W.}~\bibnamefont {Niedenzu}}, \ and\ \bibinfo
  {author} {\bibfnamefont {W.}~\bibnamefont {Lechner}},\ }\href {\doibase
  10.22331/q-2020-12-24-377} {\bibfield  {journal} {\bibinfo  {journal}
  {{Quantum}}\ }\textbf {\bibinfo {volume} {4}},\ \bibinfo {pages} {377}
  (\bibinfo {year} {2020}{\natexlab{b}})}\BibitemShut {NoStop}%
\bibitem [{\citenamefont {Revathy}\ \emph {et~al.}(2020)\citenamefont
  {Revathy}, \citenamefont {Mukherjee}, \citenamefont {Divakaran},\ and\
  \citenamefont {del Campo}}]{revathy20universal}%
  \BibitemOpen
  \bibfield  {author} {\bibinfo {author} {\bibfnamefont {B.~S.}\ \bibnamefont
  {Revathy}}, \bibinfo {author} {\bibfnamefont {V.}~\bibnamefont {Mukherjee}},
  \bibinfo {author} {\bibfnamefont {U.}~\bibnamefont {Divakaran}}, \ and\
  \bibinfo {author} {\bibfnamefont {A.}~\bibnamefont {del Campo}},\ }\href
  {\doibase 10.1103/PhysRevResearch.2.043247} {\bibfield  {journal} {\bibinfo
  {journal} {Phys. Rev. Research}\ }\textbf {\bibinfo {volume} {2}},\ \bibinfo
  {pages} {043247} (\bibinfo {year} {2020})}\BibitemShut {NoStop}%
\bibitem [{\citenamefont {Binder}\ \emph {et~al.}(2015)\citenamefont {Binder},
  \citenamefont {Vinjanampathy}, \citenamefont {Modi},\ and\ \citenamefont
  {Goold}}]{binder15quantacell}%
  \BibitemOpen
  \bibfield  {author} {\bibinfo {author} {\bibfnamefont {F.~C.}\ \bibnamefont
  {Binder}}, \bibinfo {author} {\bibfnamefont {S.}~\bibnamefont
  {Vinjanampathy}}, \bibinfo {author} {\bibfnamefont {K.}~\bibnamefont {Modi}},
  \ and\ \bibinfo {author} {\bibfnamefont {J.}~\bibnamefont {Goold}},\ }\href
  {\doibase 10.1088/1367-2630/17/7/075015} {\bibfield  {journal} {\bibinfo
  {journal} {New Journal of Physics}\ }\textbf {\bibinfo {volume} {17}},\
  \bibinfo {pages} {075015} (\bibinfo {year} {2015})}\BibitemShut {NoStop}%
\bibitem [{\citenamefont {Campaioli}\ \emph {et~al.}(2017)\citenamefont
  {Campaioli}, \citenamefont {Pollock}, \citenamefont {Binder}, \citenamefont
  {C\'eleri}, \citenamefont {Goold}, \citenamefont {Vinjanampathy},\ and\
  \citenamefont {Modi}}]{campaioli17enhancing}%
  \BibitemOpen
  \bibfield  {author} {\bibinfo {author} {\bibfnamefont {F.}~\bibnamefont
  {Campaioli}}, \bibinfo {author} {\bibfnamefont {F.~A.}\ \bibnamefont
  {Pollock}}, \bibinfo {author} {\bibfnamefont {F.~C.}\ \bibnamefont {Binder}},
  \bibinfo {author} {\bibfnamefont {L.}~\bibnamefont {C\'eleri}}, \bibinfo
  {author} {\bibfnamefont {J.}~\bibnamefont {Goold}}, \bibinfo {author}
  {\bibfnamefont {S.}~\bibnamefont {Vinjanampathy}}, \ and\ \bibinfo {author}
  {\bibfnamefont {K.}~\bibnamefont {Modi}},\ }\href {\doibase
  10.1103/PhysRevLett.118.150601} {\bibfield  {journal} {\bibinfo  {journal}
  {Phys. Rev. Lett.}\ }\textbf {\bibinfo {volume} {118}},\ \bibinfo {pages}
  {150601} (\bibinfo {year} {2017})}\BibitemShut {NoStop}%
\bibitem [{\citenamefont {Ferraro}\ \emph {et~al.}(2018)\citenamefont
  {Ferraro}, \citenamefont {Campisi}, \citenamefont {Andolina}, \citenamefont
  {Pellegrini},\ and\ \citenamefont {Polini}}]{ferraro18high}%
  \BibitemOpen
  \bibfield  {author} {\bibinfo {author} {\bibfnamefont {D.}~\bibnamefont
  {Ferraro}}, \bibinfo {author} {\bibfnamefont {M.}~\bibnamefont {Campisi}},
  \bibinfo {author} {\bibfnamefont {G.~M.}\ \bibnamefont {Andolina}}, \bibinfo
  {author} {\bibfnamefont {V.}~\bibnamefont {Pellegrini}}, \ and\ \bibinfo
  {author} {\bibfnamefont {M.}~\bibnamefont {Polini}},\ }\href {\doibase
  10.1103/PhysRevLett.120.117702} {\bibfield  {journal} {\bibinfo  {journal}
  {Phys. Rev. Lett.}\ }\textbf {\bibinfo {volume} {120}},\ \bibinfo {pages}
  {117702} (\bibinfo {year} {2018})}\BibitemShut {NoStop}%
\bibitem [{\citenamefont {Campaioli}\ \emph {et~al.}(2018)\citenamefont
  {Campaioli}, \citenamefont {Pollock},\ and\ \citenamefont
  {Vinjanampathy}}]{campaioli18quantum}%
  \BibitemOpen
  \bibfield  {author} {\bibinfo {author} {\bibfnamefont {F.}~\bibnamefont
  {Campaioli}}, \bibinfo {author} {\bibfnamefont {F.~A.}\ \bibnamefont
  {Pollock}}, \ and\ \bibinfo {author} {\bibfnamefont {S.}~\bibnamefont
  {Vinjanampathy}},\ }\enquote {\bibinfo {title} {Quantum batteries},}\ in\
  \href {\doibase 10.1007/978-3-319-99046-0_8} {\emph {\bibinfo {booktitle}
  {Thermodynamics in the Quantum Regime: Fundamental Aspects and New
  Directions}}},\ \bibinfo {editor} {edited by\ \bibinfo {editor}
  {\bibfnamefont {F.}~\bibnamefont {Binder}}, \bibinfo {editor} {\bibfnamefont
  {L.~A.}\ \bibnamefont {Correa}}, \bibinfo {editor} {\bibfnamefont
  {C.}~\bibnamefont {Gogolin}}, \bibinfo {editor} {\bibfnamefont
  {J.}~\bibnamefont {Anders}}, \ and\ \bibinfo {editor} {\bibfnamefont
  {G.}~\bibnamefont {Adesso}}}\ (\bibinfo  {publisher} {Springer International
  Publishing},\ \bibinfo {address} {Cham},\ \bibinfo {year} {2018})\ pp.\
  \bibinfo {pages} {207--225}\BibitemShut {NoStop}%
\bibitem [{\citenamefont {Le}\ \emph {et~al.}(2018)\citenamefont {Le},
  \citenamefont {Levinsen}, \citenamefont {Modi}, \citenamefont {Parish},\ and\
  \citenamefont {Pollock}}]{le18spin}%
  \BibitemOpen
  \bibfield  {author} {\bibinfo {author} {\bibfnamefont {T.~P.}\ \bibnamefont
  {Le}}, \bibinfo {author} {\bibfnamefont {J.}~\bibnamefont {Levinsen}},
  \bibinfo {author} {\bibfnamefont {K.}~\bibnamefont {Modi}}, \bibinfo {author}
  {\bibfnamefont {M.~M.}\ \bibnamefont {Parish}}, \ and\ \bibinfo {author}
  {\bibfnamefont {F.~A.}\ \bibnamefont {Pollock}},\ }\href {\doibase
  10.1103/PhysRevA.97.022106} {\bibfield  {journal} {\bibinfo  {journal} {Phys.
  Rev. A}\ }\textbf {\bibinfo {volume} {97}},\ \bibinfo {pages} {022106}
  (\bibinfo {year} {2018})}\BibitemShut {NoStop}%
\bibitem [{\citenamefont {Rossini}\ \emph {et~al.}(2020)\citenamefont
  {Rossini}, \citenamefont {Andolina}, \citenamefont {Rosa}, \citenamefont
  {Carrega},\ and\ \citenamefont {Polini}}]{rossini19quantum}%
  \BibitemOpen
  \bibfield  {author} {\bibinfo {author} {\bibfnamefont {D.}~\bibnamefont
  {Rossini}}, \bibinfo {author} {\bibfnamefont {G.~M.}\ \bibnamefont
  {Andolina}}, \bibinfo {author} {\bibfnamefont {D.}~\bibnamefont {Rosa}},
  \bibinfo {author} {\bibfnamefont {M.}~\bibnamefont {Carrega}}, \ and\
  \bibinfo {author} {\bibfnamefont {M.}~\bibnamefont {Polini}},\ }\href
  {\doibase 10.1103/PhysRevLett.125.236402} {\bibfield  {journal} {\bibinfo
  {journal} {Phys. Rev. Lett.}\ }\textbf {\bibinfo {volume} {125}},\ \bibinfo
  {pages} {236402} (\bibinfo {year} {2020})}\BibitemShut {NoStop}%
\bibitem [{\citenamefont {Andolina}\ \emph
  {et~al.}(2019{\natexlab{a}})\citenamefont {Andolina}, \citenamefont {Keck},
  \citenamefont {Mari}, \citenamefont {Campisi}, \citenamefont {Giovannetti},\
  and\ \citenamefont {Polini}}]{andolina19extractable}%
  \BibitemOpen
  \bibfield  {author} {\bibinfo {author} {\bibfnamefont {G.~M.}\ \bibnamefont
  {Andolina}}, \bibinfo {author} {\bibfnamefont {M.}~\bibnamefont {Keck}},
  \bibinfo {author} {\bibfnamefont {A.}~\bibnamefont {Mari}}, \bibinfo {author}
  {\bibfnamefont {M.}~\bibnamefont {Campisi}}, \bibinfo {author} {\bibfnamefont
  {V.}~\bibnamefont {Giovannetti}}, \ and\ \bibinfo {author} {\bibfnamefont
  {M.}~\bibnamefont {Polini}},\ }\href {\doibase
  10.1103/PhysRevLett.122.047702} {\bibfield  {journal} {\bibinfo  {journal}
  {Phys. Rev. Lett.}\ }\textbf {\bibinfo {volume} {122}},\ \bibinfo {pages}
  {047702} (\bibinfo {year} {2019}{\natexlab{a}})}\BibitemShut {NoStop}%
\bibitem [{\citenamefont {Andolina}\ \emph
  {et~al.}(2019{\natexlab{b}})\citenamefont {Andolina}, \citenamefont {Keck},
  \citenamefont {Mari}, \citenamefont {Giovannetti},\ and\ \citenamefont
  {Polini}}]{andolina19quantum}%
  \BibitemOpen
  \bibfield  {author} {\bibinfo {author} {\bibfnamefont {G.~M.}\ \bibnamefont
  {Andolina}}, \bibinfo {author} {\bibfnamefont {M.}~\bibnamefont {Keck}},
  \bibinfo {author} {\bibfnamefont {A.}~\bibnamefont {Mari}}, \bibinfo {author}
  {\bibfnamefont {V.}~\bibnamefont {Giovannetti}}, \ and\ \bibinfo {author}
  {\bibfnamefont {M.}~\bibnamefont {Polini}},\ }\href {\doibase
  10.1103/PhysRevB.99.205437} {\bibfield  {journal} {\bibinfo  {journal} {Phys.
  Rev. B}\ }\textbf {\bibinfo {volume} {99}},\ \bibinfo {pages} {205437}
  (\bibinfo {year} {2019}{\natexlab{b}})}\BibitemShut {NoStop}%
\bibitem [{\citenamefont {Rossini}\ \emph {et~al.}(2019)\citenamefont
  {Rossini}, \citenamefont {Andolina},\ and\ \citenamefont
  {Polini}}]{rossini19many}%
  \BibitemOpen
  \bibfield  {author} {\bibinfo {author} {\bibfnamefont {D.}~\bibnamefont
  {Rossini}}, \bibinfo {author} {\bibfnamefont {G.~M.}\ \bibnamefont
  {Andolina}}, \ and\ \bibinfo {author} {\bibfnamefont {M.}~\bibnamefont
  {Polini}},\ }\href {\doibase 10.1103/PhysRevB.100.115142} {\bibfield
  {journal} {\bibinfo  {journal} {Phys. Rev. B}\ }\textbf {\bibinfo {volume}
  {100}},\ \bibinfo {pages} {115142} (\bibinfo {year} {2019})}\BibitemShut
  {NoStop}%
\bibitem [{\citenamefont {Ito}\ and\ \citenamefont
  {Watanabe}(2020)}]{ito20collectively}%
  \BibitemOpen
  \bibfield  {author} {\bibinfo {author} {\bibfnamefont {K.}~\bibnamefont
  {Ito}}\ and\ \bibinfo {author} {\bibfnamefont {G.}~\bibnamefont {Watanabe}},\
  }\href@noop {} {\enquote {\bibinfo {title} {Collectively enhanced high-power
  and high-capacity charging of quantum batteries via quantum heat engines},}\
  } (\bibinfo {year} {2020}),\ \Eprint {http://arxiv.org/abs/2008.07089}
  {arXiv:2008.07089 [quant-ph]} \BibitemShut {NoStop}%
\bibitem [{\citenamefont {\ifmmode~\mbox{\c{C}}\else
  \c{C}\fi{}akmak}(2020)}]{cakmak20ergotropy}%
  \BibitemOpen
  \bibfield  {author} {\bibinfo {author} {\bibfnamefont {B.~i. e. i. f. m.~c.}\
  \bibnamefont {\ifmmode~\mbox{\c{C}}\else \c{C}\fi{}akmak}},\ }\href {\doibase
  10.1103/PhysRevE.102.042111} {\bibfield  {journal} {\bibinfo  {journal}
  {Phys. Rev. E}\ }\textbf {\bibinfo {volume} {102}},\ \bibinfo {pages}
  {042111} (\bibinfo {year} {2020})}\BibitemShut {NoStop}%
\bibitem [{\citenamefont {Carrega}\ \emph {et~al.}(2020)\citenamefont
  {Carrega}, \citenamefont {Crescente}, \citenamefont {Ferraro},\ and\
  \citenamefont {Sassetti}}]{carrega20dissipative}%
  \BibitemOpen
  \bibfield  {author} {\bibinfo {author} {\bibfnamefont {M.}~\bibnamefont
  {Carrega}}, \bibinfo {author} {\bibfnamefont {A.}~\bibnamefont {Crescente}},
  \bibinfo {author} {\bibfnamefont {D.}~\bibnamefont {Ferraro}}, \ and\
  \bibinfo {author} {\bibfnamefont {M.}~\bibnamefont {Sassetti}},\ }\href
  {\doibase 10.1088/1367-2630/abaa01} {\bibfield  {journal} {\bibinfo
  {journal} {New Journal of Physics}\ }\textbf {\bibinfo {volume} {22}},\
  \bibinfo {pages} {083085} (\bibinfo {year} {2020})}\BibitemShut {NoStop}%
\bibitem [{\citenamefont {Crescente}\ \emph
  {et~al.}(2020{\natexlab{a}})\citenamefont {Crescente}, \citenamefont
  {Carrega}, \citenamefont {Sassetti},\ and\ \citenamefont
  {Ferraro}}]{crescente20charging}%
  \BibitemOpen
  \bibfield  {author} {\bibinfo {author} {\bibfnamefont {A.}~\bibnamefont
  {Crescente}}, \bibinfo {author} {\bibfnamefont {M.}~\bibnamefont {Carrega}},
  \bibinfo {author} {\bibfnamefont {M.}~\bibnamefont {Sassetti}}, \ and\
  \bibinfo {author} {\bibfnamefont {D.}~\bibnamefont {Ferraro}},\ }\href
  {\doibase 10.1088/1367-2630/ab91fc} {\bibfield  {journal} {\bibinfo
  {journal} {New Journal of Physics}\ }\textbf {\bibinfo {volume} {22}},\
  \bibinfo {pages} {063057} (\bibinfo {year} {2020}{\natexlab{a}})}\BibitemShut
  {NoStop}%
\bibitem [{\citenamefont {Zanardi}\ \emph {et~al.}(2008)\citenamefont
  {Zanardi}, \citenamefont {Paris},\ and\ \citenamefont
  {Campos~Venuti}}]{zanardi08quantum}%
  \BibitemOpen
  \bibfield  {author} {\bibinfo {author} {\bibfnamefont {P.}~\bibnamefont
  {Zanardi}}, \bibinfo {author} {\bibfnamefont {M.~G.~A.}\ \bibnamefont
  {Paris}}, \ and\ \bibinfo {author} {\bibfnamefont {L.}~\bibnamefont
  {Campos~Venuti}},\ }\href {\doibase 10.1103/PhysRevA.78.042105} {\bibfield
  {journal} {\bibinfo  {journal} {Phys. Rev. A}\ }\textbf {\bibinfo {volume}
  {78}},\ \bibinfo {pages} {042105} (\bibinfo {year} {2008})}\BibitemShut
  {NoStop}%
\bibitem [{\citenamefont {Rams}\ \emph {et~al.}(2018)\citenamefont {Rams},
  \citenamefont {Sierant}, \citenamefont {Dutta}, \citenamefont {Horodecki},\
  and\ \citenamefont {Zakrzewski}}]{rams18at}%
  \BibitemOpen
  \bibfield  {author} {\bibinfo {author} {\bibfnamefont {M.~M.}\ \bibnamefont
  {Rams}}, \bibinfo {author} {\bibfnamefont {P.}~\bibnamefont {Sierant}},
  \bibinfo {author} {\bibfnamefont {O.}~\bibnamefont {Dutta}}, \bibinfo
  {author} {\bibfnamefont {P.}~\bibnamefont {Horodecki}}, \ and\ \bibinfo
  {author} {\bibfnamefont {J.}~\bibnamefont {Zakrzewski}},\ }\href {\doibase
  10.1103/PhysRevX.8.021022} {\bibfield  {journal} {\bibinfo  {journal} {Phys.
  Rev. X}\ }\textbf {\bibinfo {volume} {8}},\ \bibinfo {pages} {021022}
  (\bibinfo {year} {2018})}\BibitemShut {NoStop}%
\bibitem [{\citenamefont {Hovhannisyan}\ and\ \citenamefont
  {Correa}(2018)}]{hovhannisyan18measuring}%
  \BibitemOpen
  \bibfield  {author} {\bibinfo {author} {\bibfnamefont {K.~V.}\ \bibnamefont
  {Hovhannisyan}}\ and\ \bibinfo {author} {\bibfnamefont {L.~A.}\ \bibnamefont
  {Correa}},\ }\href {\doibase 10.1103/PhysRevB.98.045101} {\bibfield
  {journal} {\bibinfo  {journal} {Phys. Rev. B}\ }\textbf {\bibinfo {volume}
  {98}},\ \bibinfo {pages} {045101} (\bibinfo {year} {2018})}\BibitemShut
  {NoStop}%
\bibitem [{\citenamefont {Potts}\ \emph {et~al.}(2019)\citenamefont {Potts},
  \citenamefont {Brask},\ and\ \citenamefont {Brunner}}]{potts19fundamental}%
  \BibitemOpen
  \bibfield  {author} {\bibinfo {author} {\bibfnamefont {P.~P.}\ \bibnamefont
  {Potts}}, \bibinfo {author} {\bibfnamefont {J.~B.}\ \bibnamefont {Brask}}, \
  and\ \bibinfo {author} {\bibfnamefont {N.}~\bibnamefont {Brunner}},\ }\href
  {\doibase 10.22331/q-2019-07-09-161} {\bibfield  {journal} {\bibinfo
  {journal} {{Quantum}}\ }\textbf {\bibinfo {volume} {3}},\ \bibinfo {pages}
  {161} (\bibinfo {year} {2019})}\BibitemShut {NoStop}%
\bibitem [{\citenamefont {Mok}\ \emph {et~al.}(2020)\citenamefont {Mok},
  \citenamefont {Bharti}, \citenamefont {Kwek},\ and\ \citenamefont
  {Bayat}}]{mok20optimal}%
  \BibitemOpen
  \bibfield  {author} {\bibinfo {author} {\bibfnamefont {W.-K.}\ \bibnamefont
  {Mok}}, \bibinfo {author} {\bibfnamefont {K.}~\bibnamefont {Bharti}},
  \bibinfo {author} {\bibfnamefont {L.-C.}\ \bibnamefont {Kwek}}, \ and\
  \bibinfo {author} {\bibfnamefont {A.}~\bibnamefont {Bayat}},\ }\href@noop {}
  {\enquote {\bibinfo {title} {Optimal probes for global quantum
  thermometry},}\ } (\bibinfo {year} {2020}),\ \Eprint
  {http://arxiv.org/abs/2010.14200} {arXiv:2010.14200 [quant-ph]} \BibitemShut
  {NoStop}%
\bibitem [{\citenamefont {Mishra}\ and\ \citenamefont
  {Bayat}(2020)}]{bayat20_arxiv}%
  \BibitemOpen
  \bibfield  {author} {\bibinfo {author} {\bibfnamefont {U.}~\bibnamefont
  {Mishra}}\ and\ \bibinfo {author} {\bibfnamefont {A.}~\bibnamefont {Bayat}},\
  }\href@noop {} {\enquote {\bibinfo {title} {Driving enhanced quantum sensing
  in partially accessible many-body systems},}\ } (\bibinfo {year} {2020}),\
  \Eprint {http://arxiv.org/abs/arXiv:2010.09050} {arXiv:arXiv:2010.09050
  [Quantum Physics]} \BibitemShut {NoStop}%
\bibitem [{\citenamefont {Montenegro}\ \emph {et~al.}(2021)\citenamefont
  {Montenegro}, \citenamefont {Mishra},\ and\ \citenamefont
  {Bayat}}]{bayat21global}%
  \BibitemOpen
  \bibfield  {author} {\bibinfo {author} {\bibfnamefont {V.}~\bibnamefont
  {Montenegro}}, \bibinfo {author} {\bibfnamefont {U.}~\bibnamefont {Mishra}},
  \ and\ \bibinfo {author} {\bibfnamefont {A.}~\bibnamefont {Bayat}},\ }\href
  {\doibase 10.1103/PhysRevLett.126.200501} {\bibfield  {journal} {\bibinfo
  {journal} {Phys. Rev. Lett.}\ }\textbf {\bibinfo {volume} {126}},\ \bibinfo
  {pages} {200501} (\bibinfo {year} {2021})}\BibitemShut {NoStop}%
\bibitem [{\citenamefont {Kim}\ \emph {et~al.}(2018)\citenamefont {Kim},
  \citenamefont {Park}, \citenamefont {Kim}, \citenamefont {Sim},\ and\
  \citenamefont {Ahn}}]{kim18detailed}%
  \BibitemOpen
  \bibfield  {author} {\bibinfo {author} {\bibfnamefont {H.}~\bibnamefont
  {Kim}}, \bibinfo {author} {\bibfnamefont {Y.}~\bibnamefont {Park}}, \bibinfo
  {author} {\bibfnamefont {K.}~\bibnamefont {Kim}}, \bibinfo {author}
  {\bibfnamefont {H.-S.}\ \bibnamefont {Sim}}, \ and\ \bibinfo {author}
  {\bibfnamefont {J.}~\bibnamefont {Ahn}},\ }\href {\doibase
  10.1103/PhysRevLett.120.180502} {\bibfield  {journal} {\bibinfo  {journal}
  {Phys. Rev. Lett.}\ }\textbf {\bibinfo {volume} {120}},\ \bibinfo {pages}
  {180502} (\bibinfo {year} {2018})}\BibitemShut {NoStop}%
\bibitem [{\citenamefont {Omran}\ \emph {et~al.}(2019)\citenamefont {Omran},
  \citenamefont {Levine}, \citenamefont {Keesling}, \citenamefont {Semeghini},
  \citenamefont {Wang}, \citenamefont {Ebadi}, \citenamefont {Bernien},
  \citenamefont {Zibrov}, \citenamefont {Pichler}, \citenamefont {Choi},
  \citenamefont {Cui}, \citenamefont {Rossignolo}, \citenamefont {Rembold},
  \citenamefont {Montangero}, \citenamefont {Calarco}, \citenamefont {Endres},
  \citenamefont {Greiner}, \citenamefont {Vuleti{\'c}},\ and\ \citenamefont
  {Lukin}}]{omran19generation}%
  \BibitemOpen
  \bibfield  {author} {\bibinfo {author} {\bibfnamefont {A.}~\bibnamefont
  {Omran}}, \bibinfo {author} {\bibfnamefont {H.}~\bibnamefont {Levine}},
  \bibinfo {author} {\bibfnamefont {A.}~\bibnamefont {Keesling}}, \bibinfo
  {author} {\bibfnamefont {G.}~\bibnamefont {Semeghini}}, \bibinfo {author}
  {\bibfnamefont {T.~T.}\ \bibnamefont {Wang}}, \bibinfo {author}
  {\bibfnamefont {S.}~\bibnamefont {Ebadi}}, \bibinfo {author} {\bibfnamefont
  {H.}~\bibnamefont {Bernien}}, \bibinfo {author} {\bibfnamefont {A.~S.}\
  \bibnamefont {Zibrov}}, \bibinfo {author} {\bibfnamefont {H.}~\bibnamefont
  {Pichler}}, \bibinfo {author} {\bibfnamefont {S.}~\bibnamefont {Choi}},
  \bibinfo {author} {\bibfnamefont {J.}~\bibnamefont {Cui}}, \bibinfo {author}
  {\bibfnamefont {M.}~\bibnamefont {Rossignolo}}, \bibinfo {author}
  {\bibfnamefont {P.}~\bibnamefont {Rembold}}, \bibinfo {author} {\bibfnamefont
  {S.}~\bibnamefont {Montangero}}, \bibinfo {author} {\bibfnamefont
  {T.}~\bibnamefont {Calarco}}, \bibinfo {author} {\bibfnamefont
  {M.}~\bibnamefont {Endres}}, \bibinfo {author} {\bibfnamefont
  {M.}~\bibnamefont {Greiner}}, \bibinfo {author} {\bibfnamefont
  {V.}~\bibnamefont {Vuleti{\'c}}}, \ and\ \bibinfo {author} {\bibfnamefont
  {M.~D.}\ \bibnamefont {Lukin}},\ }\href {\doibase 10.1126/science.aax9743}
  {\bibfield  {journal} {\bibinfo  {journal} {Science}\ }\textbf {\bibinfo
  {volume} {365}},\ \bibinfo {pages} {570} (\bibinfo {year} {2019})},\ \Eprint
  {http://arxiv.org/abs/https://science.sciencemag.org/content/365/6453/570.full.pdf}
  {https://science.sciencemag.org/content/365/6453/570.full.pdf} \BibitemShut
  {NoStop}%
\bibitem [{\citenamefont {Schreiber}\ \emph {et~al.}(2015)\citenamefont
  {Schreiber}, \citenamefont {Hodgman}, \citenamefont {Bordia}, \citenamefont
  {L{\"u}schen}, \citenamefont {Fischer}, \citenamefont {Vosk}, \citenamefont
  {Altman}, \citenamefont {Schneider},\ and\ \citenamefont
  {Bloch}}]{schreiber15observation}%
  \BibitemOpen
  \bibfield  {author} {\bibinfo {author} {\bibfnamefont {M.}~\bibnamefont
  {Schreiber}}, \bibinfo {author} {\bibfnamefont {S.~S.}\ \bibnamefont
  {Hodgman}}, \bibinfo {author} {\bibfnamefont {P.}~\bibnamefont {Bordia}},
  \bibinfo {author} {\bibfnamefont {H.~P.}\ \bibnamefont {L{\"u}schen}},
  \bibinfo {author} {\bibfnamefont {M.~H.}\ \bibnamefont {Fischer}}, \bibinfo
  {author} {\bibfnamefont {R.}~\bibnamefont {Vosk}}, \bibinfo {author}
  {\bibfnamefont {E.}~\bibnamefont {Altman}}, \bibinfo {author} {\bibfnamefont
  {U.}~\bibnamefont {Schneider}}, \ and\ \bibinfo {author} {\bibfnamefont
  {I.}~\bibnamefont {Bloch}},\ }\href {\doibase 10.1126/science.aaa7432}
  {\bibfield  {journal} {\bibinfo  {journal} {Science}\ }\textbf {\bibinfo
  {volume} {349}},\ \bibinfo {pages} {842} (\bibinfo {year}
  {2015})}\BibitemShut {NoStop}%
\bibitem [{\citenamefont {Breuer}\ and\ \citenamefont
  {Petruccione}(2002)}]{breuer02}%
  \BibitemOpen
  \bibfield  {author} {\bibinfo {author} {\bibfnamefont {H.~P.}\ \bibnamefont
  {Breuer}}\ and\ \bibinfo {author} {\bibfnamefont {F.}~\bibnamefont
  {Petruccione}},\ }\href@noop {} {\emph {\bibinfo {title} {The Theory of Open
  Quantum Systems}}}\ (\bibinfo  {publisher} {Oxford University Press},\
  \bibinfo {year} {2002})\BibitemShut {NoStop}%
\bibitem [{\citenamefont {Mandel}\ and\ \citenamefont
  {Wolf}(1995)}]{mandel95optical}%
  \BibitemOpen
  \bibfield  {author} {\bibinfo {author} {\bibfnamefont {L.}~\bibnamefont
  {Mandel}}\ and\ \bibinfo {author} {\bibfnamefont {E.}~\bibnamefont {Wolf}},\
  }\href@noop {} {\emph {\bibinfo {title} {Optical coherence and quantum
  optics}}}\ (\bibinfo  {publisher} {Cambridge University Press},\ \bibinfo
  {year} {1995})\BibitemShut {NoStop}%
\bibitem [{\citenamefont {Latune}\ \emph {et~al.}(2019)\citenamefont {Latune},
  \citenamefont {Sinayskiy},\ and\ \citenamefont
  {Petruccione}}]{latune19thermodynamics}%
  \BibitemOpen
  \bibfield  {author} {\bibinfo {author} {\bibfnamefont {C.~L.}\ \bibnamefont
  {Latune}}, \bibinfo {author} {\bibfnamefont {I.}~\bibnamefont {Sinayskiy}}, \
  and\ \bibinfo {author} {\bibfnamefont {F.}~\bibnamefont {Petruccione}},\
  }\href {\doibase 10.1103/PhysRevResearch.1.033192} {\bibfield  {journal}
  {\bibinfo  {journal} {Phys. Rev. Research}\ }\textbf {\bibinfo {volume}
  {1}},\ \bibinfo {pages} {033192} (\bibinfo {year} {2019})}\BibitemShut
  {NoStop}%
\bibitem [{\citenamefont {Greiner}\ \emph {et~al.}(1995)\citenamefont
  {Greiner}, \citenamefont {Neise},\ and\ \citenamefont
  {St\"{o}cker}}]{greiner95thermodynamics}%
  \BibitemOpen
  \bibfield  {author} {\bibinfo {author} {\bibfnamefont {W.}~\bibnamefont
  {Greiner}}, \bibinfo {author} {\bibfnamefont {L.}~\bibnamefont {Neise}}, \
  and\ \bibinfo {author} {\bibfnamefont {H.}~\bibnamefont {St\"{o}cker}},\
  }\href@noop {} {\emph {\bibinfo {title} {Thermodynamics and Statistical
  Mechanics}}}\ (\bibinfo  {publisher} {Springer-Verlag New York},\ \bibinfo
  {year} {1995})\BibitemShut {NoStop}%
\bibitem [{\citenamefont {Chaikin}\ and\ \citenamefont
  {Lubensky}(1995)}]{chaikin95principles}%
  \BibitemOpen
  \bibfield  {author} {\bibinfo {author} {\bibfnamefont {P.~M.}\ \bibnamefont
  {Chaikin}}\ and\ \bibinfo {author} {\bibfnamefont {T.~C.}\ \bibnamefont
  {Lubensky}},\ }\href {\doibase 10.1017/CBO9780511813467} {\emph {\bibinfo
  {title} {Principles of Condensed Matter Physics}}}\ (\bibinfo  {publisher}
  {Cambridge University Press},\ \bibinfo {year} {1995})\BibitemShut {NoStop}%
\bibitem [{\citenamefont {Sachdev}(1999)}]{sachdev99quantum}%
  \BibitemOpen
  \bibfield  {author} {\bibinfo {author} {\bibfnamefont {S.}~\bibnamefont
  {Sachdev}},\ }\href@noop {} {\emph {\bibinfo {title} {Quantum Phase
  Transitions}}}\ (\bibinfo  {publisher} {Cambridge University Press,
  Cambridge, England},\ \bibinfo {year} {1999})\BibitemShut {NoStop}%
\bibitem [{\citenamefont {Ma}\ \emph {et~al.}(2017)\citenamefont {Ma},
  \citenamefont {Su},\ and\ \citenamefont {Sun}}]{ma17quantum}%
  \BibitemOpen
  \bibfield  {author} {\bibinfo {author} {\bibfnamefont {Y.-H.}\ \bibnamefont
  {Ma}}, \bibinfo {author} {\bibfnamefont {S.-H.}\ \bibnamefont {Su}}, \ and\
  \bibinfo {author} {\bibfnamefont {C.-P.}\ \bibnamefont {Sun}},\ }\href
  {\doibase 10.1103/PhysRevE.96.022143} {\bibfield  {journal} {\bibinfo
  {journal} {Phys. Rev. E}\ }\textbf {\bibinfo {volume} {96}},\ \bibinfo
  {pages} {022143} (\bibinfo {year} {2017})}\BibitemShut {NoStop}%
\bibitem [{\citenamefont {Fadaie}\ \emph {et~al.}(2018)\citenamefont {Fadaie},
  \citenamefont {Yunt},\ and\ \citenamefont {M\"ustecapl\ifmmode \imath \else
  \i \fi{}o\ifmmode~\breve{g}\else \u{g}\fi{}lu}}]{fadaie18topological}%
  \BibitemOpen
  \bibfield  {author} {\bibinfo {author} {\bibfnamefont {M.}~\bibnamefont
  {Fadaie}}, \bibinfo {author} {\bibfnamefont {E.}~\bibnamefont {Yunt}}, \ and\
  \bibinfo {author} {\bibfnamefont {O.~E.}\ \bibnamefont {M\"ustecapl\ifmmode
  \imath \else \i \fi{}o\ifmmode~\breve{g}\else \u{g}\fi{}lu}},\ }\href
  {\doibase 10.1103/PhysRevE.98.052124} {\bibfield  {journal} {\bibinfo
  {journal} {Phys. Rev. E}\ }\textbf {\bibinfo {volume} {98}},\ \bibinfo
  {pages} {052124} (\bibinfo {year} {2018})}\BibitemShut {NoStop}%
\bibitem [{\citenamefont {Chand}\ and\ \citenamefont
  {Biswas}(2018)}]{chand18critical}%
  \BibitemOpen
  \bibfield  {author} {\bibinfo {author} {\bibfnamefont {S.}~\bibnamefont
  {Chand}}\ and\ \bibinfo {author} {\bibfnamefont {A.}~\bibnamefont {Biswas}},\
  }\href {\doibase 10.1103/PhysRevE.98.052147} {\bibfield  {journal} {\bibinfo
  {journal} {Phys. Rev. E}\ }\textbf {\bibinfo {volume} {98}},\ \bibinfo
  {pages} {052147} (\bibinfo {year} {2018})}\BibitemShut {NoStop}%
\bibitem [{\citenamefont {Fogarty}\ and\ \citenamefont
  {Busch}(2020)}]{fogarty20a}%
  \BibitemOpen
  \bibfield  {author} {\bibinfo {author} {\bibfnamefont {T.}~\bibnamefont
  {Fogarty}}\ and\ \bibinfo {author} {\bibfnamefont {T.}~\bibnamefont
  {Busch}},\ }\href {\doibase 10.1088/2058-9565/abbc63} {\bibfield  {journal}
  {\bibinfo  {journal} {Quantum Science and Technology}\ }\textbf {\bibinfo
  {volume} {6}},\ \bibinfo {pages} {015003} (\bibinfo {year}
  {2020})}\BibitemShut {NoStop}%
\bibitem [{\citenamefont {Dziarmaga}(2010)}]{dziarmaga10dynamics}%
  \BibitemOpen
  \bibfield  {author} {\bibinfo {author} {\bibfnamefont {J.}~\bibnamefont
  {Dziarmaga}},\ }\href {\doibase 10.1080/00018732.2010.514702} {\bibfield
  {journal} {\bibinfo  {journal} {Advances in Physics}\ }\textbf {\bibinfo
  {volume} {59}},\ \bibinfo {pages} {1063} (\bibinfo {year}
  {2010})}\BibitemShut {NoStop}%
\bibitem [{\citenamefont {Dutta}\ \emph {et~al.}(2015)\citenamefont {Dutta},
  \citenamefont {Aeppli}, \citenamefont {Chakrabarti}, \citenamefont
  {Divakaran}, \citenamefont {Rosenbaum},\ and\ \citenamefont
  {Sen}}]{dutta15quantum}%
  \BibitemOpen
  \bibfield  {author} {\bibinfo {author} {\bibfnamefont {A.}~\bibnamefont
  {Dutta}}, \bibinfo {author} {\bibfnamefont {G.}~\bibnamefont {Aeppli}},
  \bibinfo {author} {\bibfnamefont {B.~K.}\ \bibnamefont {Chakrabarti}},
  \bibinfo {author} {\bibfnamefont {U.}~\bibnamefont {Divakaran}}, \bibinfo
  {author} {\bibfnamefont {T.~F.}\ \bibnamefont {Rosenbaum}}, \ and\ \bibinfo
  {author} {\bibfnamefont {D.}~\bibnamefont {Sen}},\ }\href@noop {} {\emph
  {\bibinfo {title} {Quantum phase transitions in transverse field spin models:
  from statistical physics to quantum information}}}\ (\bibinfo  {publisher}
  {Cambridge University Press, Cambridge},\ \bibinfo {year} {2015})\BibitemShut
  {NoStop}%
\bibitem [{\citenamefont {Polkovnikov}\ \emph {et~al.}(2011)\citenamefont
  {Polkovnikov}, \citenamefont {Sengupta}, \citenamefont {Silva},\ and\
  \citenamefont {Vengalattore}}]{polkovnikov11colloquium}%
  \BibitemOpen
  \bibfield  {author} {\bibinfo {author} {\bibfnamefont {A.}~\bibnamefont
  {Polkovnikov}}, \bibinfo {author} {\bibfnamefont {K.}~\bibnamefont
  {Sengupta}}, \bibinfo {author} {\bibfnamefont {A.}~\bibnamefont {Silva}}, \
  and\ \bibinfo {author} {\bibfnamefont {M.}~\bibnamefont {Vengalattore}},\
  }\href {\doibase 10.1103/RevModPhys.83.863} {\bibfield  {journal} {\bibinfo
  {journal} {Rev. Mod. Phys.}\ }\textbf {\bibinfo {volume} {83}},\ \bibinfo
  {pages} {863} (\bibinfo {year} {2011})}\BibitemShut {NoStop}%
\bibitem [{\citenamefont {Zurek}\ \emph {et~al.}(2005)\citenamefont {Zurek},
  \citenamefont {Dorner},\ and\ \citenamefont {Zoller}}]{zurek05dynamics}%
  \BibitemOpen
  \bibfield  {author} {\bibinfo {author} {\bibfnamefont {W.~H.}\ \bibnamefont
  {Zurek}}, \bibinfo {author} {\bibfnamefont {U.}~\bibnamefont {Dorner}}, \
  and\ \bibinfo {author} {\bibfnamefont {P.}~\bibnamefont {Zoller}},\ }\href
  {\doibase 10.1103/PhysRevLett.95.105701} {\bibfield  {journal} {\bibinfo
  {journal} {Phys. Rev. Lett.}\ }\textbf {\bibinfo {volume} {95}},\ \bibinfo
  {pages} {105701} (\bibinfo {year} {2005})}\BibitemShut {NoStop}%
\bibitem [{\citenamefont {Polkovnikov}(2005)}]{polkovnikov05universal}%
  \BibitemOpen
  \bibfield  {author} {\bibinfo {author} {\bibfnamefont {A.}~\bibnamefont
  {Polkovnikov}},\ }\href {\doibase 10.1103/PhysRevB.72.161201} {\bibfield
  {journal} {\bibinfo  {journal} {Phys. Rev. B}\ }\textbf {\bibinfo {volume}
  {72}},\ \bibinfo {pages} {161201} (\bibinfo {year} {2005})}\BibitemShut
  {NoStop}%
\bibitem [{\citenamefont {Fei}\ \emph {et~al.}(2020)\citenamefont {Fei},
  \citenamefont {Freitas}, \citenamefont {Cavina}, \citenamefont {Quan},\ and\
  \citenamefont {Esposito}}]{fei2020work}%
  \BibitemOpen
  \bibfield  {author} {\bibinfo {author} {\bibfnamefont {Z.}~\bibnamefont
  {Fei}}, \bibinfo {author} {\bibfnamefont {N.}~\bibnamefont {Freitas}},
  \bibinfo {author} {\bibfnamefont {V.}~\bibnamefont {Cavina}}, \bibinfo
  {author} {\bibfnamefont {H.~T.}\ \bibnamefont {Quan}}, \ and\ \bibinfo
  {author} {\bibfnamefont {M.}~\bibnamefont {Esposito}},\ }\href {\doibase
  10.1103/PhysRevLett.124.170603} {\bibfield  {journal} {\bibinfo  {journal}
  {Phys. Rev. Lett.}\ }\textbf {\bibinfo {volume} {124}},\ \bibinfo {pages}
  {170603} (\bibinfo {year} {2020})}\BibitemShut {NoStop}%
\bibitem [{\citenamefont {Lieb}\ \emph {et~al.}(1961)\citenamefont {Lieb},
  \citenamefont {Schultz},\ and\ \citenamefont {Mattis}}]{lieb61two}%
  \BibitemOpen
  \bibfield  {author} {\bibinfo {author} {\bibfnamefont {E.}~\bibnamefont
  {Lieb}}, \bibinfo {author} {\bibfnamefont {T.}~\bibnamefont {Schultz}}, \
  and\ \bibinfo {author} {\bibfnamefont {D.}~\bibnamefont {Mattis}},\ }\href
  {\doibase https://doi.org/10.1016/0003-4916(61)90115-4} {\bibfield  {journal}
  {\bibinfo  {journal} {Annals of Physics}\ }\textbf {\bibinfo {volume} {16}},\
  \bibinfo {pages} {407 } (\bibinfo {year} {1961})}\BibitemShut {NoStop}%
\bibitem [{\citenamefont {Pfeuty}(1970)}]{pfeuty70the}%
  \BibitemOpen
  \bibfield  {author} {\bibinfo {author} {\bibfnamefont {P.}~\bibnamefont
  {Pfeuty}},\ }\href {\doibase https://doi.org/10.1016/0003-4916(70)90270-8}
  {\bibfield  {journal} {\bibinfo  {journal} {Annals of Physics}\ }\textbf
  {\bibinfo {volume} {57}},\ \bibinfo {pages} {79 } (\bibinfo {year}
  {1970})}\BibitemShut {NoStop}%
\bibitem [{\citenamefont {Bunder}\ and\ \citenamefont
  {McKenzie}(1999)}]{bunder99effect}%
  \BibitemOpen
  \bibfield  {author} {\bibinfo {author} {\bibfnamefont {J.~E.}\ \bibnamefont
  {Bunder}}\ and\ \bibinfo {author} {\bibfnamefont {R.~H.}\ \bibnamefont
  {McKenzie}},\ }\href {\doibase 10.1103/PhysRevB.60.344} {\bibfield  {journal}
  {\bibinfo  {journal} {Phys. Rev. B}\ }\textbf {\bibinfo {volume} {60}},\
  \bibinfo {pages} {344} (\bibinfo {year} {1999})}\BibitemShut {NoStop}%
\bibitem [{\citenamefont {Dziarmaga}(2005)}]{dziarmaga05dynamics}%
  \BibitemOpen
  \bibfield  {author} {\bibinfo {author} {\bibfnamefont {J.}~\bibnamefont
  {Dziarmaga}},\ }\href {\doibase 10.1103/PhysRevLett.95.245701} {\bibfield
  {journal} {\bibinfo  {journal} {Phys. Rev. Lett.}\ }\textbf {\bibinfo
  {volume} {95}},\ \bibinfo {pages} {245701} (\bibinfo {year}
  {2005})}\BibitemShut {NoStop}%
\bibitem [{\citenamefont {Sengupta}\ \emph {et~al.}(2008)\citenamefont
  {Sengupta}, \citenamefont {Sen},\ and\ \citenamefont
  {Mondal}}]{sengupta08exact}%
  \BibitemOpen
  \bibfield  {author} {\bibinfo {author} {\bibfnamefont {K.}~\bibnamefont
  {Sengupta}}, \bibinfo {author} {\bibfnamefont {D.}~\bibnamefont {Sen}}, \
  and\ \bibinfo {author} {\bibfnamefont {S.}~\bibnamefont {Mondal}},\ }\href
  {\doibase 10.1103/PhysRevLett.100.077204} {\bibfield  {journal} {\bibinfo
  {journal} {Phys. Rev. Lett.}\ }\textbf {\bibinfo {volume} {100}},\ \bibinfo
  {pages} {077204} (\bibinfo {year} {2008})}\BibitemShut {NoStop}%
\bibitem [{\citenamefont {del Campo}\ \emph {et~al.}(2018)\citenamefont {del
  Campo}, \citenamefont {Chenu}, \citenamefont {Deng},\ and\ \citenamefont
  {Wu}}]{chenu18thermodynamics}%
  \BibitemOpen
  \bibfield  {author} {\bibinfo {author} {\bibfnamefont {A.}~\bibnamefont {del
  Campo}}, \bibinfo {author} {\bibfnamefont {A.}~\bibnamefont {Chenu}},
  \bibinfo {author} {\bibfnamefont {S.}~\bibnamefont {Deng}}, \ and\ \bibinfo
  {author} {\bibfnamefont {H.}~\bibnamefont {Wu}},\ }\enquote {\bibinfo {title}
  {Friction-free quantum machines},}\ in\ \href {\doibase
  10.1007/978-3-319-99046-0_5} {\emph {\bibinfo {booktitle} {Thermodynamics in
  the Quantum Regime: Fundamental Aspects and New Directions}}},\ \bibinfo
  {editor} {edited by\ \bibinfo {editor} {\bibfnamefont {F.}~\bibnamefont
  {Binder}}, \bibinfo {editor} {\bibfnamefont {L.~A.}\ \bibnamefont {Correa}},
  \bibinfo {editor} {\bibfnamefont {C.}~\bibnamefont {Gogolin}}, \bibinfo
  {editor} {\bibfnamefont {J.}~\bibnamefont {Anders}}, \ and\ \bibinfo {editor}
  {\bibfnamefont {G.}~\bibnamefont {Adesso}}}\ (\bibinfo  {publisher} {Springer
  International Publishing},\ \bibinfo {address} {Cham},\ \bibinfo {year}
  {2018})\ pp.\ \bibinfo {pages} {127--148}\BibitemShut {NoStop}%
\bibitem [{\citenamefont {Demirplak}\ and\ \citenamefont
  {Rice}(2003)}]{demirplak03adiabatic}%
  \BibitemOpen
  \bibfield  {author} {\bibinfo {author} {\bibfnamefont {M.}~\bibnamefont
  {Demirplak}}\ and\ \bibinfo {author} {\bibfnamefont {S.~A.}\ \bibnamefont
  {Rice}},\ }\href {\doibase 10.1021/jp030708a} {\bibfield  {journal} {\bibinfo
   {journal} {The Journal of Physical Chemistry A}\ }\textbf {\bibinfo {volume}
  {107}},\ \bibinfo {pages} {9937} (\bibinfo {year} {2003})}\BibitemShut
  {NoStop}%
\bibitem [{\citenamefont {Berry}(2009)}]{berry09transitionless}%
  \BibitemOpen
  \bibfield  {author} {\bibinfo {author} {\bibfnamefont {M.~V.}\ \bibnamefont
  {Berry}},\ }\href {\doibase 10.1088/1751-8113/42/36/365303} {\bibfield
  {journal} {\bibinfo  {journal} {Journal of Physics A: Mathematical and
  Theoretical}\ }\textbf {\bibinfo {volume} {42}},\ \bibinfo {pages} {365303}
  (\bibinfo {year} {2009})}\BibitemShut {NoStop}%
\bibitem [{\citenamefont {del Campo}\ \emph {et~al.}(2012)\citenamefont {del
  Campo}, \citenamefont {Rams},\ and\ \citenamefont {Zurek}}]{campo12assisted}%
  \BibitemOpen
  \bibfield  {author} {\bibinfo {author} {\bibfnamefont {A.}~\bibnamefont {del
  Campo}}, \bibinfo {author} {\bibfnamefont {M.~M.}\ \bibnamefont {Rams}}, \
  and\ \bibinfo {author} {\bibfnamefont {W.~H.}\ \bibnamefont {Zurek}},\ }\href
  {\doibase 10.1103/PhysRevLett.109.115703} {\bibfield  {journal} {\bibinfo
  {journal} {Phys. Rev. Lett.}\ }\textbf {\bibinfo {volume} {109}},\ \bibinfo
  {pages} {115703} (\bibinfo {year} {2012})}\BibitemShut {NoStop}%
\bibitem [{\citenamefont {Deffner}\ \emph {et~al.}(2014)\citenamefont
  {Deffner}, \citenamefont {Jarzynski},\ and\ \citenamefont {del
  Campo}}]{deffner14classical}%
  \BibitemOpen
  \bibfield  {author} {\bibinfo {author} {\bibfnamefont {S.}~\bibnamefont
  {Deffner}}, \bibinfo {author} {\bibfnamefont {C.}~\bibnamefont {Jarzynski}},
  \ and\ \bibinfo {author} {\bibfnamefont {A.}~\bibnamefont {del Campo}},\
  }\href {\doibase 10.1103/PhysRevX.4.021013} {\bibfield  {journal} {\bibinfo
  {journal} {Phys. Rev. X}\ }\textbf {\bibinfo {volume} {4}},\ \bibinfo {pages}
  {021013} (\bibinfo {year} {2014})}\BibitemShut {NoStop}%
\bibitem [{\citenamefont {Patra}\ and\ \citenamefont
  {Jarzynski}(2017)}]{patra17shortcuts}%
  \BibitemOpen
  \bibfield  {author} {\bibinfo {author} {\bibfnamefont {A.}~\bibnamefont
  {Patra}}\ and\ \bibinfo {author} {\bibfnamefont {C.}~\bibnamefont
  {Jarzynski}},\ }\href {\doibase 10.1088/1367-2630/aa924c} {\bibfield
  {journal} {\bibinfo  {journal} {New Journal of Physics}\ }\textbf {\bibinfo
  {volume} {19}},\ \bibinfo {pages} {125009} (\bibinfo {year}
  {2017})}\BibitemShut {NoStop}%
\bibitem [{\citenamefont {Gu\'ery-Odelin}\ \emph {et~al.}(2019)\citenamefont
  {Gu\'ery-Odelin}, \citenamefont {Ruschhaupt}, \citenamefont {Kiely},
  \citenamefont {Torrontegui}, \citenamefont {Mart\'{\i}nez-Garaot},\ and\
  \citenamefont {Muga}}]{odelin19shortcuts}%
  \BibitemOpen
  \bibfield  {author} {\bibinfo {author} {\bibfnamefont {D.}~\bibnamefont
  {Gu\'ery-Odelin}}, \bibinfo {author} {\bibfnamefont {A.}~\bibnamefont
  {Ruschhaupt}}, \bibinfo {author} {\bibfnamefont {A.}~\bibnamefont {Kiely}},
  \bibinfo {author} {\bibfnamefont {E.}~\bibnamefont {Torrontegui}}, \bibinfo
  {author} {\bibfnamefont {S.}~\bibnamefont {Mart\'{\i}nez-Garaot}}, \ and\
  \bibinfo {author} {\bibfnamefont {J.~G.}\ \bibnamefont {Muga}},\ }\href
  {\doibase 10.1103/RevModPhys.91.045001} {\bibfield  {journal} {\bibinfo
  {journal} {Rev. Mod. Phys.}\ }\textbf {\bibinfo {volume} {91}},\ \bibinfo
  {pages} {045001} (\bibinfo {year} {2019})}\BibitemShut {NoStop}%
\bibitem [{\citenamefont {Xu}\ \emph {et~al.}(2020)\citenamefont {Xu},
  \citenamefont {Li}, \citenamefont {Busch}, \citenamefont {Chen},\ and\
  \citenamefont {Fogarty}}]{xu20effects}%
  \BibitemOpen
  \bibfield  {author} {\bibinfo {author} {\bibfnamefont {T.-N.}\ \bibnamefont
  {Xu}}, \bibinfo {author} {\bibfnamefont {J.}~\bibnamefont {Li}}, \bibinfo
  {author} {\bibfnamefont {T.}~\bibnamefont {Busch}}, \bibinfo {author}
  {\bibfnamefont {X.}~\bibnamefont {Chen}}, \ and\ \bibinfo {author}
  {\bibfnamefont {T.}~\bibnamefont {Fogarty}},\ }\href {\doibase
  10.1103/PhysRevResearch.2.023125} {\bibfield  {journal} {\bibinfo  {journal}
  {Phys. Rev. Research}\ }\textbf {\bibinfo {volume} {2}},\ \bibinfo {pages}
  {023125} (\bibinfo {year} {2020})}\BibitemShut {NoStop}%
\bibitem [{\citenamefont {Patra}\ and\ \citenamefont
  {Jarzynski}(2021)}]{patra21semiclassical}%
  \BibitemOpen
  \bibfield  {author} {\bibinfo {author} {\bibfnamefont {A.}~\bibnamefont
  {Patra}}\ and\ \bibinfo {author} {\bibfnamefont {C.}~\bibnamefont
  {Jarzynski}},\ }\href@noop {} {\enquote {\bibinfo {title} {Semiclassical
  fast-forward shortcuts to adiabaticity},}\ } (\bibinfo {year} {2021}),\
  \Eprint {http://arxiv.org/abs/2101.05901} {arXiv:2101.05901 [quant-ph]}
  \BibitemShut {NoStop}%
\bibitem [{\citenamefont {Vacanti}\ \emph {et~al.}(2014)\citenamefont
  {Vacanti}, \citenamefont {Fazio}, \citenamefont {Montangero}, \citenamefont
  {Palma}, \citenamefont {Paternostro},\ and\ \citenamefont
  {Vedral}}]{vacanti14transitionless}%
  \BibitemOpen
  \bibfield  {author} {\bibinfo {author} {\bibfnamefont {G.}~\bibnamefont
  {Vacanti}}, \bibinfo {author} {\bibfnamefont {R.}~\bibnamefont {Fazio}},
  \bibinfo {author} {\bibfnamefont {S.}~\bibnamefont {Montangero}}, \bibinfo
  {author} {\bibfnamefont {G.~M.}\ \bibnamefont {Palma}}, \bibinfo {author}
  {\bibfnamefont {M.}~\bibnamefont {Paternostro}}, \ and\ \bibinfo {author}
  {\bibfnamefont {V.}~\bibnamefont {Vedral}},\ }\href {\doibase
  10.1088/1367-2630/16/5/053017} {\bibfield  {journal} {\bibinfo  {journal}
  {New Journal of Physics}\ }\textbf {\bibinfo {volume} {16}},\ \bibinfo
  {pages} {053017} (\bibinfo {year} {2014})}\BibitemShut {NoStop}%
\bibitem [{\citenamefont {Alipour}\ \emph {et~al.}(2020)\citenamefont
  {Alipour}, \citenamefont {Chenu}, \citenamefont {Rezakhani},\ and\
  \citenamefont {del Campo}}]{alipour20shortcuts}%
  \BibitemOpen
  \bibfield  {author} {\bibinfo {author} {\bibfnamefont {S.}~\bibnamefont
  {Alipour}}, \bibinfo {author} {\bibfnamefont {A.}~\bibnamefont {Chenu}},
  \bibinfo {author} {\bibfnamefont {A.~T.}\ \bibnamefont {Rezakhani}}, \ and\
  \bibinfo {author} {\bibfnamefont {A.}~\bibnamefont {del Campo}},\ }\href
  {\doibase 10.22331/q-2020-09-28-336} {\bibfield  {journal} {\bibinfo
  {journal} {{Quantum}}\ }\textbf {\bibinfo {volume} {4}},\ \bibinfo {pages}
  {336} (\bibinfo {year} {2020})}\BibitemShut {NoStop}%
\bibitem [{\citenamefont {An}\ \emph {et~al.}(2016)\citenamefont {An},
  \citenamefont {Lv}, \citenamefont {del Campo},\ and\ \citenamefont
  {Kim}}]{an16shortcuts}%
  \BibitemOpen
  \bibfield  {author} {\bibinfo {author} {\bibfnamefont {S.}~\bibnamefont
  {An}}, \bibinfo {author} {\bibfnamefont {D.}~\bibnamefont {Lv}}, \bibinfo
  {author} {\bibfnamefont {A.}~\bibnamefont {del Campo}}, \ and\ \bibinfo
  {author} {\bibfnamefont {K.}~\bibnamefont {Kim}},\ }\href {\doibase
  10.1038/ncomms12999} {\bibfield  {journal} {\bibinfo  {journal} {Nature
  Communications}\ }\textbf {\bibinfo {volume} {7}},\ \bibinfo {pages} {12999}
  (\bibinfo {year} {2016})}\BibitemShut {NoStop}%
\bibitem [{\citenamefont {del Campo}\ \emph {et~al.}(2014)\citenamefont {del
  Campo}, \citenamefont {Goold},\ and\ \citenamefont
  {Paternostro}}]{delcampo14more}%
  \BibitemOpen
  \bibfield  {author} {\bibinfo {author} {\bibfnamefont {A.}~\bibnamefont {del
  Campo}}, \bibinfo {author} {\bibfnamefont {J.}~\bibnamefont {Goold}}, \ and\
  \bibinfo {author} {\bibfnamefont {M.}~\bibnamefont {Paternostro}},\ }\href
  {\doibase 10.1038/srep06208} {\bibfield  {journal} {\bibinfo  {journal} {Sci.
  Rep.}\ }\textbf {\bibinfo {volume} {4}},\ \bibinfo {pages} {6208} (\bibinfo
  {year} {2014})}\BibitemShut {NoStop}%
\bibitem [{\citenamefont {\ifmmode~\mbox{\c{C}}\else \c{C}\fi{}akmak}\ and\
  \citenamefont {M\"ustecapl\ifmmode \imath \else \i
  \fi{}o\ifmmode~\breve{g}\else \u{g}\fi{}lu}(2019)}]{cakmak19spin}%
  \BibitemOpen
  \bibfield  {author} {\bibinfo {author} {\bibfnamefont {B.}~\bibnamefont
  {\ifmmode~\mbox{\c{C}}\else \c{C}\fi{}akmak}}\ and\ \bibinfo {author}
  {\bibfnamefont {O.~E.}\ \bibnamefont {M\"ustecapl\ifmmode \imath \else \i
  \fi{}o\ifmmode~\breve{g}\else \u{g}\fi{}lu}},\ }\href {\doibase
  10.1103/PhysRevE.99.032108} {\bibfield  {journal} {\bibinfo  {journal} {Phys.
  Rev. E}\ }\textbf {\bibinfo {volume} {99}},\ \bibinfo {pages} {032108}
  (\bibinfo {year} {2019})}\BibitemShut {NoStop}%
\bibitem [{\citenamefont {Li}\ \emph {et~al.}(2018)\citenamefont {Li},
  \citenamefont {Fogarty}, \citenamefont {Campbell}, \citenamefont {Chen},\
  and\ \citenamefont {Busch}}]{li18an}%
  \BibitemOpen
  \bibfield  {author} {\bibinfo {author} {\bibfnamefont {J.}~\bibnamefont
  {Li}}, \bibinfo {author} {\bibfnamefont {T.}~\bibnamefont {Fogarty}},
  \bibinfo {author} {\bibfnamefont {S.}~\bibnamefont {Campbell}}, \bibinfo
  {author} {\bibfnamefont {X.}~\bibnamefont {Chen}}, \ and\ \bibinfo {author}
  {\bibfnamefont {T.}~\bibnamefont {Busch}},\ }\href {\doibase
  10.1088/1367-2630/aa9cd8} {\bibfield  {journal} {\bibinfo  {journal} {New
  Journal of Physics}\ }\textbf {\bibinfo {volume} {20}},\ \bibinfo {pages}
  {015005} (\bibinfo {year} {2018})}\BibitemShut {NoStop}%
\bibitem [{\citenamefont {Abah}\ \emph {et~al.}(2020)\citenamefont {Abah},
  \citenamefont {Paternostro},\ and\ \citenamefont {Lutz}}]{abah20shortcut}%
  \BibitemOpen
  \bibfield  {author} {\bibinfo {author} {\bibfnamefont {O.}~\bibnamefont
  {Abah}}, \bibinfo {author} {\bibfnamefont {M.}~\bibnamefont {Paternostro}}, \
  and\ \bibinfo {author} {\bibfnamefont {E.}~\bibnamefont {Lutz}},\ }\href
  {\doibase 10.1103/PhysRevResearch.2.023120} {\bibfield  {journal} {\bibinfo
  {journal} {Phys. Rev. Research}\ }\textbf {\bibinfo {volume} {2}},\ \bibinfo
  {pages} {023120} (\bibinfo {year} {2020})}\BibitemShut {NoStop}%
\bibitem [{\citenamefont {Campbell}\ and\ \citenamefont
  {Deffner}(2017)}]{campbell17trade}%
  \BibitemOpen
  \bibfield  {author} {\bibinfo {author} {\bibfnamefont {S.}~\bibnamefont
  {Campbell}}\ and\ \bibinfo {author} {\bibfnamefont {S.}~\bibnamefont
  {Deffner}},\ }\href {\doibase 10.1103/PhysRevLett.118.100601} {\bibfield
  {journal} {\bibinfo  {journal} {Phys. Rev. Lett.}\ }\textbf {\bibinfo
  {volume} {118}},\ \bibinfo {pages} {100601} (\bibinfo {year}
  {2017})}\BibitemShut {NoStop}%
\bibitem [{\citenamefont {Abah}\ and\ \citenamefont
  {Lutz}(2018)}]{abah18performance}%
  \BibitemOpen
  \bibfield  {author} {\bibinfo {author} {\bibfnamefont {O.}~\bibnamefont
  {Abah}}\ and\ \bibinfo {author} {\bibfnamefont {E.}~\bibnamefont {Lutz}},\
  }\href {\doibase 10.1103/PhysRevE.98.032121} {\bibfield  {journal} {\bibinfo
  {journal} {Phys. Rev. E}\ }\textbf {\bibinfo {volume} {98}},\ \bibinfo
  {pages} {032121} (\bibinfo {year} {2018})}\BibitemShut {NoStop}%
\bibitem [{\citenamefont {Abah}\ and\ \citenamefont
  {Paternostro}(2019)}]{abah19shortcut}%
  \BibitemOpen
  \bibfield  {author} {\bibinfo {author} {\bibfnamefont {O.}~\bibnamefont
  {Abah}}\ and\ \bibinfo {author} {\bibfnamefont {M.}~\bibnamefont
  {Paternostro}},\ }\href {\doibase 10.1103/PhysRevE.99.022110} {\bibfield
  {journal} {\bibinfo  {journal} {Phys. Rev. E}\ }\textbf {\bibinfo {volume}
  {99}},\ \bibinfo {pages} {022110} (\bibinfo {year} {2019})}\BibitemShut
  {NoStop}%
\bibitem [{\citenamefont {Mukherjee}\ \emph {et~al.}(2016)\citenamefont
  {Mukherjee}, \citenamefont {Montangero},\ and\ \citenamefont
  {Fazio}}]{mukherjee16local}%
  \BibitemOpen
  \bibfield  {author} {\bibinfo {author} {\bibfnamefont {V.}~\bibnamefont
  {Mukherjee}}, \bibinfo {author} {\bibfnamefont {S.}~\bibnamefont
  {Montangero}}, \ and\ \bibinfo {author} {\bibfnamefont {R.}~\bibnamefont
  {Fazio}},\ }\href {\doibase 10.1103/PhysRevA.93.062108} {\bibfield  {journal}
  {\bibinfo  {journal} {Phys. Rev. A}\ }\textbf {\bibinfo {volume} {93}},\
  \bibinfo {pages} {062108} (\bibinfo {year} {2016})}\BibitemShut {NoStop}%
\bibitem [{\citenamefont {Sels}\ and\ \citenamefont
  {Polkovnikov}(2017)}]{sels17minimizing}%
  \BibitemOpen
  \bibfield  {author} {\bibinfo {author} {\bibfnamefont {D.}~\bibnamefont
  {Sels}}\ and\ \bibinfo {author} {\bibfnamefont {A.}~\bibnamefont
  {Polkovnikov}},\ }\href {\doibase 10.1073/pnas.1619826114} {\bibfield
  {journal} {\bibinfo  {journal} {Proceedings of the National Academy of
  Sciences}\ }\textbf {\bibinfo {volume} {114}},\ \bibinfo {pages} {E3909}
  (\bibinfo {year} {2017})},\ \Eprint
  {http://arxiv.org/abs/https://www.pnas.org/content/114/20/E3909.full.pdf}
  {https://www.pnas.org/content/114/20/E3909.full.pdf} \BibitemShut {NoStop}%
\bibitem [{\citenamefont {Kolodrubetz}\ \emph {et~al.}(2017)\citenamefont
  {Kolodrubetz}, \citenamefont {Sels}, \citenamefont {Mehta},\ and\
  \citenamefont {Polkovnikov}}]{kolodrubetz17geometry}%
  \BibitemOpen
  \bibfield  {author} {\bibinfo {author} {\bibfnamefont {M.}~\bibnamefont
  {Kolodrubetz}}, \bibinfo {author} {\bibfnamefont {D.}~\bibnamefont {Sels}},
  \bibinfo {author} {\bibfnamefont {P.}~\bibnamefont {Mehta}}, \ and\ \bibinfo
  {author} {\bibfnamefont {A.}~\bibnamefont {Polkovnikov}},\ }\href {\doibase
  https://doi.org/10.1016/j.physrep.2017.07.001} {\bibfield  {journal}
  {\bibinfo  {journal} {Physics Reports}\ }\textbf {\bibinfo {volume} {697}},\
  \bibinfo {pages} {1 } (\bibinfo {year} {2017})},\ \bibinfo {note} {geometry
  and non-adiabatic response in quantum and classical systems}\BibitemShut
  {NoStop}%
\bibitem [{\citenamefont {Claeys}\ \emph {et~al.}(2019)\citenamefont {Claeys},
  \citenamefont {Pandey}, \citenamefont {Sels},\ and\ \citenamefont
  {Polkovnikov}}]{claeys19floquet}%
  \BibitemOpen
  \bibfield  {author} {\bibinfo {author} {\bibfnamefont {P.~W.}\ \bibnamefont
  {Claeys}}, \bibinfo {author} {\bibfnamefont {M.}~\bibnamefont {Pandey}},
  \bibinfo {author} {\bibfnamefont {D.}~\bibnamefont {Sels}}, \ and\ \bibinfo
  {author} {\bibfnamefont {A.}~\bibnamefont {Polkovnikov}},\ }\href {\doibase
  10.1103/PhysRevLett.123.090602} {\bibfield  {journal} {\bibinfo  {journal}
  {Phys. Rev. Lett.}\ }\textbf {\bibinfo {volume} {123}},\ \bibinfo {pages}
  {090602} (\bibinfo {year} {2019})}\BibitemShut {NoStop}%
\bibitem [{\citenamefont {Beau}\ \emph {et~al.}(2016)\citenamefont {Beau},
  \citenamefont {Jaramillo},\ and\ \citenamefont {del Campo}}]{beau16scaling}%
  \BibitemOpen
  \bibfield  {author} {\bibinfo {author} {\bibfnamefont {M.}~\bibnamefont
  {Beau}}, \bibinfo {author} {\bibfnamefont {J.}~\bibnamefont {Jaramillo}}, \
  and\ \bibinfo {author} {\bibfnamefont {A.}~\bibnamefont {del Campo}},\ }\href
  {\doibase 10.3390/e18050168} {\bibfield  {journal} {\bibinfo  {journal}
  {Entropy}\ }\textbf {\bibinfo {volume} {18}},\ \bibinfo {pages} {168}
  (\bibinfo {year} {2016})}\BibitemShut {NoStop}%
\bibitem [{\citenamefont {Alicki}(1979)}]{alicki79the}%
  \BibitemOpen
  \bibfield  {author} {\bibinfo {author} {\bibfnamefont {R.}~\bibnamefont
  {Alicki}},\ }\href@noop {} {\bibfield  {journal} {\bibinfo  {journal}
  {Journal of Physics A: Mathematical and General}\ }\textbf {\bibinfo {volume}
  {12}},\ \bibinfo {pages} {L103} (\bibinfo {year} {1979})}\BibitemShut
  {NoStop}%
\bibitem [{\citenamefont {Alipour}\ \emph {et~al.}(2019)\citenamefont
  {Alipour}, \citenamefont {Rezakhani}, \citenamefont {Chenu}, \citenamefont
  {del Campo},\ and\ \citenamefont {Ala-Nissila}}]{alipour19unambiguous}%
  \BibitemOpen
  \bibfield  {author} {\bibinfo {author} {\bibfnamefont {S.}~\bibnamefont
  {Alipour}}, \bibinfo {author} {\bibfnamefont {A.~T.}\ \bibnamefont
  {Rezakhani}}, \bibinfo {author} {\bibfnamefont {A.}~\bibnamefont {Chenu}},
  \bibinfo {author} {\bibfnamefont {A.}~\bibnamefont {del Campo}}, \ and\
  \bibinfo {author} {\bibfnamefont {T.}~\bibnamefont {Ala-Nissila}},\
  }\href@noop {} {\enquote {\bibinfo {title} {Unambiguous formulation for heat
  and work in arbitrary quantum evolution},}\ } (\bibinfo {year} {2019}),\
  \Eprint {http://arxiv.org/abs/1912.01939} {arXiv:1912.01939 [quant-ph]}
  \BibitemShut {NoStop}%
\bibitem [{\citenamefont {Whitney}(2018)}]{whitney18non}%
  \BibitemOpen
  \bibfield  {author} {\bibinfo {author} {\bibfnamefont {R.~S.}\ \bibnamefont
  {Whitney}},\ }\href {\doibase 10.1103/PhysRevB.98.085415} {\bibfield
  {journal} {\bibinfo  {journal} {Phys. Rev. B}\ }\textbf {\bibinfo {volume}
  {98}},\ \bibinfo {pages} {085415} (\bibinfo {year} {2018})}\BibitemShut
  {NoStop}%
\bibitem [{\citenamefont {Niedenzu}\ \emph {et~al.}(2018)\citenamefont
  {Niedenzu}, \citenamefont {Mukherjee}, \citenamefont {Ghosh}, \citenamefont
  {Kofman},\ and\ \citenamefont {Kurizki}}]{niedenzu18quantum}%
  \BibitemOpen
  \bibfield  {author} {\bibinfo {author} {\bibfnamefont {W.}~\bibnamefont
  {Niedenzu}}, \bibinfo {author} {\bibfnamefont {V.}~\bibnamefont {Mukherjee}},
  \bibinfo {author} {\bibfnamefont {A.}~\bibnamefont {Ghosh}}, \bibinfo
  {author} {\bibfnamefont {A.~G.}\ \bibnamefont {Kofman}}, \ and\ \bibinfo
  {author} {\bibfnamefont {G.}~\bibnamefont {Kurizki}},\ }\href {\doibase
  10.1038/s41467-017-01991-6} {\bibfield  {journal} {\bibinfo  {journal}
  {Nature Communications}\ }\textbf {\bibinfo {volume} {9}},\ \bibinfo {pages}
  {165} (\bibinfo {year} {2018})}\BibitemShut {NoStop}%
\bibitem [{\citenamefont {Quan}\ \emph {et~al.}(2007)\citenamefont {Quan},
  \citenamefont {Liu}, \citenamefont {Sun},\ and\ \citenamefont
  {Nori}}]{quan07quantum}%
  \BibitemOpen
  \bibfield  {author} {\bibinfo {author} {\bibfnamefont {H.~T.}\ \bibnamefont
  {Quan}}, \bibinfo {author} {\bibfnamefont {Y.-x.}\ \bibnamefont {Liu}},
  \bibinfo {author} {\bibfnamefont {C.~P.}\ \bibnamefont {Sun}}, \ and\
  \bibinfo {author} {\bibfnamefont {F.}~\bibnamefont {Nori}},\ }\href {\doibase
  10.1103/PhysRevE.76.031105} {\bibfield  {journal} {\bibinfo  {journal} {Phys.
  Rev. E}\ }\textbf {\bibinfo {volume} {76}},\ \bibinfo {pages} {031105}
  (\bibinfo {year} {2007})}\BibitemShut {NoStop}%
\bibitem [{\citenamefont {Cleuren}\ \emph {et~al.}(2012)\citenamefont
  {Cleuren}, \citenamefont {Rutten},\ and\ \citenamefont {Van~den
  Broeck}}]{cleuren12cooling}%
  \BibitemOpen
  \bibfield  {author} {\bibinfo {author} {\bibfnamefont {B.}~\bibnamefont
  {Cleuren}}, \bibinfo {author} {\bibfnamefont {B.}~\bibnamefont {Rutten}}, \
  and\ \bibinfo {author} {\bibfnamefont {C.}~\bibnamefont {Van~den Broeck}},\
  }\href {\doibase 10.1103/PhysRevLett.108.120603} {\bibfield  {journal}
  {\bibinfo  {journal} {Phys. Rev. Lett.}\ }\textbf {\bibinfo {volume} {108}},\
  \bibinfo {pages} {120603} (\bibinfo {year} {2012})}\BibitemShut {NoStop}%
\bibitem [{\citenamefont {Kol\'a\ifmmode~\check{r}\else \v{r}\fi{}}\ \emph
  {et~al.}(2012)\citenamefont {Kol\'a\ifmmode~\check{r}\else \v{r}\fi{}},
  \citenamefont {Gelbwaser-Klimovsky}, \citenamefont {Alicki},\ and\
  \citenamefont {Kurizki}}]{kolar12quantum}%
  \BibitemOpen
  \bibfield  {author} {\bibinfo {author} {\bibfnamefont {M.}~\bibnamefont
  {Kol\'a\ifmmode~\check{r}\else \v{r}\fi{}}}, \bibinfo {author} {\bibfnamefont
  {D.}~\bibnamefont {Gelbwaser-Klimovsky}}, \bibinfo {author} {\bibfnamefont
  {R.}~\bibnamefont {Alicki}}, \ and\ \bibinfo {author} {\bibfnamefont
  {G.}~\bibnamefont {Kurizki}},\ }\href {\doibase
  10.1103/PhysRevLett.109.090601} {\bibfield  {journal} {\bibinfo  {journal}
  {Phys. Rev. Lett.}\ }\textbf {\bibinfo {volume} {109}},\ \bibinfo {pages}
  {090601} (\bibinfo {year} {2012})}\BibitemShut {NoStop}%
\bibitem [{\citenamefont {Ro\ss{}nagel}\ \emph {et~al.}(2014)\citenamefont
  {Ro\ss{}nagel}, \citenamefont {Abah}, \citenamefont {Schmidt-Kaler},
  \citenamefont {Singer},\ and\ \citenamefont {Lutz}}]{rossnagel14nanoscale}%
  \BibitemOpen
  \bibfield  {author} {\bibinfo {author} {\bibfnamefont {J.}~\bibnamefont
  {Ro\ss{}nagel}}, \bibinfo {author} {\bibfnamefont {O.}~\bibnamefont {Abah}},
  \bibinfo {author} {\bibfnamefont {F.}~\bibnamefont {Schmidt-Kaler}}, \bibinfo
  {author} {\bibfnamefont {K.}~\bibnamefont {Singer}}, \ and\ \bibinfo {author}
  {\bibfnamefont {E.}~\bibnamefont {Lutz}},\ }\href {\doibase
  10.1103/PhysRevLett.112.030602} {\bibfield  {journal} {\bibinfo  {journal}
  {Phys. Rev. Lett.}\ }\textbf {\bibinfo {volume} {112}},\ \bibinfo {pages}
  {030602} (\bibinfo {year} {2014})}\BibitemShut {NoStop}%
\bibitem [{\citenamefont {Uzdin}\ \emph {et~al.}(2015)\citenamefont {Uzdin},
  \citenamefont {Levy},\ and\ \citenamefont {Kosloff}}]{uzdin15equivalence}%
  \BibitemOpen
  \bibfield  {author} {\bibinfo {author} {\bibfnamefont {R.}~\bibnamefont
  {Uzdin}}, \bibinfo {author} {\bibfnamefont {A.}~\bibnamefont {Levy}}, \ and\
  \bibinfo {author} {\bibfnamefont {R.}~\bibnamefont {Kosloff}},\ }\href
  {\doibase 10.1103/PhysRevX.5.031044} {\bibfield  {journal} {\bibinfo
  {journal} {Phys. Rev. X}\ }\textbf {\bibinfo {volume} {5}},\ \bibinfo {pages}
  {031044} (\bibinfo {year} {2015})}\BibitemShut {NoStop}%
\bibitem [{\citenamefont {Watanabe}\ \emph {et~al.}(2017)\citenamefont
  {Watanabe}, \citenamefont {Venkatesh}, \citenamefont {Talkner},\ and\
  \citenamefont {del Campo}}]{watanabe17quantum}%
  \BibitemOpen
  \bibfield  {author} {\bibinfo {author} {\bibfnamefont {G.}~\bibnamefont
  {Watanabe}}, \bibinfo {author} {\bibfnamefont {B.~P.}\ \bibnamefont
  {Venkatesh}}, \bibinfo {author} {\bibfnamefont {P.}~\bibnamefont {Talkner}},
  \ and\ \bibinfo {author} {\bibfnamefont {A.}~\bibnamefont {del Campo}},\
  }\href {\doibase 10.1103/PhysRevLett.118.050601} {\bibfield  {journal}
  {\bibinfo  {journal} {Phys. Rev. Lett.}\ }\textbf {\bibinfo {volume} {118}},\
  \bibinfo {pages} {050601} (\bibinfo {year} {2017})}\BibitemShut {NoStop}%
\bibitem [{\citenamefont {Friedenberger}\ and\ \citenamefont
  {Lutz}(2017)}]{friedenberger17when}%
  \BibitemOpen
  \bibfield  {author} {\bibinfo {author} {\bibfnamefont {A.}~\bibnamefont
  {Friedenberger}}\ and\ \bibinfo {author} {\bibfnamefont {E.}~\bibnamefont
  {Lutz}},\ }\href {\doibase 10.1209/0295-5075/120/10002} {\bibfield  {journal}
  {\bibinfo  {journal} {{EPL} (Europhysics Letters)}\ }\textbf {\bibinfo
  {volume} {120}},\ \bibinfo {pages} {10002} (\bibinfo {year}
  {2017})}\BibitemShut {NoStop}%
\bibitem [{\citenamefont {Freitas}\ and\ \citenamefont
  {Paz}(2017)}]{paz17fundamental}%
  \BibitemOpen
  \bibfield  {author} {\bibinfo {author} {\bibfnamefont {N.}~\bibnamefont
  {Freitas}}\ and\ \bibinfo {author} {\bibfnamefont {J.~P.}\ \bibnamefont
  {Paz}},\ }\href {\doibase 10.1103/PhysRevE.95.012146} {\bibfield  {journal}
  {\bibinfo  {journal} {Phys. Rev. E}\ }\textbf {\bibinfo {volume} {95}},\
  \bibinfo {pages} {012146} (\bibinfo {year} {2017})}\BibitemShut {NoStop}%
\bibitem [{\citenamefont {Ghosh}\ \emph {et~al.}(2018)\citenamefont {Ghosh},
  \citenamefont {Gelbwaser-Klimovsky}, \citenamefont {Niedenzu}, \citenamefont
  {Lvovsky}, \citenamefont {Mazets}, \citenamefont {Scully},\ and\
  \citenamefont {Kurizki}}]{ghosh18two}%
  \BibitemOpen
  \bibfield  {author} {\bibinfo {author} {\bibfnamefont {A.}~\bibnamefont
  {Ghosh}}, \bibinfo {author} {\bibfnamefont {D.}~\bibnamefont
  {Gelbwaser-Klimovsky}}, \bibinfo {author} {\bibfnamefont {W.}~\bibnamefont
  {Niedenzu}}, \bibinfo {author} {\bibfnamefont {A.~I.}\ \bibnamefont
  {Lvovsky}}, \bibinfo {author} {\bibfnamefont {I.}~\bibnamefont {Mazets}},
  \bibinfo {author} {\bibfnamefont {M.~O.}\ \bibnamefont {Scully}}, \ and\
  \bibinfo {author} {\bibfnamefont {G.}~\bibnamefont {Kurizki}},\ }\href
  {\doibase 10.1073/pnas.1805354115} {\bibfield  {journal} {\bibinfo  {journal}
  {Proceedings of the National Academy of Sciences}\ } (\bibinfo {year}
  {2018}),\ 10.1073/pnas.1805354115}\BibitemShut {NoStop}%
\bibitem [{\citenamefont {Ghosh}\ \emph {et~al.}(2019)\citenamefont {Ghosh},
  \citenamefont {Mukherjee}, \citenamefont {Niedenzu},\ and\ \citenamefont
  {Kurizki}}]{ghosh19are}%
  \BibitemOpen
  \bibfield  {author} {\bibinfo {author} {\bibfnamefont {A.}~\bibnamefont
  {Ghosh}}, \bibinfo {author} {\bibfnamefont {V.}~\bibnamefont {Mukherjee}},
  \bibinfo {author} {\bibfnamefont {W.}~\bibnamefont {Niedenzu}}, \ and\
  \bibinfo {author} {\bibfnamefont {G.}~\bibnamefont {Kurizki}},\ }\href
  {\doibase 10.1140/epjst/e2019-800060-7} {\bibfield  {journal} {\bibinfo
  {journal} {Eur. Phys. J. Special Topics}\ }\textbf {\bibinfo {volume}
  {227}},\ \bibinfo {pages} {2043} (\bibinfo {year} {2019})}\BibitemShut
  {NoStop}%
\bibitem [{\citenamefont {Erdman}\ \emph {et~al.}(2019)\citenamefont {Erdman},
  \citenamefont {Cavina}, \citenamefont {Fazio}, \citenamefont {Taddei},\ and\
  \citenamefont {Giovannetti}}]{erdman19maximum}%
  \BibitemOpen
  \bibfield  {author} {\bibinfo {author} {\bibfnamefont {P.~A.}\ \bibnamefont
  {Erdman}}, \bibinfo {author} {\bibfnamefont {V.}~\bibnamefont {Cavina}},
  \bibinfo {author} {\bibfnamefont {R.}~\bibnamefont {Fazio}}, \bibinfo
  {author} {\bibfnamefont {F.}~\bibnamefont {Taddei}}, \ and\ \bibinfo {author}
  {\bibfnamefont {V.}~\bibnamefont {Giovannetti}},\ }\href {\doibase
  10.1088/1367-2630/ab4dca} {\bibfield  {journal} {\bibinfo  {journal} {New
  Journal of Physics}\ }\textbf {\bibinfo {volume} {21}},\ \bibinfo {pages}
  {103049} (\bibinfo {year} {2019})}\BibitemShut {NoStop}%
\bibitem [{\citenamefont {Mukherjee}\ \emph {et~al.}(2020)\citenamefont
  {Mukherjee}, \citenamefont {Kofman},\ and\ \citenamefont
  {Kurizki}}]{mukherjee20anti}%
  \BibitemOpen
  \bibfield  {author} {\bibinfo {author} {\bibfnamefont {V.}~\bibnamefont
  {Mukherjee}}, \bibinfo {author} {\bibfnamefont {A.~G.}\ \bibnamefont
  {Kofman}}, \ and\ \bibinfo {author} {\bibfnamefont {G.}~\bibnamefont
  {Kurizki}},\ }\href {\doibase 10.1038/s42005-019-0272-z} {\bibfield
  {journal} {\bibinfo  {journal} {Communications Physics}\ }\textbf {\bibinfo
  {volume} {3}},\ \bibinfo {pages} {8} (\bibinfo {year} {2020})}\BibitemShut
  {NoStop}%
\bibitem [{\citenamefont {Alicki}(2014)}]{alicki14}%
  \BibitemOpen
  \bibfield  {author} {\bibinfo {author} {\bibfnamefont {R.}~\bibnamefont
  {Alicki}},\ }\href@noop {} {\bibfield  {journal} {\bibinfo  {journal} {Open
  Systems And Information Dynamics}\ }\textbf {\bibinfo {volume} {21}},\
  \bibinfo {pages} {1440002} (\bibinfo {year} {2014})}\BibitemShut {NoStop}%
\bibitem [{\citenamefont {Gelbwaser-Klimovsky}\ \emph
  {et~al.}(2013)\citenamefont {Gelbwaser-Klimovsky}, \citenamefont {Alicki},\
  and\ \citenamefont {Kurizki}}]{klimovsky13minimal}%
  \BibitemOpen
  \bibfield  {author} {\bibinfo {author} {\bibfnamefont {D.}~\bibnamefont
  {Gelbwaser-Klimovsky}}, \bibinfo {author} {\bibfnamefont {R.}~\bibnamefont
  {Alicki}}, \ and\ \bibinfo {author} {\bibfnamefont {G.}~\bibnamefont
  {Kurizki}},\ }\href {\doibase 10.1103/PhysRevE.87.012140} {\bibfield
  {journal} {\bibinfo  {journal} {Phys. Rev. E}\ }\textbf {\bibinfo {volume}
  {87}},\ \bibinfo {pages} {012140} (\bibinfo {year} {2013})}\BibitemShut
  {NoStop}%
\bibitem [{\citenamefont {Alicki}\ \emph {et~al.}(2012)\citenamefont {Alicki},
  \citenamefont {Gelbwaser-Klimovsky},\ and\ \citenamefont
  {Kurizki}}]{alicki12periodically}%
  \BibitemOpen
  \bibfield  {author} {\bibinfo {author} {\bibfnamefont {R.}~\bibnamefont
  {Alicki}}, \bibinfo {author} {\bibfnamefont {D.}~\bibnamefont
  {Gelbwaser-Klimovsky}}, \ and\ \bibinfo {author} {\bibfnamefont
  {G.}~\bibnamefont {Kurizki}},\ }\href@noop {} {\bibfield  {journal} {\bibinfo
   {journal} {arXiv:1205.4552}\ } (\bibinfo {year} {2012})}\BibitemShut
  {NoStop}%
\bibitem [{\citenamefont {Zheng}\ and\ \citenamefont
  {Poletti}(2015)}]{zheng15quantum}%
  \BibitemOpen
  \bibfield  {author} {\bibinfo {author} {\bibfnamefont {Y.}~\bibnamefont
  {Zheng}}\ and\ \bibinfo {author} {\bibfnamefont {D.}~\bibnamefont
  {Poletti}},\ }\href {\doibase 10.1103/PhysRevE.92.012110} {\bibfield
  {journal} {\bibinfo  {journal} {Phys. Rev. E}\ }\textbf {\bibinfo {volume}
  {92}},\ \bibinfo {pages} {012110} (\bibinfo {year} {2015})}\BibitemShut
  {NoStop}%
\bibitem [{\citenamefont {Watanabe}\ \emph {et~al.}(2020)\citenamefont
  {Watanabe}, \citenamefont {Venkatesh}, \citenamefont {Talkner}, \citenamefont
  {Hwang},\ and\ \citenamefont {del Campo}}]{watanabe20quantum}%
  \BibitemOpen
  \bibfield  {author} {\bibinfo {author} {\bibfnamefont {G.}~\bibnamefont
  {Watanabe}}, \bibinfo {author} {\bibfnamefont {B.~P.}\ \bibnamefont
  {Venkatesh}}, \bibinfo {author} {\bibfnamefont {P.}~\bibnamefont {Talkner}},
  \bibinfo {author} {\bibfnamefont {M.-J.}\ \bibnamefont {Hwang}}, \ and\
  \bibinfo {author} {\bibfnamefont {A.}~\bibnamefont {del Campo}},\ }\href
  {\doibase 10.1103/PhysRevLett.124.210603} {\bibfield  {journal} {\bibinfo
  {journal} {Phys. Rev. Lett.}\ }\textbf {\bibinfo {volume} {124}},\ \bibinfo
  {pages} {210603} (\bibinfo {year} {2020})}\BibitemShut {NoStop}%
\bibitem [{\citenamefont {Carollo}\ \emph
  {et~al.}(2020{\natexlab{a}})\citenamefont {Carollo}, \citenamefont
  {Gambetta}, \citenamefont {Brandner}, \citenamefont {Garrahan},\ and\
  \citenamefont {Lesanovsky}}]{carollo20nonequilibrium}%
  \BibitemOpen
  \bibfield  {author} {\bibinfo {author} {\bibfnamefont {F.}~\bibnamefont
  {Carollo}}, \bibinfo {author} {\bibfnamefont {F.~M.}\ \bibnamefont
  {Gambetta}}, \bibinfo {author} {\bibfnamefont {K.}~\bibnamefont {Brandner}},
  \bibinfo {author} {\bibfnamefont {J.~P.}\ \bibnamefont {Garrahan}}, \ and\
  \bibinfo {author} {\bibfnamefont {I.}~\bibnamefont {Lesanovsky}},\ }\href
  {\doibase 10.1103/PhysRevLett.124.170602} {\bibfield  {journal} {\bibinfo
  {journal} {Phys. Rev. Lett.}\ }\textbf {\bibinfo {volume} {124}},\ \bibinfo
  {pages} {170602} (\bibinfo {year} {2020}{\natexlab{a}})}\BibitemShut
  {NoStop}%
\bibitem [{\citenamefont {Azimi}\ \emph {et~al.}(2014)\citenamefont {Azimi},
  \citenamefont {Chotorlishvili}, \citenamefont {Mishra}, \citenamefont
  {Vekua}, \citenamefont {Hübner},\ and\ \citenamefont
  {Berakdar}}]{azimi14quantum}%
  \BibitemOpen
  \bibfield  {author} {\bibinfo {author} {\bibfnamefont {M.}~\bibnamefont
  {Azimi}}, \bibinfo {author} {\bibfnamefont {L.}~\bibnamefont
  {Chotorlishvili}}, \bibinfo {author} {\bibfnamefont {S.~K.}\ \bibnamefont
  {Mishra}}, \bibinfo {author} {\bibfnamefont {T.}~\bibnamefont {Vekua}},
  \bibinfo {author} {\bibfnamefont {W.}~\bibnamefont {Hübner}}, \ and\
  \bibinfo {author} {\bibfnamefont {J.}~\bibnamefont {Berakdar}},\ }\href
  {\doibase 10.1088/1367-2630/16/6/063018} {\bibfield  {journal} {\bibinfo
  {journal} {New Journal of Physics}\ }\textbf {\bibinfo {volume} {16}},\
  \bibinfo {pages} {063018} (\bibinfo {year} {2014})}\BibitemShut {NoStop}%
\bibitem [{\citenamefont {Yunt}\ \emph {et~al.}(2020)\citenamefont {Yunt},
  \citenamefont {Fadaie}, \citenamefont {M\"ustecapl\ifmmode \imath \else \i
  \fi{}o\ifmmode~\breve{g}\else \u{g}\fi{}lu},\ and\ \citenamefont
  {Smith}}]{yunt20topological}%
  \BibitemOpen
  \bibfield  {author} {\bibinfo {author} {\bibfnamefont {E.}~\bibnamefont
  {Yunt}}, \bibinfo {author} {\bibfnamefont {M.}~\bibnamefont {Fadaie}},
  \bibinfo {author} {\bibfnamefont {O.~E.}\ \bibnamefont {M\"ustecapl\ifmmode
  \imath \else \i \fi{}o\ifmmode~\breve{g}\else \u{g}\fi{}lu}}, \ and\ \bibinfo
  {author} {\bibfnamefont {C.~M.}\ \bibnamefont {Smith}},\ }\href {\doibase
  10.1103/PhysRevB.102.155423} {\bibfield  {journal} {\bibinfo  {journal}
  {Phys. Rev. B}\ }\textbf {\bibinfo {volume} {102}},\ \bibinfo {pages}
  {155423} (\bibinfo {year} {2020})}\BibitemShut {NoStop}%
\bibitem [{\citenamefont {Kumar}\ and\ \citenamefont
  {Benjamin}(2020)}]{kumar20a}%
  \BibitemOpen
  \bibfield  {author} {\bibinfo {author} {\bibfnamefont {A.}~\bibnamefont
  {Kumar}}\ and\ \bibinfo {author} {\bibfnamefont {C.}~\bibnamefont
  {Benjamin}},\ }\href@noop {} {\enquote {\bibinfo {title} {A thermodynamic
  probe of the topological phase transition in a floquet topological
  insulator},}\ } (\bibinfo {year} {2020}),\ \Eprint
  {http://arxiv.org/abs/2012.02172} {arXiv:2012.02172 [cond-mat.mes-hall]}
  \BibitemShut {NoStop}%
\bibitem [{\citenamefont {Carollo}\ \emph
  {et~al.}(2020{\natexlab{b}})\citenamefont {Carollo}, \citenamefont
  {Brandner},\ and\ \citenamefont
  {Lesanovsky}}]{carollo20nonequilibrium_arxiv}%
  \BibitemOpen
  \bibfield  {author} {\bibinfo {author} {\bibfnamefont {F.}~\bibnamefont
  {Carollo}}, \bibinfo {author} {\bibfnamefont {K.}~\bibnamefont {Brandner}}, \
  and\ \bibinfo {author} {\bibfnamefont {I.}~\bibnamefont {Lesanovsky}},\
  }\href@noop {} {\enquote {\bibinfo {title} {Nonequilibrium many-body quantum
  engine driven by time-translation symmetry breaking},}\ } (\bibinfo {year}
  {2020}{\natexlab{b}}),\ \Eprint {http://arxiv.org/abs/2007.00690}
  {arXiv:2007.00690 [cond-mat.stat-mech]} \BibitemShut {NoStop}%
\bibitem [{\citenamefont {Fusco}\ \emph {et~al.}(2016)\citenamefont {Fusco},
  \citenamefont {Paternostro},\ and\ \citenamefont {De~Chiara}}]{fusco16work}%
  \BibitemOpen
  \bibfield  {author} {\bibinfo {author} {\bibfnamefont {L.}~\bibnamefont
  {Fusco}}, \bibinfo {author} {\bibfnamefont {M.}~\bibnamefont {Paternostro}},
  \ and\ \bibinfo {author} {\bibfnamefont {G.}~\bibnamefont {De~Chiara}},\
  }\href {\doibase 10.1103/PhysRevE.94.052122} {\bibfield  {journal} {\bibinfo
  {journal} {Phys. Rev. E}\ }\textbf {\bibinfo {volume} {94}},\ \bibinfo
  {pages} {052122} (\bibinfo {year} {2016})}\BibitemShut {NoStop}%
\bibitem [{\citenamefont {Abiuso}\ and\ \citenamefont
  {Perarnau-Llobet}(2020)}]{abiuso20optimal}%
  \BibitemOpen
  \bibfield  {author} {\bibinfo {author} {\bibfnamefont {P.}~\bibnamefont
  {Abiuso}}\ and\ \bibinfo {author} {\bibfnamefont {M.}~\bibnamefont
  {Perarnau-Llobet}},\ }\href {\doibase 10.1103/PhysRevLett.124.110606}
  {\bibfield  {journal} {\bibinfo  {journal} {Phys. Rev. Lett.}\ }\textbf
  {\bibinfo {volume} {124}},\ \bibinfo {pages} {110606} (\bibinfo {year}
  {2020})}\BibitemShut {NoStop}%
\bibitem [{\citenamefont {Holubec}\ and\ \citenamefont
  {Ryabov}(2017)}]{holubec17work}%
  \BibitemOpen
  \bibfield  {author} {\bibinfo {author} {\bibfnamefont {V.}~\bibnamefont
  {Holubec}}\ and\ \bibinfo {author} {\bibfnamefont {A.}~\bibnamefont
  {Ryabov}},\ }\href {\doibase 10.1103/PhysRevE.96.030102} {\bibfield
  {journal} {\bibinfo  {journal} {Phys. Rev. E}\ }\textbf {\bibinfo {volume}
  {96}},\ \bibinfo {pages} {030102} (\bibinfo {year} {2017})}\BibitemShut
  {NoStop}%
\bibitem [{\citenamefont {Pietzonka}\ and\ \citenamefont
  {Seifert}(2018)}]{pietzonka18universal}%
  \BibitemOpen
  \bibfield  {author} {\bibinfo {author} {\bibfnamefont {P.}~\bibnamefont
  {Pietzonka}}\ and\ \bibinfo {author} {\bibfnamefont {U.}~\bibnamefont
  {Seifert}},\ }\href {\doibase 10.1103/PhysRevLett.120.190602} {\bibfield
  {journal} {\bibinfo  {journal} {Phys. Rev. Lett.}\ }\textbf {\bibinfo
  {volume} {120}},\ \bibinfo {pages} {190602} (\bibinfo {year}
  {2018})}\BibitemShut {NoStop}%
\bibitem [{\citenamefont {Patan\`e}\ \emph {et~al.}(2008)\citenamefont
  {Patan\`e}, \citenamefont {Silva}, \citenamefont {Amico}, \citenamefont
  {Fazio},\ and\ \citenamefont {Santoro}}]{patane08adiabatic}%
  \BibitemOpen
  \bibfield  {author} {\bibinfo {author} {\bibfnamefont {D.}~\bibnamefont
  {Patan\`e}}, \bibinfo {author} {\bibfnamefont {A.}~\bibnamefont {Silva}},
  \bibinfo {author} {\bibfnamefont {L.}~\bibnamefont {Amico}}, \bibinfo
  {author} {\bibfnamefont {R.}~\bibnamefont {Fazio}}, \ and\ \bibinfo {author}
  {\bibfnamefont {G.~E.}\ \bibnamefont {Santoro}},\ }\href {\doibase
  10.1103/PhysRevLett.101.175701} {\bibfield  {journal} {\bibinfo  {journal}
  {Phys. Rev. Lett.}\ }\textbf {\bibinfo {volume} {101}},\ \bibinfo {pages}
  {175701} (\bibinfo {year} {2008})}\BibitemShut {NoStop}%
\bibitem [{\citenamefont {Kibble}(1980)}]{kibble80some}%
  \BibitemOpen
  \bibfield  {author} {\bibinfo {author} {\bibfnamefont {T.}~\bibnamefont
  {Kibble}},\ }\href {\doibase https://doi.org/10.1016/0370-1573(80)90091-5}
  {\bibfield  {journal} {\bibinfo  {journal} {Physics Reports}\ }\textbf
  {\bibinfo {volume} {67}},\ \bibinfo {pages} {183 } (\bibinfo {year}
  {1980})}\BibitemShut {NoStop}%
\bibitem [{\citenamefont {Damski}\ and\ \citenamefont
  {Zurek}(2006)}]{damski06adiabatic}%
  \BibitemOpen
  \bibfield  {author} {\bibinfo {author} {\bibfnamefont {B.}~\bibnamefont
  {Damski}}\ and\ \bibinfo {author} {\bibfnamefont {W.~H.}\ \bibnamefont
  {Zurek}},\ }\href {\doibase 10.1103/PhysRevA.73.063405} {\bibfield  {journal}
  {\bibinfo  {journal} {Phys. Rev. A}\ }\textbf {\bibinfo {volume} {73}},\
  \bibinfo {pages} {063405} (\bibinfo {year} {2006})}\BibitemShut {NoStop}%
\bibitem [{\citenamefont {Lipkin}\ \emph {et~al.}(1965)\citenamefont {Lipkin},
  \citenamefont {Meshkov},\ and\ \citenamefont {Glick}}]{lipkin65validity}%
  \BibitemOpen
  \bibfield  {author} {\bibinfo {author} {\bibfnamefont {H.}~\bibnamefont
  {Lipkin}}, \bibinfo {author} {\bibfnamefont {N.}~\bibnamefont {Meshkov}}, \
  and\ \bibinfo {author} {\bibfnamefont {A.}~\bibnamefont {Glick}},\ }\href
  {\doibase https://doi.org/10.1016/0029-5582(65)90862-X} {\bibfield  {journal}
  {\bibinfo  {journal} {Nuclear Physics}\ }\textbf {\bibinfo {volume} {62}},\
  \bibinfo {pages} {188 } (\bibinfo {year} {1965})}\BibitemShut {NoStop}%
\bibitem [{\citenamefont {Meshkov}\ \emph {et~al.}(1965)\citenamefont
  {Meshkov}, \citenamefont {Glick},\ and\ \citenamefont
  {Lipkin}}]{meshkov65validity}%
  \BibitemOpen
  \bibfield  {author} {\bibinfo {author} {\bibfnamefont {N.}~\bibnamefont
  {Meshkov}}, \bibinfo {author} {\bibfnamefont {A.}~\bibnamefont {Glick}}, \
  and\ \bibinfo {author} {\bibfnamefont {H.}~\bibnamefont {Lipkin}},\ }\href
  {\doibase https://doi.org/10.1016/0029-5582(65)90863-1} {\bibfield  {journal}
  {\bibinfo  {journal} {Nuclear Physics}\ }\textbf {\bibinfo {volume} {62}},\
  \bibinfo {pages} {199 } (\bibinfo {year} {1965})}\BibitemShut {NoStop}%
\bibitem [{\citenamefont {Glick}\ \emph {et~al.}(1965)\citenamefont {Glick},
  \citenamefont {Lipkin},\ and\ \citenamefont {Meshkov}}]{glick65validity}%
  \BibitemOpen
  \bibfield  {author} {\bibinfo {author} {\bibfnamefont {A.}~\bibnamefont
  {Glick}}, \bibinfo {author} {\bibfnamefont {H.}~\bibnamefont {Lipkin}}, \
  and\ \bibinfo {author} {\bibfnamefont {N.}~\bibnamefont {Meshkov}},\ }\href
  {\doibase https://doi.org/10.1016/0029-5582(65)90864-3} {\bibfield  {journal}
  {\bibinfo  {journal} {Nuclear Physics}\ }\textbf {\bibinfo {volume} {62}},\
  \bibinfo {pages} {211 } (\bibinfo {year} {1965})}\BibitemShut {NoStop}%
\bibitem [{\citenamefont {Feldmann}\ and\ \citenamefont
  {Kosloff}(2006)}]{feldmann06quantum}%
  \BibitemOpen
  \bibfield  {author} {\bibinfo {author} {\bibfnamefont {T.}~\bibnamefont
  {Feldmann}}\ and\ \bibinfo {author} {\bibfnamefont {R.}~\bibnamefont
  {Kosloff}},\ }\href {\doibase 10.1103/PhysRevE.73.025107} {\bibfield
  {journal} {\bibinfo  {journal} {Phys. Rev. E}\ }\textbf {\bibinfo {volume}
  {73}},\ \bibinfo {pages} {025107} (\bibinfo {year} {2006})}\BibitemShut
  {NoStop}%
\bibitem [{\citenamefont {Dann}\ \emph {et~al.}(2020)\citenamefont {Dann},
  \citenamefont {Kosloff},\ and\ \citenamefont {Salamon}}]{dann20quantum}%
  \BibitemOpen
  \bibfield  {author} {\bibinfo {author} {\bibfnamefont {R.}~\bibnamefont
  {Dann}}, \bibinfo {author} {\bibfnamefont {R.}~\bibnamefont {Kosloff}}, \
  and\ \bibinfo {author} {\bibfnamefont {P.}~\bibnamefont {Salamon}},\ }\href
  {\doibase 10.3390/e22111255} {\bibfield  {journal} {\bibinfo  {journal}
  {Entropy}\ }\textbf {\bibinfo {volume} {22}} (\bibinfo {year} {2020}),\
  10.3390/e22111255}\BibitemShut {NoStop}%
\bibitem [{\citenamefont {Aleiner}\ \emph {et~al.}(2010)\citenamefont
  {Aleiner}, \citenamefont {Altshuler},\ and\ \citenamefont
  {Shlyapnikov}}]{aleiner10a}%
  \BibitemOpen
  \bibfield  {author} {\bibinfo {author} {\bibfnamefont {I.~L.}\ \bibnamefont
  {Aleiner}}, \bibinfo {author} {\bibfnamefont {B.~L.}\ \bibnamefont
  {Altshuler}}, \ and\ \bibinfo {author} {\bibfnamefont {G.~V.}\ \bibnamefont
  {Shlyapnikov}},\ }\href {\doibase 10.1038/nphys1758} {\bibfield  {journal}
  {\bibinfo  {journal} {Nature Physics}\ }\textbf {\bibinfo {volume} {6}},\
  \bibinfo {pages} {900} (\bibinfo {year} {2010})}\BibitemShut {NoStop}%
\bibitem [{\citenamefont {Nandkishore}\ and\ \citenamefont
  {Huse}(2015)}]{nandkishore15many}%
  \BibitemOpen
  \bibfield  {author} {\bibinfo {author} {\bibfnamefont {R.}~\bibnamefont
  {Nandkishore}}\ and\ \bibinfo {author} {\bibfnamefont {D.~A.}\ \bibnamefont
  {Huse}},\ }\href {\doibase 10.1146/annurev-conmatphys-031214-014726}
  {\bibfield  {journal} {\bibinfo  {journal} {Annual Review of Condensed Matter
  Physics}\ }\textbf {\bibinfo {volume} {6}},\ \bibinfo {pages} {15} (\bibinfo
  {year} {2015})},\ \Eprint
  {http://arxiv.org/abs/https://doi.org/10.1146/annurev-conmatphys-031214-014726}
  {https://doi.org/10.1146/annurev-conmatphys-031214-014726} \BibitemShut
  {NoStop}%
\bibitem [{\citenamefont {Alet}\ and\ \citenamefont
  {Laflorencie}(2018)}]{alet18many}%
  \BibitemOpen
  \bibfield  {author} {\bibinfo {author} {\bibfnamefont {F.}~\bibnamefont
  {Alet}}\ and\ \bibinfo {author} {\bibfnamefont {N.}~\bibnamefont
  {Laflorencie}},\ }\href {\doibase https://doi.org/10.1016/j.crhy.2018.03.003}
  {\bibfield  {journal} {\bibinfo  {journal} {Comptes Rendus Physique}\
  }\textbf {\bibinfo {volume} {19}},\ \bibinfo {pages} {498 } (\bibinfo {year}
  {2018})},\ \bibinfo {note} {quantum simulation / Simulation
  quantique}\BibitemShut {NoStop}%
\bibitem [{\citenamefont {Decker}\ \emph {et~al.}(2020)\citenamefont {Decker},
  \citenamefont {Karrasch}, \citenamefont {Eisert},\ and\ \citenamefont
  {Kennes}}]{decker20floquet}%
  \BibitemOpen
  \bibfield  {author} {\bibinfo {author} {\bibfnamefont {K.~S.~C.}\
  \bibnamefont {Decker}}, \bibinfo {author} {\bibfnamefont {C.}~\bibnamefont
  {Karrasch}}, \bibinfo {author} {\bibfnamefont {J.}~\bibnamefont {Eisert}}, \
  and\ \bibinfo {author} {\bibfnamefont {D.~M.}\ \bibnamefont {Kennes}},\
  }\href {\doibase 10.1103/PhysRevLett.124.190601} {\bibfield  {journal}
  {\bibinfo  {journal} {Phys. Rev. Lett.}\ }\textbf {\bibinfo {volume} {124}},\
  \bibinfo {pages} {190601} (\bibinfo {year} {2020})}\BibitemShut {NoStop}%
\bibitem [{\citenamefont {Rigol}\ \emph {et~al.}(2007)\citenamefont {Rigol},
  \citenamefont {Dunjko}, \citenamefont {Yurovsky},\ and\ \citenamefont
  {Olshanii}}]{rigol07relaxation}%
  \BibitemOpen
  \bibfield  {author} {\bibinfo {author} {\bibfnamefont {M.}~\bibnamefont
  {Rigol}}, \bibinfo {author} {\bibfnamefont {V.}~\bibnamefont {Dunjko}},
  \bibinfo {author} {\bibfnamefont {V.}~\bibnamefont {Yurovsky}}, \ and\
  \bibinfo {author} {\bibfnamefont {M.}~\bibnamefont {Olshanii}},\ }\href
  {\doibase 10.1103/PhysRevLett.98.050405} {\bibfield  {journal} {\bibinfo
  {journal} {Phys. Rev. Lett.}\ }\textbf {\bibinfo {volume} {98}},\ \bibinfo
  {pages} {050405} (\bibinfo {year} {2007})}\BibitemShut {NoStop}%
\bibitem [{\citenamefont {Rigol}\ \emph {et~al.}(2008)\citenamefont {Rigol},
  \citenamefont {Dunjko1},\ and\ \citenamefont
  {Olshanii}}]{rigol08thermalization}%
  \BibitemOpen
  \bibfield  {author} {\bibinfo {author} {\bibfnamefont {M.}~\bibnamefont
  {Rigol}}, \bibinfo {author} {\bibfnamefont {V.}~\bibnamefont {Dunjko1}}, \
  and\ \bibinfo {author} {\bibfnamefont {M.}~\bibnamefont {Olshanii}},\
  }\href@noop {} {\bibfield  {journal} {\bibinfo  {journal} {Nature}\ }\textbf
  {\bibinfo {volume} {452}},\ \bibinfo {pages} {854} (\bibinfo {year}
  {2008})}\BibitemShut {NoStop}%
\bibitem [{\citenamefont {Yunger~Halpern}\ \emph {et~al.}(2019)\citenamefont
  {Yunger~Halpern}, \citenamefont {White}, \citenamefont {Gopalakrishnan},\
  and\ \citenamefont {Refael}}]{halpern2019quantum}%
  \BibitemOpen
  \bibfield  {author} {\bibinfo {author} {\bibfnamefont {N.}~\bibnamefont
  {Yunger~Halpern}}, \bibinfo {author} {\bibfnamefont {C.~D.}\ \bibnamefont
  {White}}, \bibinfo {author} {\bibfnamefont {S.}~\bibnamefont
  {Gopalakrishnan}}, \ and\ \bibinfo {author} {\bibfnamefont {G.}~\bibnamefont
  {Refael}},\ }\href {\doibase 10.1103/PhysRevB.99.024203} {\bibfield
  {journal} {\bibinfo  {journal} {Phys. Rev. B}\ }\textbf {\bibinfo {volume}
  {99}},\ \bibinfo {pages} {024203} (\bibinfo {year} {2019})}\BibitemShut
  {NoStop}%
\bibitem [{\citenamefont {Oganesyan}\ and\ \citenamefont
  {Huse}(2007)}]{oganesyan07localization}%
  \BibitemOpen
  \bibfield  {author} {\bibinfo {author} {\bibfnamefont {V.}~\bibnamefont
  {Oganesyan}}\ and\ \bibinfo {author} {\bibfnamefont {D.~A.}\ \bibnamefont
  {Huse}},\ }\href {\doibase 10.1103/PhysRevB.75.155111} {\bibfield  {journal}
  {\bibinfo  {journal} {Phys. Rev. B}\ }\textbf {\bibinfo {volume} {75}},\
  \bibinfo {pages} {155111} (\bibinfo {year} {2007})}\BibitemShut {NoStop}%
\bibitem [{\citenamefont {Pal}\ and\ \citenamefont {Huse}(2010)}]{pal10many}%
  \BibitemOpen
  \bibfield  {author} {\bibinfo {author} {\bibfnamefont {A.}~\bibnamefont
  {Pal}}\ and\ \bibinfo {author} {\bibfnamefont {D.~A.}\ \bibnamefont {Huse}},\
  }\href {\doibase 10.1103/PhysRevB.82.174411} {\bibfield  {journal} {\bibinfo
  {journal} {Phys. Rev. B}\ }\textbf {\bibinfo {volume} {82}},\ \bibinfo
  {pages} {174411} (\bibinfo {year} {2010})}\BibitemShut {NoStop}%
\bibitem [{\citenamefont {Chiaracane}\ \emph {et~al.}(2020)\citenamefont
  {Chiaracane}, \citenamefont {Mitchison}, \citenamefont {Purkayastha},
  \citenamefont {Haack},\ and\ \citenamefont
  {Goold}}]{chiaracane20quasiperiodic}%
  \BibitemOpen
  \bibfield  {author} {\bibinfo {author} {\bibfnamefont {C.}~\bibnamefont
  {Chiaracane}}, \bibinfo {author} {\bibfnamefont {M.~T.}\ \bibnamefont
  {Mitchison}}, \bibinfo {author} {\bibfnamefont {A.}~\bibnamefont
  {Purkayastha}}, \bibinfo {author} {\bibfnamefont {G.}~\bibnamefont {Haack}},
  \ and\ \bibinfo {author} {\bibfnamefont {J.}~\bibnamefont {Goold}},\ }\href
  {\doibase 10.1103/PhysRevResearch.2.013093} {\bibfield  {journal} {\bibinfo
  {journal} {Phys. Rev. Research}\ }\textbf {\bibinfo {volume} {2}},\ \bibinfo
  {pages} {013093} (\bibinfo {year} {2020})}\BibitemShut {NoStop}%
\bibitem [{\citenamefont {Ganeshan}\ \emph {et~al.}(2015)\citenamefont
  {Ganeshan}, \citenamefont {Pixley},\ and\ \citenamefont
  {Das~Sarma}}]{ganeshan15nearest}%
  \BibitemOpen
  \bibfield  {author} {\bibinfo {author} {\bibfnamefont {S.}~\bibnamefont
  {Ganeshan}}, \bibinfo {author} {\bibfnamefont {J.~H.}\ \bibnamefont
  {Pixley}}, \ and\ \bibinfo {author} {\bibfnamefont {S.}~\bibnamefont
  {Das~Sarma}},\ }\href {\doibase 10.1103/PhysRevLett.114.146601} {\bibfield
  {journal} {\bibinfo  {journal} {Phys. Rev. Lett.}\ }\textbf {\bibinfo
  {volume} {114}},\ \bibinfo {pages} {146601} (\bibinfo {year}
  {2015})}\BibitemShut {NoStop}%
\bibitem [{\citenamefont {Purkayastha}\ \emph {et~al.}(2017)\citenamefont
  {Purkayastha}, \citenamefont {Dhar},\ and\ \citenamefont
  {Kulkarni}}]{archak17quasi}%
  \BibitemOpen
  \bibfield  {author} {\bibinfo {author} {\bibfnamefont {A.}~\bibnamefont
  {Purkayastha}}, \bibinfo {author} {\bibfnamefont {A.}~\bibnamefont {Dhar}}, \
  and\ \bibinfo {author} {\bibfnamefont {M.}~\bibnamefont {Kulkarni}},\ }\href
  {\doibase 10.1103/PhysRevB.96.180204} {\bibfield  {journal} {\bibinfo
  {journal} {Phys. Rev. B}\ }\textbf {\bibinfo {volume} {96}},\ \bibinfo
  {pages} {180204} (\bibinfo {year} {2017})}\BibitemShut {NoStop}%
\bibitem [{\citenamefont {Kim}\ \emph {et~al.}(2011)\citenamefont {Kim},
  \citenamefont {Sagawa}, \citenamefont {De~Liberato},\ and\ \citenamefont
  {Ueda}}]{kim11quantum}%
  \BibitemOpen
  \bibfield  {author} {\bibinfo {author} {\bibfnamefont {S.~W.}\ \bibnamefont
  {Kim}}, \bibinfo {author} {\bibfnamefont {T.}~\bibnamefont {Sagawa}},
  \bibinfo {author} {\bibfnamefont {S.}~\bibnamefont {De~Liberato}}, \ and\
  \bibinfo {author} {\bibfnamefont {M.}~\bibnamefont {Ueda}},\ }\href {\doibase
  10.1103/PhysRevLett.106.070401} {\bibfield  {journal} {\bibinfo  {journal}
  {Phys. Rev. Lett.}\ }\textbf {\bibinfo {volume} {106}},\ \bibinfo {pages}
  {070401} (\bibinfo {year} {2011})}\BibitemShut {NoStop}%
\bibitem [{\citenamefont {Lutz}\ and\ \citenamefont
  {Ciliberto}(2015)}]{lutz15information}%
  \BibitemOpen
  \bibfield  {author} {\bibinfo {author} {\bibfnamefont {E.}~\bibnamefont
  {Lutz}}\ and\ \bibinfo {author} {\bibfnamefont {S.}~\bibnamefont
  {Ciliberto}},\ }\href {\doibase 10.1063/PT.3.2912} {\bibfield  {journal}
  {\bibinfo  {journal} {Physics Today}\ }\textbf {\bibinfo {volume} {68}},\
  \bibinfo {pages} {30} (\bibinfo {year} {2015})},\ \Eprint
  {http://arxiv.org/abs/https://doi.org/10.1063/PT.3.2912}
  {https://doi.org/10.1063/PT.3.2912} \BibitemShut {NoStop}%
\bibitem [{\citenamefont {Mohammady}\ and\ \citenamefont
  {Anders}(2017)}]{mohammady17a}%
  \BibitemOpen
  \bibfield  {author} {\bibinfo {author} {\bibfnamefont {M.~H.}\ \bibnamefont
  {Mohammady}}\ and\ \bibinfo {author} {\bibfnamefont {J.}~\bibnamefont
  {Anders}},\ }\href {\doibase 10.1088/1367-2630/aa8ba1} {\bibfield  {journal}
  {\bibinfo  {journal} {New Journal of Physics}\ }\textbf {\bibinfo {volume}
  {19}},\ \bibinfo {pages} {113026} (\bibinfo {year} {2017})}\BibitemShut
  {NoStop}%
\bibitem [{\citenamefont {Bengtsson}\ \emph
  {et~al.}(2018{\natexlab{a}})\citenamefont {Bengtsson}, \citenamefont
  {Tengstrand}, \citenamefont {Wacker}, \citenamefont {Samuelsson},
  \citenamefont {Ueda}, \citenamefont {Linke},\ and\ \citenamefont
  {Reimann}}]{bengtsson18quantum}%
  \BibitemOpen
  \bibfield  {author} {\bibinfo {author} {\bibfnamefont {J.}~\bibnamefont
  {Bengtsson}}, \bibinfo {author} {\bibfnamefont {M.~N.}\ \bibnamefont
  {Tengstrand}}, \bibinfo {author} {\bibfnamefont {A.}~\bibnamefont {Wacker}},
  \bibinfo {author} {\bibfnamefont {P.}~\bibnamefont {Samuelsson}}, \bibinfo
  {author} {\bibfnamefont {M.}~\bibnamefont {Ueda}}, \bibinfo {author}
  {\bibfnamefont {H.}~\bibnamefont {Linke}}, \ and\ \bibinfo {author}
  {\bibfnamefont {S.~M.}\ \bibnamefont {Reimann}},\ }\href {\doibase
  10.1103/PhysRevLett.120.100601} {\bibfield  {journal} {\bibinfo  {journal}
  {Phys. Rev. Lett.}\ }\textbf {\bibinfo {volume} {120}},\ \bibinfo {pages}
  {100601} (\bibinfo {year} {2018}{\natexlab{a}})}\BibitemShut {NoStop}%
\bibitem [{\citenamefont {Bengtsson}\ \emph
  {et~al.}(2018{\natexlab{b}})\citenamefont {Bengtsson}, \citenamefont
  {Tengstrand},\ and\ \citenamefont {Reimann}}]{bengtsson18bosonic}%
  \BibitemOpen
  \bibfield  {author} {\bibinfo {author} {\bibfnamefont {J.}~\bibnamefont
  {Bengtsson}}, \bibinfo {author} {\bibfnamefont {M.~N.}\ \bibnamefont
  {Tengstrand}}, \ and\ \bibinfo {author} {\bibfnamefont {S.~M.}\ \bibnamefont
  {Reimann}},\ }\href {\doibase 10.1103/PhysRevA.97.062128} {\bibfield
  {journal} {\bibinfo  {journal} {Phys. Rev. A}\ }\textbf {\bibinfo {volume}
  {97}},\ \bibinfo {pages} {062128} (\bibinfo {year}
  {2018}{\natexlab{b}})}\BibitemShut {NoStop}%
\bibitem [{\citenamefont {B\'{e}rut}\ \emph {et~al.}(2012)\citenamefont
  {B\'{e}rut}, \citenamefont {Arakelyan}, \citenamefont {Petrosyan},
  \citenamefont {Ciliberto}, \citenamefont {Dillenschneider},\ and\
  \citenamefont {Lutz}}]{berut12experimental}%
  \BibitemOpen
  \bibfield  {author} {\bibinfo {author} {\bibfnamefont {A.}~\bibnamefont
  {B\'{e}rut}}, \bibinfo {author} {\bibfnamefont {A.}~\bibnamefont
  {Arakelyan}}, \bibinfo {author} {\bibfnamefont {A.}~\bibnamefont
  {Petrosyan}}, \bibinfo {author} {\bibfnamefont {S.}~\bibnamefont
  {Ciliberto}}, \bibinfo {author} {\bibfnamefont {R.}~\bibnamefont
  {Dillenschneider}}, \ and\ \bibinfo {author} {\bibfnamefont {E.}~\bibnamefont
  {Lutz}},\ }\href@noop {} {\bibfield  {journal} {\bibinfo  {journal} {Nature}\
  }\textbf {\bibinfo {volume} {403}},\ \bibinfo {pages} {187} (\bibinfo {year}
  {2012})}\BibitemShut {NoStop}%
\bibitem [{\citenamefont {Toyabe}\ \emph {et~al.}(2010)\citenamefont {Toyabe},
  \citenamefont {Sagawa}, \citenamefont {Ueda}, \citenamefont {Muneyuki},\ and\
  \citenamefont {Sano}}]{toyabe10experimental}%
  \BibitemOpen
  \bibfield  {author} {\bibinfo {author} {\bibfnamefont {S.}~\bibnamefont
  {Toyabe}}, \bibinfo {author} {\bibfnamefont {T.}~\bibnamefont {Sagawa}},
  \bibinfo {author} {\bibfnamefont {M.}~\bibnamefont {Ueda}}, \bibinfo {author}
  {\bibfnamefont {E.}~\bibnamefont {Muneyuki}}, \ and\ \bibinfo {author}
  {\bibfnamefont {M.}~\bibnamefont {Sano}},\ }\href {\doibase
  10.1038/nphys1821} {\bibfield  {journal} {\bibinfo  {journal} {Nature
  Physics}\ }\textbf {\bibinfo {volume} {6}},\ \bibinfo {pages} {988} (\bibinfo
  {year} {2010})}\BibitemShut {NoStop}%
\bibitem [{\citenamefont {Rold{\'a}n}\ \emph {et~al.}(2014)\citenamefont
  {Rold{\'a}n}, \citenamefont {Mart{\'i}nez}, \citenamefont {Parrondo},\ and\
  \citenamefont {Petrov}}]{roldan14universal}%
  \BibitemOpen
  \bibfield  {author} {\bibinfo {author} {\bibfnamefont {{\'E}.}~\bibnamefont
  {Rold{\'a}n}}, \bibinfo {author} {\bibfnamefont {I.~A.}\ \bibnamefont
  {Mart{\'i}nez}}, \bibinfo {author} {\bibfnamefont {J.~M.~R.}\ \bibnamefont
  {Parrondo}}, \ and\ \bibinfo {author} {\bibfnamefont {D.}~\bibnamefont
  {Petrov}},\ }\href {\doibase 10.1038/nphys2940} {\bibfield  {journal}
  {\bibinfo  {journal} {Nature Physics}\ }\textbf {\bibinfo {volume} {10}},\
  \bibinfo {pages} {457} (\bibinfo {year} {2014})}\BibitemShut {NoStop}%
\bibitem [{\citenamefont {Koski}\ \emph {et~al.}(2015)\citenamefont {Koski},
  \citenamefont {Kutvonen}, \citenamefont {Khaymovich}, \citenamefont
  {Ala-Nissila},\ and\ \citenamefont {Pekola}}]{koski15on}%
  \BibitemOpen
  \bibfield  {author} {\bibinfo {author} {\bibfnamefont {J.~V.}\ \bibnamefont
  {Koski}}, \bibinfo {author} {\bibfnamefont {A.}~\bibnamefont {Kutvonen}},
  \bibinfo {author} {\bibfnamefont {I.~M.}\ \bibnamefont {Khaymovich}},
  \bibinfo {author} {\bibfnamefont {T.}~\bibnamefont {Ala-Nissila}}, \ and\
  \bibinfo {author} {\bibfnamefont {J.~P.}\ \bibnamefont {Pekola}},\ }\href
  {\doibase 10.1103/PhysRevLett.115.260602} {\bibfield  {journal} {\bibinfo
  {journal} {Phys. Rev. Lett.}\ }\textbf {\bibinfo {volume} {115}},\ \bibinfo
  {pages} {260602} (\bibinfo {year} {2015})}\BibitemShut {NoStop}%
\bibitem [{\citenamefont {Bernien}\ \emph {et~al.}(2017)\citenamefont
  {Bernien}, \citenamefont {Schwartz}, \citenamefont {Keesling}, \citenamefont
  {Levine}, \citenamefont {Omran}, \citenamefont {Pichler}, \citenamefont
  {Choi}, \citenamefont {Zibrov}, \citenamefont {Endres}, \citenamefont
  {Greiner}, \citenamefont {Vuleti\'{c}},\ and\ \citenamefont
  {Lukin}}]{bernien17probing}%
  \BibitemOpen
  \bibfield  {author} {\bibinfo {author} {\bibfnamefont {H.}~\bibnamefont
  {Bernien}}, \bibinfo {author} {\bibfnamefont {S.}~\bibnamefont {Schwartz}},
  \bibinfo {author} {\bibfnamefont {A.}~\bibnamefont {Keesling}}, \bibinfo
  {author} {\bibfnamefont {H.}~\bibnamefont {Levine}}, \bibinfo {author}
  {\bibfnamefont {A.}~\bibnamefont {Omran}}, \bibinfo {author} {\bibfnamefont
  {H.}~\bibnamefont {Pichler}}, \bibinfo {author} {\bibfnamefont
  {S.}~\bibnamefont {Choi}}, \bibinfo {author} {\bibfnamefont {A.~S.}\
  \bibnamefont {Zibrov}}, \bibinfo {author} {\bibfnamefont {M.}~\bibnamefont
  {Endres}}, \bibinfo {author} {\bibfnamefont {M.}~\bibnamefont {Greiner}},
  \bibinfo {author} {\bibfnamefont {V.}~\bibnamefont {Vuleti\'{c}}}, \ and\
  \bibinfo {author} {\bibfnamefont {M.~D.}\ \bibnamefont {Lukin}},\ }\href
  {\doibase 10.1038/nature24622} {\bibfield  {journal} {\bibinfo  {journal}
  {Nature}\ }\textbf {\bibinfo {volume} {551}},\ \bibinfo {pages} {579}
  (\bibinfo {year} {2017})}\BibitemShut {NoStop}%
\bibitem [{\citenamefont {Ebadi}\ \emph {et~al.}(2020)\citenamefont {Ebadi},
  \citenamefont {Wang}, \citenamefont {Levine}, \citenamefont {Keesling},
  \citenamefont {Semeghini}, \citenamefont {Omran}, \citenamefont {Bluvstein},
  \citenamefont {Samajdar}, \citenamefont {Pichler}, \citenamefont {Ho},
  \citenamefont {Choi}, \citenamefont {Sachdev}, \citenamefont {Greiner},
  \citenamefont {Vuletic},\ and\ \citenamefont {Lukin}}]{ebadi20quantum}%
  \BibitemOpen
  \bibfield  {author} {\bibinfo {author} {\bibfnamefont {S.}~\bibnamefont
  {Ebadi}}, \bibinfo {author} {\bibfnamefont {T.~T.}\ \bibnamefont {Wang}},
  \bibinfo {author} {\bibfnamefont {H.}~\bibnamefont {Levine}}, \bibinfo
  {author} {\bibfnamefont {A.}~\bibnamefont {Keesling}}, \bibinfo {author}
  {\bibfnamefont {G.}~\bibnamefont {Semeghini}}, \bibinfo {author}
  {\bibfnamefont {A.}~\bibnamefont {Omran}}, \bibinfo {author} {\bibfnamefont
  {D.}~\bibnamefont {Bluvstein}}, \bibinfo {author} {\bibfnamefont
  {R.}~\bibnamefont {Samajdar}}, \bibinfo {author} {\bibfnamefont
  {H.}~\bibnamefont {Pichler}}, \bibinfo {author} {\bibfnamefont {W.~W.}\
  \bibnamefont {Ho}}, \bibinfo {author} {\bibfnamefont {S.}~\bibnamefont
  {Choi}}, \bibinfo {author} {\bibfnamefont {S.}~\bibnamefont {Sachdev}},
  \bibinfo {author} {\bibfnamefont {M.}~\bibnamefont {Greiner}}, \bibinfo
  {author} {\bibfnamefont {V.}~\bibnamefont {Vuletic}}, \ and\ \bibinfo
  {author} {\bibfnamefont {M.~D.}\ \bibnamefont {Lukin}},\ }\href@noop {}
  {\enquote {\bibinfo {title} {Quantum phases of matter on a 256-atom
  programmable quantum simulator},}\ } (\bibinfo {year} {2020}),\ \Eprint
  {http://arxiv.org/abs/2012.12281} {arXiv:2012.12281 [quant-ph]} \BibitemShut
  {NoStop}%
\bibitem [{\citenamefont {Valagiannopoulos}(2020)}]{constantinos20perfect}%
  \BibitemOpen
  \bibfield  {author} {\bibinfo {author} {\bibfnamefont {C.}~\bibnamefont
  {Valagiannopoulos}},\ }\href {\doibase 10.1103/PhysRevB.101.195301}
  {\bibfield  {journal} {\bibinfo  {journal} {Phys. Rev. B}\ }\textbf {\bibinfo
  {volume} {101}},\ \bibinfo {pages} {195301} (\bibinfo {year}
  {2020})}\BibitemShut {NoStop}%
\bibitem [{\citenamefont {Giovannetti}\ \emph {et~al.}(2011)\citenamefont
  {Giovannetti}, \citenamefont {Lloyd},\ and\ \citenamefont
  {Maccone}}]{giovannetti11}%
  \BibitemOpen
  \bibfield  {author} {\bibinfo {author} {\bibfnamefont {V.}~\bibnamefont
  {Giovannetti}}, \bibinfo {author} {\bibfnamefont {S.}~\bibnamefont {Lloyd}},
  \ and\ \bibinfo {author} {\bibfnamefont {L.}~\bibnamefont {Maccone}},\ }\href
  {\doibase 10.1038/nphoton.2011.35} {\bibfield  {journal} {\bibinfo  {journal}
  {Nature Photonics}\ }\textbf {\bibinfo {volume} {5}},\ \bibinfo {pages} {222}
  (\bibinfo {year} {2011})}\BibitemShut {NoStop}%
\bibitem [{\citenamefont {Joulain}\ \emph {et~al.}(2016)\citenamefont
  {Joulain}, \citenamefont {Drevillon}, \citenamefont {Ezzahri},\ and\
  \citenamefont {Ordonez-Miranda}}]{joulain16quantum}%
  \BibitemOpen
  \bibfield  {author} {\bibinfo {author} {\bibfnamefont {K.}~\bibnamefont
  {Joulain}}, \bibinfo {author} {\bibfnamefont {J.}~\bibnamefont {Drevillon}},
  \bibinfo {author} {\bibfnamefont {Y.}~\bibnamefont {Ezzahri}}, \ and\
  \bibinfo {author} {\bibfnamefont {J.}~\bibnamefont {Ordonez-Miranda}},\
  }\href {\doibase 10.1103/PhysRevLett.116.200601} {\bibfield  {journal}
  {\bibinfo  {journal} {Phys. Rev. Lett.}\ }\textbf {\bibinfo {volume} {116}},\
  \bibinfo {pages} {200601} (\bibinfo {year} {2016})}\BibitemShut {NoStop}%
\bibitem [{\citenamefont {Silva}\ \emph {et~al.}(2020)\citenamefont {Silva},
  \citenamefont {Landi}, \citenamefont {Drumond},\ and\ \citenamefont
  {Pereira}}]{silva20heat}%
  \BibitemOpen
  \bibfield  {author} {\bibinfo {author} {\bibfnamefont {S.~H.~S.}\
  \bibnamefont {Silva}}, \bibinfo {author} {\bibfnamefont {G.~T.}\ \bibnamefont
  {Landi}}, \bibinfo {author} {\bibfnamefont {R.~C.}\ \bibnamefont {Drumond}},
  \ and\ \bibinfo {author} {\bibfnamefont {E.}~\bibnamefont {Pereira}},\
  }\href@noop {} {\enquote {\bibinfo {title} {Heat rectitication on the xx
  chain},}\ } (\bibinfo {year} {2020}),\ \Eprint
  {http://arxiv.org/abs/2012.04811} {arXiv:2012.04811 [quant-ph]} \BibitemShut
  {NoStop}%
\bibitem [{\citenamefont {Bu\ifmmode~\check{z}\else \v{z}\fi{}ek}\ \emph
  {et~al.}(1999)\citenamefont {Bu\ifmmode~\check{z}\else \v{z}\fi{}ek},
  \citenamefont {Derka},\ and\ \citenamefont {Massar}}]{buzek99optimal}%
  \BibitemOpen
  \bibfield  {author} {\bibinfo {author} {\bibfnamefont {V.}~\bibnamefont
  {Bu\ifmmode~\check{z}\else \v{z}\fi{}ek}}, \bibinfo {author} {\bibfnamefont
  {R.}~\bibnamefont {Derka}}, \ and\ \bibinfo {author} {\bibfnamefont
  {S.}~\bibnamefont {Massar}},\ }\href {\doibase 10.1103/PhysRevLett.82.2207}
  {\bibfield  {journal} {\bibinfo  {journal} {Phys. Rev. Lett.}\ }\textbf
  {\bibinfo {volume} {82}},\ \bibinfo {pages} {2207} (\bibinfo {year}
  {1999})}\BibitemShut {NoStop}%
\bibitem [{\citenamefont {Erker}\ \emph {et~al.}(2017)\citenamefont {Erker},
  \citenamefont {Mitchison}, \citenamefont {Silva}, \citenamefont {Woods},
  \citenamefont {Brunner},\ and\ \citenamefont {Huber}}]{erker17autonomous}%
  \BibitemOpen
  \bibfield  {author} {\bibinfo {author} {\bibfnamefont {P.}~\bibnamefont
  {Erker}}, \bibinfo {author} {\bibfnamefont {M.~T.}\ \bibnamefont
  {Mitchison}}, \bibinfo {author} {\bibfnamefont {R.}~\bibnamefont {Silva}},
  \bibinfo {author} {\bibfnamefont {M.~P.}\ \bibnamefont {Woods}}, \bibinfo
  {author} {\bibfnamefont {N.}~\bibnamefont {Brunner}}, \ and\ \bibinfo
  {author} {\bibfnamefont {M.}~\bibnamefont {Huber}},\ }\href {\doibase
  10.1103/PhysRevX.7.031022} {\bibfield  {journal} {\bibinfo  {journal} {Phys.
  Rev. X}\ }\textbf {\bibinfo {volume} {7}},\ \bibinfo {pages} {031022}
  (\bibinfo {year} {2017})}\BibitemShut {NoStop}%
\bibitem [{\citenamefont {Pusz}\ and\ \citenamefont
  {Woronowicz}(1978)}]{pusz78}%
  \BibitemOpen
  \bibfield  {author} {\bibinfo {author} {\bibfnamefont {W.}~\bibnamefont
  {Pusz}}\ and\ \bibinfo {author} {\bibfnamefont {S.~L.}\ \bibnamefont
  {Woronowicz}},\ }\href@noop {} {\bibfield  {journal} {\bibinfo  {journal}
  {Communications in Mathematical Physics}\ }\textbf {\bibinfo {volume} {58}},\
  \bibinfo {pages} {273} (\bibinfo {year} {1978})}\BibitemShut {NoStop}%
\bibitem [{\citenamefont {Lenard}(1978)}]{lenard78}%
  \BibitemOpen
  \bibfield  {author} {\bibinfo {author} {\bibfnamefont {A.}~\bibnamefont
  {Lenard}},\ }\href@noop {} {\bibfield  {journal} {\bibinfo  {journal}
  {Journal of Statistical Physics}\ }\textbf {\bibinfo {volume} {19}},\
  \bibinfo {pages} {575} (\bibinfo {year} {1978})}\BibitemShut {NoStop}%
\bibitem [{\citenamefont {Crescente}\ \emph
  {et~al.}(2020{\natexlab{b}})\citenamefont {Crescente}, \citenamefont
  {Carrega}, \citenamefont {Sassetti},\ and\ \citenamefont
  {Ferraro}}]{crescente20ultrafast}%
  \BibitemOpen
  \bibfield  {author} {\bibinfo {author} {\bibfnamefont {A.}~\bibnamefont
  {Crescente}}, \bibinfo {author} {\bibfnamefont {M.}~\bibnamefont {Carrega}},
  \bibinfo {author} {\bibfnamefont {M.}~\bibnamefont {Sassetti}}, \ and\
  \bibinfo {author} {\bibfnamefont {D.}~\bibnamefont {Ferraro}},\ }\href
  {\doibase 10.1103/PhysRevB.102.245407} {\bibfield  {journal} {\bibinfo
  {journal} {Phys. Rev. B}\ }\textbf {\bibinfo {volume} {102}},\ \bibinfo
  {pages} {245407} (\bibinfo {year} {2020}{\natexlab{b}})}\BibitemShut
  {NoStop}%
\bibitem [{\citenamefont {Alicki}\ and\ \citenamefont
  {Fannes}(2013)}]{alicki13entanglement}%
  \BibitemOpen
  \bibfield  {author} {\bibinfo {author} {\bibfnamefont {R.}~\bibnamefont
  {Alicki}}\ and\ \bibinfo {author} {\bibfnamefont {M.}~\bibnamefont
  {Fannes}},\ }\href {\doibase 10.1103/PhysRevE.87.042123} {\bibfield
  {journal} {\bibinfo  {journal} {Phys. Rev. E}\ }\textbf {\bibinfo {volume}
  {87}},\ \bibinfo {pages} {042123} (\bibinfo {year} {2013})}\BibitemShut
  {NoStop}%
\bibitem [{\citenamefont {Rosa}\ \emph {et~al.}(2020)\citenamefont {Rosa},
  \citenamefont {Rossini}, \citenamefont {Andolina}, \citenamefont {Polini},\
  and\ \citenamefont {Carrega}}]{rosa20ultra}%
  \BibitemOpen
  \bibfield  {author} {\bibinfo {author} {\bibfnamefont {D.}~\bibnamefont
  {Rosa}}, \bibinfo {author} {\bibfnamefont {D.}~\bibnamefont {Rossini}},
  \bibinfo {author} {\bibfnamefont {G.~M.}\ \bibnamefont {Andolina}}, \bibinfo
  {author} {\bibfnamefont {M.}~\bibnamefont {Polini}}, \ and\ \bibinfo {author}
  {\bibfnamefont {M.}~\bibnamefont {Carrega}},\ }\href {\doibase
  10.1007/JHEP11(2020)067} {\bibfield  {journal} {\bibinfo  {journal} {Journal
  of High Energy Physics}\ }\textbf {\bibinfo {volume} {2020}},\ \bibinfo
  {pages} {67} (\bibinfo {year} {2020})}\BibitemShut {NoStop}%
\bibitem [{\citenamefont {PARIS}(2009)}]{paris09quantum}%
  \BibitemOpen
  \bibfield  {author} {\bibinfo {author} {\bibfnamefont {M.~G.~A.}\
  \bibnamefont {PARIS}},\ }\href {\doibase 10.1142/S0219749909004839}
  {\bibfield  {journal} {\bibinfo  {journal} {International Journal of Quantum
  Information}\ }\textbf {\bibinfo {volume} {07}},\ \bibinfo {pages} {125}
  (\bibinfo {year} {2009})}\BibitemShut {NoStop}%
\bibitem [{\citenamefont {Brunelli}\ \emph {et~al.}(2011)\citenamefont
  {Brunelli}, \citenamefont {Olivares},\ and\ \citenamefont
  {Paris}}]{brunelli11qubit}%
  \BibitemOpen
  \bibfield  {author} {\bibinfo {author} {\bibfnamefont {M.}~\bibnamefont
  {Brunelli}}, \bibinfo {author} {\bibfnamefont {S.}~\bibnamefont {Olivares}},
  \ and\ \bibinfo {author} {\bibfnamefont {M.~G.~A.}\ \bibnamefont {Paris}},\
  }\href@noop {} {\bibfield  {journal} {\bibinfo  {journal} {Phys. Rev. A}\
  }\textbf {\bibinfo {volume} {84}},\ \bibinfo {pages} {032105} (\bibinfo
  {year} {2011})}\BibitemShut {NoStop}%
\bibitem [{\citenamefont {Correa}\ \emph {et~al.}(2015)\citenamefont {Correa},
  \citenamefont {Mehboudi}, \citenamefont {Adesso},\ and\ \citenamefont
  {Sanpera}}]{correa15individual}%
  \BibitemOpen
  \bibfield  {author} {\bibinfo {author} {\bibfnamefont {L.~A.}\ \bibnamefont
  {Correa}}, \bibinfo {author} {\bibfnamefont {M.}~\bibnamefont {Mehboudi}},
  \bibinfo {author} {\bibfnamefont {G.}~\bibnamefont {Adesso}}, \ and\ \bibinfo
  {author} {\bibfnamefont {A.}~\bibnamefont {Sanpera}},\ }\href {\doibase
  10.1103/PhysRevLett.114.220405} {\bibfield  {journal} {\bibinfo  {journal}
  {Phys. Rev. Lett.}\ }\textbf {\bibinfo {volume} {114}},\ \bibinfo {pages}
  {220405} (\bibinfo {year} {2015})}\BibitemShut {NoStop}%
\bibitem [{\citenamefont {Correa}\ \emph {et~al.}(2017)\citenamefont {Correa},
  \citenamefont {Perarnau-Llobet}, \citenamefont {Hovhannisyan}, \citenamefont
  {Hern\'andez-Santana}, \citenamefont {Mehboudi},\ and\ \citenamefont
  {Sanpera}}]{correa17enhancement}%
  \BibitemOpen
  \bibfield  {author} {\bibinfo {author} {\bibfnamefont {L.~A.}\ \bibnamefont
  {Correa}}, \bibinfo {author} {\bibfnamefont {M.}~\bibnamefont
  {Perarnau-Llobet}}, \bibinfo {author} {\bibfnamefont {K.~V.}\ \bibnamefont
  {Hovhannisyan}}, \bibinfo {author} {\bibfnamefont {S.}~\bibnamefont
  {Hern\'andez-Santana}}, \bibinfo {author} {\bibfnamefont {M.}~\bibnamefont
  {Mehboudi}}, \ and\ \bibinfo {author} {\bibfnamefont {A.}~\bibnamefont
  {Sanpera}},\ }\href {\doibase 10.1103/PhysRevA.96.062103} {\bibfield
  {journal} {\bibinfo  {journal} {Phys. Rev. A}\ }\textbf {\bibinfo {volume}
  {96}},\ \bibinfo {pages} {062103} (\bibinfo {year} {2017})}\BibitemShut
  {NoStop}%
\bibitem [{\citenamefont {Hofer}\ \emph {et~al.}(2017)\citenamefont {Hofer},
  \citenamefont {Brask}, \citenamefont {Perarnau-Llobet},\ and\ \citenamefont
  {Brunner}}]{brunner17quantum}%
  \BibitemOpen
  \bibfield  {author} {\bibinfo {author} {\bibfnamefont {P.~P.}\ \bibnamefont
  {Hofer}}, \bibinfo {author} {\bibfnamefont {J.~B.}\ \bibnamefont {Brask}},
  \bibinfo {author} {\bibfnamefont {M.}~\bibnamefont {Perarnau-Llobet}}, \ and\
  \bibinfo {author} {\bibfnamefont {N.}~\bibnamefont {Brunner}},\ }\href
  {\doibase 10.1103/PhysRevLett.119.090603} {\bibfield  {journal} {\bibinfo
  {journal} {Phys. Rev. Lett.}\ }\textbf {\bibinfo {volume} {119}},\ \bibinfo
  {pages} {090603} (\bibinfo {year} {2017})}\BibitemShut {NoStop}%
\bibitem [{\citenamefont {De~Pasquale}\ and\ \citenamefont
  {Stace}(2018)}]{depasquale18quantum}%
  \BibitemOpen
  \bibfield  {author} {\bibinfo {author} {\bibfnamefont {A.}~\bibnamefont
  {De~Pasquale}}\ and\ \bibinfo {author} {\bibfnamefont {T.~M.}\ \bibnamefont
  {Stace}},\ }\enquote {\bibinfo {title} {Quantum thermometry},}\ in\ \href
  {\doibase 10.1007/978-3-319-99046-0_21} {\emph {\bibinfo {booktitle}
  {Thermodynamics in the Quantum Regime: Fundamental Aspects and New
  Directions}}},\ \bibinfo {editor} {edited by\ \bibinfo {editor}
  {\bibfnamefont {F.}~\bibnamefont {Binder}}, \bibinfo {editor} {\bibfnamefont
  {L.~A.}\ \bibnamefont {Correa}}, \bibinfo {editor} {\bibfnamefont
  {C.}~\bibnamefont {Gogolin}}, \bibinfo {editor} {\bibfnamefont
  {J.}~\bibnamefont {Anders}}, \ and\ \bibinfo {editor} {\bibfnamefont
  {G.}~\bibnamefont {Adesso}}}\ (\bibinfo  {publisher} {Springer International
  Publishing},\ \bibinfo {address} {Cham},\ \bibinfo {year} {2018})\ pp.\
  \bibinfo {pages} {503--527}\BibitemShut {NoStop}%
\bibitem [{\citenamefont {Mukherjee}\ \emph {et~al.}(2019)\citenamefont
  {Mukherjee}, \citenamefont {Zwick}, \citenamefont {Ghosh}, \citenamefont
  {Chen},\ and\ \citenamefont {Kurizki}}]{mukherjee19enhanced}%
  \BibitemOpen
  \bibfield  {author} {\bibinfo {author} {\bibfnamefont {V.}~\bibnamefont
  {Mukherjee}}, \bibinfo {author} {\bibfnamefont {A.}~\bibnamefont {Zwick}},
  \bibinfo {author} {\bibfnamefont {A.}~\bibnamefont {Ghosh}}, \bibinfo
  {author} {\bibfnamefont {X.}~\bibnamefont {Chen}}, \ and\ \bibinfo {author}
  {\bibfnamefont {G.}~\bibnamefont {Kurizki}},\ }\href {\doibase
  10.1038/s42005-019-0265-y} {\bibfield  {journal} {\bibinfo  {journal}
  {Communications Physics}\ }\textbf {\bibinfo {volume} {2}},\ \bibinfo {pages}
  {162} (\bibinfo {year} {2019})}\BibitemShut {NoStop}%
\bibitem [{\citenamefont {Bhattacharjee}\ \emph {et~al.}(2020)\citenamefont
  {Bhattacharjee}, \citenamefont {Bhattacharya}, \citenamefont {Niedenzu},
  \citenamefont {Mukherjee},\ and\ \citenamefont
  {Dutta}}]{bhattacharjee20quantum}%
  \BibitemOpen
  \bibfield  {author} {\bibinfo {author} {\bibfnamefont {S.}~\bibnamefont
  {Bhattacharjee}}, \bibinfo {author} {\bibfnamefont {U.}~\bibnamefont
  {Bhattacharya}}, \bibinfo {author} {\bibfnamefont {W.}~\bibnamefont
  {Niedenzu}}, \bibinfo {author} {\bibfnamefont {V.}~\bibnamefont {Mukherjee}},
  \ and\ \bibinfo {author} {\bibfnamefont {A.}~\bibnamefont {Dutta}},\ }\href
  {\doibase 10.1088/1367-2630/ab61d6} {\bibfield  {journal} {\bibinfo
  {journal} {New Journal of Physics}\ }\textbf {\bibinfo {volume} {22}},\
  \bibinfo {pages} {013024} (\bibinfo {year} {2020})}\BibitemShut {NoStop}%
\bibitem [{\citenamefont {Levy}\ \emph {et~al.}(2020)\citenamefont {Levy},
  \citenamefont {Göb}, \citenamefont {Deng}, \citenamefont {Singer},
  \citenamefont {Torrontegui},\ and\ \citenamefont {Wang}}]{levy20single}%
  \BibitemOpen
  \bibfield  {author} {\bibinfo {author} {\bibfnamefont {A.}~\bibnamefont
  {Levy}}, \bibinfo {author} {\bibfnamefont {M.}~\bibnamefont {Göb}}, \bibinfo
  {author} {\bibfnamefont {B.}~\bibnamefont {Deng}}, \bibinfo {author}
  {\bibfnamefont {K.}~\bibnamefont {Singer}}, \bibinfo {author} {\bibfnamefont
  {E.}~\bibnamefont {Torrontegui}}, \ and\ \bibinfo {author} {\bibfnamefont
  {D.}~\bibnamefont {Wang}},\ }\href {\doibase 10.1088/1367-2630/abad7f}
  {\bibfield  {journal} {\bibinfo  {journal} {New Journal of Physics}\ }\textbf
  {\bibinfo {volume} {22}},\ \bibinfo {pages} {093020} (\bibinfo {year}
  {2020})}\BibitemShut {NoStop}%
\bibitem [{\citenamefont {Campbell}\ \emph {et~al.}(2018)\citenamefont
  {Campbell}, \citenamefont {Genoni},\ and\ \citenamefont
  {Deffner}}]{campbell18precision}%
  \BibitemOpen
  \bibfield  {author} {\bibinfo {author} {\bibfnamefont {S.}~\bibnamefont
  {Campbell}}, \bibinfo {author} {\bibfnamefont {M.~G.}\ \bibnamefont
  {Genoni}}, \ and\ \bibinfo {author} {\bibfnamefont {S.}~\bibnamefont
  {Deffner}},\ }\href@noop {} {\bibfield  {journal} {\bibinfo  {journal}
  {Quantum Science and Technology}\ }\textbf {\bibinfo {volume} {3}},\ \bibinfo
  {pages} {025002} (\bibinfo {year} {2018})}\BibitemShut {NoStop}%
\bibitem [{\citenamefont {Peres}(1980)}]{peres80measurement}%
  \BibitemOpen
  \bibfield  {author} {\bibinfo {author} {\bibfnamefont {A.}~\bibnamefont
  {Peres}},\ }\href {\doibase 10.1119/1.12061} {\bibfield  {journal} {\bibinfo
  {journal} {American Journal of Physics}\ }\textbf {\bibinfo {volume} {48}},\
  \bibinfo {pages} {552} (\bibinfo {year} {1980})},\ \Eprint
  {http://arxiv.org/abs/https://doi.org/10.1119/1.12061}
  {https://doi.org/10.1119/1.12061} \BibitemShut {NoStop}%
\bibitem [{\citenamefont {Zhang}\ \emph {et~al.}(2017)\citenamefont {Zhang},
  \citenamefont {Pagano}, \citenamefont {Hess}, \citenamefont {Kyprianidis},
  \citenamefont {Becker}, \citenamefont {Kaplan}, \citenamefont {Gorshkov},
  \citenamefont {Gong},\ and\ \citenamefont {Monroe}}]{zhang17observation}%
  \BibitemOpen
  \bibfield  {author} {\bibinfo {author} {\bibfnamefont {J.}~\bibnamefont
  {Zhang}}, \bibinfo {author} {\bibfnamefont {G.}~\bibnamefont {Pagano}},
  \bibinfo {author} {\bibfnamefont {P.~W.}\ \bibnamefont {Hess}}, \bibinfo
  {author} {\bibfnamefont {A.}~\bibnamefont {Kyprianidis}}, \bibinfo {author}
  {\bibfnamefont {P.}~\bibnamefont {Becker}}, \bibinfo {author} {\bibfnamefont
  {H.}~\bibnamefont {Kaplan}}, \bibinfo {author} {\bibfnamefont {A.~V.}\
  \bibnamefont {Gorshkov}}, \bibinfo {author} {\bibfnamefont {Z.-X.}\
  \bibnamefont {Gong}}, \ and\ \bibinfo {author} {\bibfnamefont
  {C.}~\bibnamefont {Monroe}},\ }\href {\doibase 10.1038/nature24654}
  {\bibfield  {journal} {\bibinfo  {journal} {Nature}\ }\textbf {\bibinfo
  {volume} {551}},\ \bibinfo {pages} {601} (\bibinfo {year}
  {2017})}\BibitemShut {NoStop}%
\bibitem [{\citenamefont {Cui}\ \emph {et~al.}(2016)\citenamefont {Cui},
  \citenamefont {Huang}, \citenamefont {Wang}, \citenamefont {Cao},
  \citenamefont {Wang}, \citenamefont {Lv}, \citenamefont {Luo}, \citenamefont
  {del Campo}, \citenamefont {Han}, \citenamefont {Li},\ and\ \citenamefont
  {Guo}}]{cui2016experimental}%
  \BibitemOpen
  \bibfield  {author} {\bibinfo {author} {\bibfnamefont {J.-M.}\ \bibnamefont
  {Cui}}, \bibinfo {author} {\bibfnamefont {Y.-F.}\ \bibnamefont {Huang}},
  \bibinfo {author} {\bibfnamefont {Z.}~\bibnamefont {Wang}}, \bibinfo {author}
  {\bibfnamefont {D.-Y.}\ \bibnamefont {Cao}}, \bibinfo {author} {\bibfnamefont
  {J.}~\bibnamefont {Wang}}, \bibinfo {author} {\bibfnamefont {W.-M.}\
  \bibnamefont {Lv}}, \bibinfo {author} {\bibfnamefont {L.}~\bibnamefont
  {Luo}}, \bibinfo {author} {\bibfnamefont {A.}~\bibnamefont {del Campo}},
  \bibinfo {author} {\bibfnamefont {Y.-J.}\ \bibnamefont {Han}}, \bibinfo
  {author} {\bibfnamefont {C.-F.}\ \bibnamefont {Li}}, \ and\ \bibinfo {author}
  {\bibfnamefont {G.-C.}\ \bibnamefont {Guo}},\ }\href {\doibase
  10.1038/srep33381} {\bibfield  {journal} {\bibinfo  {journal} {Scientific
  Reports}\ }\textbf {\bibinfo {volume} {6}},\ \bibinfo {pages} {33381}
  (\bibinfo {year} {2016})}\BibitemShut {NoStop}%
\bibitem [{\citenamefont {Bando}\ \emph {et~al.}(2020)\citenamefont {Bando},
  \citenamefont {Susa}, \citenamefont {Oshiyama}, \citenamefont {Shibata},
  \citenamefont {Ohzeki}, \citenamefont {G\'omez-Ruiz}, \citenamefont {Lidar},
  \citenamefont {Suzuki}, \citenamefont {del Campo},\ and\ \citenamefont
  {Nishimori}}]{bando20probing}%
  \BibitemOpen
  \bibfield  {author} {\bibinfo {author} {\bibfnamefont {Y.}~\bibnamefont
  {Bando}}, \bibinfo {author} {\bibfnamefont {Y.}~\bibnamefont {Susa}},
  \bibinfo {author} {\bibfnamefont {H.}~\bibnamefont {Oshiyama}}, \bibinfo
  {author} {\bibfnamefont {N.}~\bibnamefont {Shibata}}, \bibinfo {author}
  {\bibfnamefont {M.}~\bibnamefont {Ohzeki}}, \bibinfo {author} {\bibfnamefont
  {F.~J.}\ \bibnamefont {G\'omez-Ruiz}}, \bibinfo {author} {\bibfnamefont
  {D.~A.}\ \bibnamefont {Lidar}}, \bibinfo {author} {\bibfnamefont
  {S.}~\bibnamefont {Suzuki}}, \bibinfo {author} {\bibfnamefont
  {A.}~\bibnamefont {del Campo}}, \ and\ \bibinfo {author} {\bibfnamefont
  {H.}~\bibnamefont {Nishimori}},\ }\href {\doibase
  10.1103/PhysRevResearch.2.033369} {\bibfield  {journal} {\bibinfo  {journal}
  {Phys. Rev. Research}\ }\textbf {\bibinfo {volume} {2}},\ \bibinfo {pages}
  {033369} (\bibinfo {year} {2020})}\BibitemShut {NoStop}%
\bibitem [{\citenamefont {Kaufman}\ \emph {et~al.}(2016)\citenamefont
  {Kaufman}, \citenamefont {Tai}, \citenamefont {Lukin}, \citenamefont
  {Rispoli}, \citenamefont {Schittko}, \citenamefont {Preiss},\ and\
  \citenamefont {Greiner}}]{kaufman16}%
  \BibitemOpen
  \bibfield  {author} {\bibinfo {author} {\bibfnamefont {A.~M.}\ \bibnamefont
  {Kaufman}}, \bibinfo {author} {\bibfnamefont {M.~E.}\ \bibnamefont {Tai}},
  \bibinfo {author} {\bibfnamefont {A.}~\bibnamefont {Lukin}}, \bibinfo
  {author} {\bibfnamefont {M.}~\bibnamefont {Rispoli}}, \bibinfo {author}
  {\bibfnamefont {R.}~\bibnamefont {Schittko}}, \bibinfo {author}
  {\bibfnamefont {P.~M.}\ \bibnamefont {Preiss}}, \ and\ \bibinfo {author}
  {\bibfnamefont {M.}~\bibnamefont {Greiner}},\ }\href {\doibase
  10.1126/science.aaf6725} {\bibfield  {journal} {\bibinfo  {journal}
  {Science}\ }\textbf {\bibinfo {volume} {353}},\ \bibinfo {pages} {794}
  (\bibinfo {year} {2016})}\BibitemShut {NoStop}%
\bibitem [{\citenamefont {Norcia}\ \emph {et~al.}(2018)\citenamefont {Norcia},
  \citenamefont {Lewis-Swan}, \citenamefont {Cline}, \citenamefont {Zhu},
  \citenamefont {Rey},\ and\ \citenamefont {Thompson}}]{norcia18cavity}%
  \BibitemOpen
  \bibfield  {author} {\bibinfo {author} {\bibfnamefont {M.~A.}\ \bibnamefont
  {Norcia}}, \bibinfo {author} {\bibfnamefont {R.~J.}\ \bibnamefont
  {Lewis-Swan}}, \bibinfo {author} {\bibfnamefont {J.~R.~K.}\ \bibnamefont
  {Cline}}, \bibinfo {author} {\bibfnamefont {B.}~\bibnamefont {Zhu}}, \bibinfo
  {author} {\bibfnamefont {A.~M.}\ \bibnamefont {Rey}}, \ and\ \bibinfo
  {author} {\bibfnamefont {J.~K.}\ \bibnamefont {Thompson}},\ }\href {\doibase
  10.1126/science.aar3102} {\bibfield  {journal} {\bibinfo  {journal}
  {Science}\ }\textbf {\bibinfo {volume} {361}},\ \bibinfo {pages} {259}
  (\bibinfo {year} {2018})},\ \Eprint
  {http://arxiv.org/abs/https://science.sciencemag.org/content/361/6399/259.full.pdf}
  {https://science.sciencemag.org/content/361/6399/259.full.pdf} \BibitemShut
  {NoStop}%
\bibitem [{\citenamefont {Barberena}\ \emph {et~al.}(2019)\citenamefont
  {Barberena}, \citenamefont {Lewis-Swan}, \citenamefont {Thompson},\ and\
  \citenamefont {Rey}}]{barberena19driven}%
  \BibitemOpen
  \bibfield  {author} {\bibinfo {author} {\bibfnamefont {D.}~\bibnamefont
  {Barberena}}, \bibinfo {author} {\bibfnamefont {R.~J.}\ \bibnamefont
  {Lewis-Swan}}, \bibinfo {author} {\bibfnamefont {J.~K.}\ \bibnamefont
  {Thompson}}, \ and\ \bibinfo {author} {\bibfnamefont {A.~M.}\ \bibnamefont
  {Rey}},\ }\href {\doibase 10.1103/PhysRevA.99.053411} {\bibfield  {journal}
  {\bibinfo  {journal} {Phys. Rev. A}\ }\textbf {\bibinfo {volume} {99}},\
  \bibinfo {pages} {053411} (\bibinfo {year} {2019})}\BibitemShut {NoStop}%
\bibitem [{\citenamefont {Tucker}\ \emph {et~al.}(2020)\citenamefont {Tucker},
  \citenamefont {Barberena}, \citenamefont {Lewis-Swan}, \citenamefont
  {Thompson}, \citenamefont {Restrepo},\ and\ \citenamefont
  {Rey}}]{tucker20facilitating}%
  \BibitemOpen
  \bibfield  {author} {\bibinfo {author} {\bibfnamefont {K.}~\bibnamefont
  {Tucker}}, \bibinfo {author} {\bibfnamefont {D.}~\bibnamefont {Barberena}},
  \bibinfo {author} {\bibfnamefont {R.~J.}\ \bibnamefont {Lewis-Swan}},
  \bibinfo {author} {\bibfnamefont {J.~K.}\ \bibnamefont {Thompson}}, \bibinfo
  {author} {\bibfnamefont {J.~G.}\ \bibnamefont {Restrepo}}, \ and\ \bibinfo
  {author} {\bibfnamefont {A.~M.}\ \bibnamefont {Rey}},\ }\href {\doibase
  10.1103/PhysRevA.102.051701} {\bibfield  {journal} {\bibinfo  {journal}
  {Phys. Rev. A}\ }\textbf {\bibinfo {volume} {102}},\ \bibinfo {pages}
  {051701} (\bibinfo {year} {2020})}\BibitemShut {NoStop}%
\bibitem [{\citenamefont {Karg}\ \emph {et~al.}(2020)\citenamefont {Karg},
  \citenamefont {Gouraud}, \citenamefont {Ngai}, \citenamefont {Schmid},
  \citenamefont {Hammerer},\ and\ \citenamefont {Treutlein}}]{karg20light}%
  \BibitemOpen
  \bibfield  {author} {\bibinfo {author} {\bibfnamefont {T.~M.}\ \bibnamefont
  {Karg}}, \bibinfo {author} {\bibfnamefont {B.}~\bibnamefont {Gouraud}},
  \bibinfo {author} {\bibfnamefont {C.~T.}\ \bibnamefont {Ngai}}, \bibinfo
  {author} {\bibfnamefont {G.-L.}\ \bibnamefont {Schmid}}, \bibinfo {author}
  {\bibfnamefont {K.}~\bibnamefont {Hammerer}}, \ and\ \bibinfo {author}
  {\bibfnamefont {P.}~\bibnamefont {Treutlein}},\ }\href {\doibase
  10.1126/science.abb0328} {\bibfield  {journal} {\bibinfo  {journal}
  {Science}\ }\textbf {\bibinfo {volume} {369}},\ \bibinfo {pages} {174}
  (\bibinfo {year} {2020})},\ \Eprint
  {http://arxiv.org/abs/https://science.sciencemag.org/content/369/6500/174.full.pdf}
  {https://science.sciencemag.org/content/369/6500/174.full.pdf} \BibitemShut
  {NoStop}%
\end{thebibliography}

%
\end{document}